\documentclass[a4paper,10pt]{article}
\usepackage[dvips]{graphicx}
\usepackage[dvipdfm]{hyperref}
\usepackage{latexsym}
\usepackage{amsmath,amssymb,epsfig,float}
\newcommand{\nn}{\nonumber}
\newcommand{\be}{\begin{equation}}
\newcommand{\ee}{\end{equation}}
\newcommand{\ba}{\begin{array}}
\newcommand{\bqa}{\begin{eqnarray}}
\newcommand{\eqa}{\end{eqnarray}}
\newcommand{\cO}{{\cal O}}
\newcommand{\mL}{\mathcal{L}}

\newcommand{\ket}{\,\rangle}
\newcommand{\bra}{\langle \,}

\begin{document}
\title{ \bf \boldmath Chiral dynamics in form factors, spectral-function sum rules, meson-meson scattering and  
semilocal duality }
  \author{Zhi-Hui~Guo$^{a,b,}$\thanks{guo@um.es}, J.~A.~Oller$^{b,}$\thanks{oller@um.es}  and J.~Ruiz de Elvira$^{c,}$\thanks{jacobore@rect.ucm.es} 
  \vspace{0.3cm} \\
  {\footnotesize $^a$ Department of Physics, Hebei Normal University, 050024 Shijiazhuang, P.~R.~China. }
  \\ {\footnotesize $^b$ Departamento de F\'isica, Universidad de Murcia,  E-30071 Murcia, Spain. }
 \\ {\footnotesize $^c$ Departamento de F\'isica Te\'orica II, Universidad Complutense de Madrid, E-28040 Madrid, Spain. }
 }

\date{}
\maketitle

\begin{abstract}
In this work, we perform the one-loop calculation of the scalar and pseudoscalar form factors 
in the framework of $U(3)$ chiral perturbation theory with explicit tree level exchanges of resonances. 
The meson-meson scattering calculation from Ref.~\cite{guo11prd} is extended as well. 
The spectral functions of the nonet scalar-scalar ($SS$) and 
pseudoscalar-pseudoscalar ($PP$) correlators are constructed by using the corresponding form factors. 
After fitting the unknown parameters to the scattering
data, we discuss the resonance content of the resulting scattering amplitudes. 
We also study  spectral-function sum rules in the $SS$-$SS$, $PP$-$PP$ and $SS$-$PP$ 
sectors as well as semilocal duality from scattering. The former relate the scalar and pseudoscalar spectra between themselves while 
the latter mainly connects the scalar spectrum with the vector one.  Finally we investigate these items as a function of $N_C$ 
for  $N_C > 3$. All these results pose strong constraints on the scalar dynamics and spectroscopy that are discussed. 
They are successfully fulfilled by our meson-meson scattering amplitudes and spectral functions. 
\end{abstract}


%%%%%%%%%%%%%%%%%%%%%%%%%%%%%%%%%%%%%%%%%%%%%%%%%%%%%%%%%%%%%%%%%%%%%%%%%%%%%%%%%%%%%%%%%%%%%%%%%%%%%%%%
\newpage

\tableofcontents

\newpage

\section{Introduction}

Spectral functions of current-current correlators are interesting objects in hadron physics. 
They are sensitive to both low and high energy dynamics of QCD, which can be evaluated in two 
reliable approaches: chiral perturbation theory ($\chi$PT) \cite{weinberg79,gasser84,gasser85} in the low-energy region 
and the operator product expansion (OPE) \cite{wi69,pascual} for the higher one.   
The celebrated Weinberg sum rules are in fact derived by studying 
the differences of the vector-vector and axial-vector-axial-vector spectral functions~\cite{weinberg67prl}, where 
not only the low and asymptotic energy regions but also the resonance region are considered. 
After then, great progresses, both on the experimental and  theoretical sides~\cite{davier06rmp}, 
have been made along this research line. The vector and axial-vector spectral functions are experimentally measurable quantities,  
mainly through the $\tau$ decays, due to the noticeable $V-A$ nature of the Standard Model in the energy region well below 
the $W$ boson mass.

Concerning the scalar and pseudoscalar spectral functions, there are no direct experimental data. 
However, great efforts have been made on the theoretical sides both in  
perturbative QCD calculations~\cite{narison83npb,jamin92zpc,baikov06prl}, which are important to reduce the QCD background in 
the search of the Higgs particle, and in the nonperturbative QCD region 
~\cite{gasser84,moussallam00epjc,moussallam00jhep,peris98jhep,bijnens01jhep,bijnens03jhep,juanjo10prd,jamin02epjc,dominguez08plb,golterman00prd}. 
Through the matching between the low and high energy behaviors of the spectral functions, valuable information 
can be obtained: light quark masses and Cabibbo angle $|V_{us}|$ ~\cite{narison83npb,jamin02epjc,dominguez08plb}, 
determination of the low energy constants (LECs) in $\chi$PT~\cite{moussallam00epjc,moussallam00jhep,bijnens03jhep,juanjo10prd,jamin04} and 
constraints on resonance parameters~\cite{peris98jhep,bijnens03jhep,juanjo10prd,golterman00prd,jamin00npb,jamin02npb}.

On the other hand, in the past decade scalar dynamics in the nonperturbative QCD region has been greatly put forward 
thanks to the combination of $\chi$PT and nonperturbative approaches from the $S$-matrix theory
~\cite{kaiser95,oller97npa,oller99prd,madrid,pavon1,leutwyler06prl,descotes06epjc,zheng01npa,zheng04npa,zhou11prd,nieves11prd,albaladejo08prl}. 
One of the main results is the reappearance of the broad $f_0(500)$ (traditionally called $\sigma$), the lightest resonance in QCD spectrum, after its long absence in 
the PDG list~\cite{pdg}. Later on the $K^*_0(800)$ or $\kappa$ resonance in $K\pi$ scattering was also confirmed~\cite{pdg}. 
Though the mass and width of these broad scalar resonances predicted by different groups are rather consistent among each other, 
other properties about the resonances in the scalar family are still under a vivid debate (e.g. their nature).

The scalar-scalar ($SS$) and pseudoscalar-pseudoscalar ($PP$) correlation functions provide us
another theoretical tool to further study the scalar resonances~\cite{guo12plb}, 
once the phase shifts and inelasticities in meson-meson scattering are well 
reproduced  \cite{oller97npa,oller99prd,madrid,leutwyler06prl,descotes06epjc,zheng01npa,zheng04npa,zhou11prd,nieves11prd,albaladejo08prl,albaladejo12}. 
As derived in Refs.~\cite{BW,moussallam00epjc,moussallam00jhep,LSR,golterman00prd} 
these correlator functions should fulfill a set of spectral-function sum rules that imply tight constraints to the scalar 
and pseudoscalar spectra, involving nontrivial relations between them. 

We take, and further study,  the meson-meson scattering amplitudes of Ref.~\cite{guo11prd} that are calculated in one-loop $U(3)$ $\chi$PT plus the explicit 
scalar and vector resonance exchanges at tree level within the framework of resonance chiral theory (R$\chi$T)~\cite{ecker89npb}.
As a novelty, we include the contributions from the tree level exchange of pseudoscalar resonances in this work so that a new fit to data is also discussed. 
The unitarized $U(3)$ $\chi$PT amplitudes are more appropriate to incorporate the heavier $f_0$ scalar 
resonances since the channels involving the $\eta$ and $\eta'$ mesons can be taken into account (while $SU(3)$ $\chi$PT~\cite{gasser85} only 
contains the pure octet $\eta_8$ field).  As noticed in Ref.~\cite{oller99prd},  further developed later 
in Refs.~\cite{pelaezprl1,pelaezprl2,sun07mpl,xiao07mpa,guo07jhep,nieves09prd,Dai:2011bs,nieves11prd,guo11prd}, 
the study of the resonance pole trajectories with increasing number of colors of QCD, $N_C$, is of interest 
to discern possible natures of the scalar resonances. E.g. a $\bar{q}q$ resonance evolves with a mass $M\sim {\cal O}(N_C^0)$, 
while its width decreases as $\Gamma\sim {\cal O}(N_C^{-1})$. For the case of a glueball  its mass behaves also as $M \sim {\cal O}(N_C^0)$ 
with a width that decreases faster with $N_C$ as $\Gamma\sim {\cal O}(N_C^{-2})$ \cite{largenc,manohar}. 
In this respect $U(3)$ $\chi$PT is also more adequate  than the $SU(3)$ version to discuss the large $N_C$ limit because   
 $U(3)$ $\chi$PT includes not only the pseudo-Goldstone octet, as $SU(3)$ does, but also the singlet $\eta_1$ that in the chiral limit is the 
ninth Goldstone boson at large $N_C$~\cite{ua1innc}. In this way, $U(3)$ $\chi$PT has the appropriate low energy degrees of freedom in the chiral limit at  
large $N_C$. Additionally, as stressed in Ref.~\cite{guo11prd}, the $\eta$ meson quickly becomes 
much lighter  with increasing $N_C$ \cite{wein2}. This fact has a strong impact on the dependence with $N_C$ of the 
pole trajectories of the scalar resonances, except for the $f_0(500)$, because of its weak  couplings to the channels with $\eta$ and $\eta'$ \cite{guo11prd}. 
We show below the $N_C$ pole trajectories for the different resonances involved in our study from the new fit to data.

Another interesting property of strong interactions that can be used to restrict meson-meson scattering 
is average or semilocal duality \cite{collinsbook}.
It allows one to relate the crossed channel dynamics with the $s$-channel one, 
which supplies another stringent constraint on the resonance properties involved in scattering. 
In a recent work~\cite{pelaez11prd}, Pel\'aez {\it et al.} have investigated how semilocal duality between the light resonances of QCD and 
Regge theory can be satisfied in $\pi\pi$ scattering. These authors employ the Inverse Amplitude Method (IAM) 
\cite{truong,dobado,dobado97,pelaezoller} to unitarize the $SU(3)$ one-loop and $SU(2)$ two-loop $\chi$PT amplitudes. 
The main conclusion of Ref.~\cite{pelaez11prd} is that in the physical case, i.e. when $N_C=3$, the $f_0(500)$ resonance, with its pole position around 
$440 - i\, 250$~MeV, plays a crucial role to oppose the vector strength from the $\rho(770)$ resonance, which guarantees the fulfillment of 
semilocal duality. 

Having introduced the main research topics considered here for the sake of clarity we summarize our work and briefly comment on the main points and results obtained by considering these topics. A complete one-loop calculation of
the scalar and pseudoscalar form factors within  $U(3)$ unitary $\chi$PT, including the tree-level exchange of scalar, pseudoscalar and vector resonances, is undertaken.
The spectral functions of the $SS$ and $PP$ two-point correlators are constructed by using the resulting scalar and pseudoscalar form factors, respectively, which are unitarized for the case of the scalar ones.
After updating the fit in Ref.~\cite{guo11prd}, which is also extended by including the explicit exchange of pseudoscalar resonances,
we study the resonance spectroscopy, quadratic pion scalar radius, the fulfillment of semilocal duality 
and the spectral function sum rules in the $SS-SS$, $PP-PP$ and $SS-PP$ cases, which are remarkably well satisfied simultaneously.
We show that it is important to take under consideration the high energy constraint for the coupling of the vector resonances to pseudo-Goldstone bosons, in order to keep semilocal duality when varying $N_C$. An interesting interplay between different resonances when studying the $N_C$ evolution of semilocal duality and the spectral function sum rules is revealed. In the former case the scalar and vector spectra appear  tightly related and in the latter one the same can be stated for the scalar and pseudoscalar ones.  This issue is closely connected with the evolution of the resonance pole position with varying $N_C$ of the pseudoscalar, vector and scalar resonances.

In this respect, for larger $N_C$, the authors of Ref.~\cite{pelaez11prd} revealed that a $\bar{q}q$ subdominant component for the $f_0(500)$, with a mass around 
1 GeV and obtained for some sets of LECs in both one-loop and two-loop calculations, is needed in order to fulfill semilocal duality. 
For those solutions, the $f_0(500)$ pole moves away from the 400-700~MeV region of the real axis, 
but it turns around moving back towards the real axis above 1 GeV as $N_C$ becomes larger than 8 or so.
The interpretation of these results was studied preliminarily in Ref.~\cite{LlanesEstrada:2011kz}, 
where the $f_0(500)$ is considered as a combination of several states and it was found that the weight of the subdominant $\bar qq$ component in the 
$f_0(500)$ grows with increasing $N_C$.  Immediately, one can ask a question: apart from the $f_0(500)$ resonance, which is only marginally contributed by this $\bar{q}q$ component, are there any other more obvious effects from the $\bar{q}q$ seed at $N_C=3$? If there are, what kind of role do these effects play 
in the fulfillment of semilocal duality?

Concerning the first question, we provide here, confirming Refs.~\cite{oller99prd,guo11prd}, a scenario for an affirmative answer. 
 A bare singlet scalar resonance $S_1$, with its bare mass around 1 GeV, is found significant \cite{guo11prd} to reproduce 
the $\pi\pi$ and $K\bar{K}$ scattering data around the $f_0(980)$ energy region and, thus, becomes an important part of 
the $f_0(980)$ resonance in the physical case, i.e. at $N_C=3$. When increasing $N_C$, the physical $f_0(980)$ resonance, gradually evolves 
to the singlet $S_1$, which acts like the role of the $\bar{q}q$ seed described in Refs.~\cite{pelaez11prd,LlanesEstrada:2011kz}. 
Apparently, the scenario that we provide now is different from that in Ref.~\cite{pelaez11prd}, since 
in the latter reference the signal of the $f_0(980)$ resonance gradually disappears when increasing $N_C$.\footnote{See the left panel of Fig.1 in Ref.~\cite{pelaez11prd}. It is verified that the peak around 1 GeV$^2$ finally disappears with larger values of $N_C$.} Regarding the second question raised in the previous paragraph we find here that its role for fulfilling the requirements of semilocal duality becomes more important with increasing $N_C$, being only of little importance at $N_C=3$. This behavior is in qualitative agreement with the results of \cite{pelaez11prd}.

In addition to the results presented in our short letter~\cite{guo12plb}, 
we show our theoretical calculations in detail and discuss more phenomenological materials in this work, 
such as the fit results, resonance pole positions and their coupling strengths, $N_C$ trajectories for all the relevant resonances and so on. 
The contents of this article are organized as follows. Section \ref{sect.chirallag} is devoted to 
the introduction of the relevant chiral Lagrangian. In Sections \ref{sect.specfandff} and \ref{sect.semilocalintro}, 
we present the details of the calculations of the spectral functions and the quantities to measure 
the degree of fulfillment of semilocal duality, respectively. The phenomenological discussions are given in Section \ref{sect.discussion}, which 
include the fit of unknown parameters to the experimental data and its consequences on the form factors, 
spectral functions, spectral-function sum rules and  semilocal duality. 
Section \ref{sect.ncrunning} is then devoted to the study of the $N_C$ evolution of these quantities. 
Finally, we conclude in Section \ref{sect.conclusion}.

%%%%%%%%%%%%%%%%%%%%%%%%%%%%%%%%%%%%%%%%%%%%%%%%%%%%%%%%%%%%%%%%%%%%%%%%%%%%%
\section{Chiral Lagrangian}\label{sect.chirallag}

The theoretical framework that we use in the present work is  $U(3)$ $\chi$PT~\cite{oriua,ohta}, 
 the tree level resonance exchanges from chiral invariant Lagrangians \cite{ecker89npb} and unitary $\chi$PT 
\cite{oller99prd,ollermeissnerplb}.
  The $U(3)\otimes U(3)$ chiral symmetry in $u$, $d$, $s$ massless QCD is broken because of quantum effects 
that violate the conservation of the 
singlet axial-vector current by the  $U_A(1)$ anomaly \cite{bardeen,fuji,adler}. 
As a result the  pseudoscalar $\eta'$ is not a pseudo-Goldstone boson \cite{wein23}. Nevertheless, 
from the large $N_C$ QCD point of view the quark loop responsible for the $U_A(1)$ anomaly \cite{bardeen} is $1/N_C$ suppressed.
In this way, the $\eta'$ becomes the ninth pseudo-Goldstone boson in the large $N_C$ limit~\cite{ua1innc}. 
In  large $N_C$ QCD the singlet  field $\eta_1$ can be conveniently incorporated into the 
effective field theory   by enlarging the number of degrees of freedom of the theory from an octet  to a nonet of pseudo-Goldstone 
bosons, which is usually called $U(3)$ $\chi$PT \cite{oriua}. In this theory there are  three expansion
parameters: momentum, quark masses and $1/N_C$, giving rise to a joint triple expansion $\delta \sim p^2 \sim m_q \sim 1/N_C$. 
We briefly recapitulate the relevant chiral Lagrangians for our work.

The leading order Lagrangian in $U(3)$ $\chi$PT reads \cite{oriua} 
\begin{eqnarray} \label{lolagrangian}
\mL_{\chi}=\frac{ F^2}{4}\langle u_\mu u^\mu \rangle+
\frac{F^2}{4}\langle \chi_+ \rangle
+ \frac{F^2}{3}M_0^2 \ln^2{\det u}\,,
\end{eqnarray}
where $\langle \ldots \rangle$ stands for the trace in flavor space and the last 
operator is the $U_A(1)$ anomaly term that gives rise to the singlet $\eta_1$ mass. 
The definitions for the chiral building blocks are   
\begin{align}
  u_\mu &= i u^+ D_\mu U u^+  \,, \nn\\
 \chi_{\pm} &= u^+ \chi u^+ {\pm} u \chi^+ u \,,\nn\\
U &=  u^2 = e^{i\frac{ \sqrt2\Phi}{ F}} \,, \nn\\
D_\mu U &= \partial_\mu U - i r_\mu U + i U l_\mu \,, \nn\\
\chi &= 2 B (s + i p) \,,
\end{align}
where $r_\mu\,, l_\mu \,, s\,, p$ are external sources \cite{gasser84} and 
the pseudo-Goldstone bosons are collected in the matrix  
\begin{equation}\label{phi1}
\Phi \,=\, \left( \begin{array}{ccc}
\frac{1}{\sqrt{2}} \pi^0+\frac{1}{\sqrt{6}}\eta_8+\frac{1}{\sqrt{3}} \eta_1 & \pi^+ & K^+ \\ \pi^- &
\frac{-1}{\sqrt{2}} \pi^0+\frac{1}{\sqrt{6}}\eta_8+\frac{1}{\sqrt{3}} \eta_1   & K^0 \\  K^- & \bar{K}^0 &
\frac{-2}{\sqrt{6}}\eta_8+\frac{1}{\sqrt{3}} \eta_1 
\end{array} \right)\,.
\end{equation}
The axial decay constant of the pseudo-Goldstone bosons in the simultaneous chiral and large $N_C$ limit is denoted by $F$. 
In the same limit the parameter $B$ is related to the quark condensate through 
$\langle 0|\bar{q}^iq^j|0\rangle =- F^2 B\delta^{ij}$.  
We do not consider isospin ($I$) breaking effects throughout.

In the present work,  we exploit the assumption on the resonance saturation of the LECs \cite{ecker89npb,ecker89plb}, so that, 
 instead of local chiral terms contributing to meson-meson scattering, we take the tree level exchanges of  the scalar, pseudoscalar and vector resonances.
 In this way we keep all local contributions to meson-meson scattering up to and including ${\cal O}(\delta^3)$, 
while also generating higher order ones. We also include the one-loop contributions  that count one order higher in $\delta$, as calculated in Ref.~\cite{guo11prd}. 

The terms that describe the interactions between scalar resonances and pseudo-Goldstone bosons in R$\chi$T are~\footnote{The terms $\hat{c}_d\bra S_9 u_\mu\ket \bra u_\mu\ket+\hat{c}_m S_1 \ln^2 \text{det}u$ were also introduced in Ref.~\cite{guo11prd} 
but found  phenomenologically irrelevant. This is in agreement with the fact that from the exchange of scalar resonances 
they start  to generate tree level meson-meson contributions that are of higher order, ${\cal O}(\delta^4)$.}~\cite{ecker89npb} 
\begin{align}\label{lagscalar}
\mL_{S}&= c_d\bra S_8 u_\mu u^\mu \ket + c_m \bra S_8 \chi_+ \ket
 + \widetilde{c}_d S_1 \bra u_\mu u^\mu \ket
+ \widetilde{c}_m  S_1 \bra  \chi_+ \ket \,,
\end{align}
and for the vector resonances it reads
\begin{eqnarray}\label{lagvector}
\mL_{V}= \frac{i G_V}{2\sqrt2}\langle V_{\mu\nu}[ u^\mu, u^\nu]\rangle \,,
\end{eqnarray}
where the anti-symmetric tensor formalism is used to describe the vector resonances~\cite{ecker89npb}. 
In addition, we also include the pseudoscalar resonances in this work, which are not 
considered in Ref.~\cite{guo11prd}. The relevant Lagrangian reads ~\cite{ecker89npb}
\begin{eqnarray}\label{lagpscalar}
\mL_{P} = i d_m \bra P_8 \chi_- \ket + i \widetilde{d}_m  P_1 \bra \chi_- \ket\,.
\end{eqnarray}
The corresponding kinetic terms for the resonance fields are~\cite{ecker89npb}
\begin{align}\label{kinerv}
\mathcal{L}_{\rm kin}^V &= -\frac{1}{2} \bra \nabla^\lambda
V_{\lambda\mu}\nabla_\nu V^{\nu\mu}
-\frac{1}{2}M^2_V V_{\mu\nu}V^{\mu\nu} \ket \,, 
\end{align}
\begin{align}\label{kinersp}
\mathcal{L}_{\rm kin}^R &= \frac{1}{2} \bra \nabla^\mu R_8 \nabla_\mu R_8
-M^2_{R_8} R_8^{2} \ket +\frac{1}{2}\big(\partial^\mu R_{1}
\partial_\mu R_{1}-M^2_{R_1} R_1^{2}\big)~,
\end{align}
where the symbol $R$ refers to either scalar or pseudoscalar resonances.

One should notice that the operators appearing in Eq.~\eqref{lagpscalar} cause mixing terms proportional to the quark masses 
 between the pseudoscalar resonances and the pseudo-Goldstone bosons. 
Though it is not a problem to consider this effect in the calculation of Feynman diagrams, it would be  
convenient to eliminate the mixing terms at the Lagrangian level. Indeed this can be accomplished by performing 
the following field redefinition in a chiral covariant way:
\begin{align}\label{redefpscalar}
P_8 & \rightarrow  \overline{P}_8 + i \frac{d_m}{M_{P_8}^2} \bigg(\chi_- - \frac{1}{3}\bra \chi_- \ket I_{3\times 3} \bigg) \,, \nonumber \\
P_1 & \rightarrow  \overline{P}_1 + i \frac{\widetilde{d}_m}{M_{P_1}^2} \bra \chi_- \ket \,,
\end{align}
with $I_{3\times 3}$ the unit $3\times 3$ matrix. Substituting Eq.~\eqref{redefpscalar} into 
Eqs.~\eqref{lagpscalar} and \eqref{kinersp}, we have the new Lagrangian
\begin{align}\label{lagpscalarshift}
 \mL_{\overline{P}} & =  \frac{1}{2} \bra \nabla^\mu \overline{P}_8 \nabla_\mu \overline{P}_8
-M^2_{P_8} \overline{P}_8^{2} \ket +\frac{1}{2}\big(\partial^\mu \overline{P}_{1}
\partial_\mu \overline{P}_{1}-M^2_{P_1} \overline{P}_1^{2}\big)   
\nonumber \\ & 
+ i\frac{d_m }{M^2_{P_8}} \bra \nabla_\mu \overline{P}_8 \,\nabla^\mu \chi_- \ket 
+ i\frac{\widetilde{d}_m }{M^2_{P_1}} \nabla_\mu \overline{P}_1 \, \nabla^\mu \bra \chi_- \ket  
\nonumber  \\ &
-\frac{d_m^2}{2 M^2_{P_8}} \bra \chi_- \chi_- \ket 
+  \big( \frac{d_m^2}{6 M^2_{P_8}} -\frac{\widetilde{d}_m^2}{2M^2_{P_1}} \big) \bra \chi_- \ket \bra \chi_- \ket 
+ \cdots \,,
\end{align}
where the omitted terms, represented by the dots in the last line, denote the local chiral operators that describe the interactions
between the pseudo-Goldstone bosons at $\cO(p^6)$ that we disregard throughout. 
We point out that the above procedure only eliminates the mixing terms between 
the pseudoscalar resonances and  pseudo-Goldstone bosons that are linearly proportional to quark masses. The new mixing terms 
with higher power of quark masses that arise through the operators in the second line of Eq.~\eqref{lagpscalarshift} 
 are $\cO(p^6)$ contributions, which are then not considered in this work.   
It is interesting to point out that after the field redefinition of Eq.~\eqref{redefpscalar}, two new operators 
describing interactions between pseudo-Goldstone bosons appear,
i.e. the ones in the third line of Eq.~\eqref{lagpscalarshift}. They coincide with the $\cO(p^4)$ terms that result from the integration 
of the pseudoscalar resonances \cite{ecker89npb}.

We remind that the nature of the pseudoscalar resonances is still a somewhat debated issue and 
also that typically the parameters of those resonances are not determined accurately ~\cite{pdg}. 
In Ref.~\cite{albaladejo10prd} the pseudoscalar resonances are generated dynamically due to the interactions 
between the scalar resonances and the pseudo-Goldstone bosons, instead of introducing them as basic degrees of freedom 
at the Lagrangian level. In order to compensate for the uncertainties arising from the not well settled properties of the pseudoscalar resonances, as well 
as from the simple tree level exchanges with bare propagators that we take here to describe them (neglecting more involved contributions as those pointed out 
in Ref.~\cite{albaladejo10prd}),    
we introduce an $L_8$-like operator~\cite{gasser85} in our study
\begin{equation}\label{lagdeltal8}
\frac{\delta L_8}{2} \bra \chi_+\chi_+ + \chi_-\chi_- \ket \,.
\end{equation}
The reason behind is that only $L_8$ and $L_7$ could receive contributions at the $\cO({p^4}$) level 
after the integration of the pseudoscalar resonances from Eqs.~\eqref{lagpscalar} and ~\eqref{kinersp}. 
Nevertheless the $L_7$ term vanishes if one further imposes the large $N_C$ relations for the pseudoscalar resonances (as we take here) 
\begin{align} 
\label{lnc4dmt}
\widetilde{d}_m &= \frac{d_m}{\sqrt3}\,, \\ 
\label{lnc4mps1}
\qquad M_{P_1} &= M_{P_8}\,.
\end{align}
We stress that $\delta L_8$ is different from $L_8$ in the $\chi$PT Lagrangian~\cite{gasser85} and 
their relation can be written as 
\begin{equation}\label{relatel8chptanddeltal8}
L_8^{\chi {\rm PT}} = L_8^{\rm Resonances} + \delta L_8\,.
\end{equation}
So that we interpret $\delta L_8$  as the contributions from some remnant pieces that are sub-leading in $1/N_C$, 
minding that the leading $N_C$ contributions are already included in the resonance part.

Finally, there are two relevant terms at ${\cal O}(\delta)$ that only incorporate pseudo-Goldstone bosons \cite{kaiser00epjc} and 
cannot be generated from the exchange of the explicit resonance fields discussed above. These contributions are: 
\begin{align}\label{laglam}
\mathcal{L}_{\Lambda} & = \Lambda_1\frac{F^2}{12}D_\mu\psi D^\mu\psi -i\Lambda_2\frac{F^2}{12} \bra U^+ \chi - \chi^+ U \ket \psi ~,\nn\\
\psi&=-i\ln \det U~,\nn\\
D_\mu\psi&=\partial_\mu \psi-2\langle a_\mu\rangle~,
\end{align}
with $a_\mu=(r_\mu-l_\mu)/2$. As we commented in Ref.~\cite{guo11prd}
 the $\Lambda_1$ parameter only affects the calculation of the masses, scattering amplitudes 
and also the scalar and pseudoscalar form factors in an indirect way, i.e. through the renormalization of the $\eta_1$ field and its effect 
in the global fit is tiny. We explicitly check that if we include $\Lambda_1$ in our discussion, the fitted value 
for this parameter approaches to zero. So that this operator is not considered in the following. 
 
For the remaining definitions of the basic chiral building blocks, the reader is referred to Ref.\cite{guo11prd} and references therein for further details.  

%%%%%%%%%%%%%%%%%%%%%%%%%%%%%%%%%%%%%%%%%%%%%%%%%%%%%%%%%%%%%%%%%%%%%%%%%%%%%%%%%%%%%%%%%%%%%%%%%%%%
%%%%%%%%%%%%%%%%%%%%%%%%%%%%%%%%%%%%%%%%%%%%%%%%%%%%%%%%%%%%%%%%%%%%%%%%%%%%%%%%%%%%%%%%%%%%%%%%%%%%
\section{Spectral Functions and Form Factors}
\label{sect.specfandff}

The two-point correlation function is defined as 
\begin{equation}
\delta^{ab} \,\Pi_R(p^2) = i \int d^4 x \, e^{i p \cdot x} <0| T [R^a(x) R^b(0)] |0>\,,
\end{equation}
where the scalar and pseudoscalar densities correspond to $R^a \equiv S^a = \bar{q} \lambda_a q$ 
and   $R^a \equiv P^a= i \bar{q} \gamma_5  \lambda_a q$, in order, 
 with $\lambda_a$ the Gell-Mann matrices for $a=1,\ldots,8$ and $\lambda_0=I_{3\times 3}\sqrt{2/3}$ for $a=0$. 
The spectral-function sum rule in the chiral limit can be then represented as  
\begin{equation}\label{defweinbergsr}
 \int_0^{\infty} \big[ {\rm Im}\,\Pi_R(s) - {\rm Im}\,\Pi_{R'}(s) \big]\,  d s = 0 \,, 
\end{equation}
where $R$ and $R'$ are different scalar  or pseudoscalar densities  mentioned above.  
The imaginary part of the two-point correlation function 
${\rm Im}\,\Pi_R$, also called the spectral function, 
is one of the key quantities that we calculate in this work. 

%%%%%%%%%%%%%%%%%%%%%%%%%%%%%%%%%%%%%%%%%%%%%%%%%%%%%%%%%%%%%%%%%%%%%%%%%%%%%%%%%%
\subsection{Scalar sector}

The scalar form factor of a pseudo-Goldstone boson pair $PQ$ is defined as
\begin{equation}\label{defsff}
 F_{PQ}^a(s) = \frac{1}{B} \bra 0|\, \bar{q} \lambda_a  q \,|\, PQ \,\ket\,.
\end{equation}
In the present work, we focus on the chiral dynamics for the components with $a=0,1,2,3,8$, which preserve the strangeness. 
By imposing the isospin symmetry, only three of the five components are independent.  
We take the neutral ones, i.e. $a=0,8,3$, which correspond to the  
singlet and $I=0$,~1 octet $SU(3)$ densities, respectively. The scalar form factors with components 
 $a=4,$ 5, 6, 7 correspond to the strangeness changing ones with 
 $I= \frac{1}{2}$ and $\frac{3}{2}$, that were studied in \cite{jamin02npb}.

The scalar spectral function is related to the scalar form factors through
\begin{eqnarray} \label{defscspecf}
 {\rm Im}\, \Pi_{S^a} (s) = \sum_{i} \rho_i(s)\, \left| F_i^a(s) \right|^2 \, \theta(s-s_i^{\rm th})\,,
\end{eqnarray}
where $\theta(x)$ is the Heaviside step function 
 and the sum on $i$ extends over the different pseudo-Goldstone boson pairs. 
In addition, $s$ is the energy squared in the center of mass frame, 
$s_i^{\rm th} = (m_A + m_B)^2$  is the threshold of the $i_{\rm th}$ channel 
and $m_A,$ $m_B$ are the masses of the corresponding two particles. 
In the previous equation  only two-body intermediate states are considered, the same ones as taken in Ref.~\cite{guo11prd} to study meson-meson scattering. 
We point out that the unitarized scalar form factors, among others,  include the contributions of the single resonance exchanges to the spectral functions. 
We proof below, when discussing the  evolution with $N_C$, that 
the two-point correlator in our analysis reduces to the single resonance exchange diagram 
at large $N_C$ \cite{pich08jhep}. 

The phase space factor for the $i_{\rm th}$ channel in Eq.~\eqref{defscspecf} is  
\begin{equation}\label{defkineticsigma}
 \rho_i(s) = \frac{\sqrt{[s-(m_A + m_B)^2][s-(m_A - m_B)^2]}}{16 \pi\, s} = \frac{q_i}{8\pi\sqrt{s}}\,,
\end{equation}
being  $q_i$ the three momentum in center of mass frame.

For the isoscalar case, there are five two-particle intermediate states made by a pseudo-Goldstone pair 
in $U(3)$ $\chi$PT, namely, $\pi\pi$, $K\bar{K}$, $\eta\eta$, $\eta\eta'$ and $\eta'\eta'$. 
For the isovector case, there are three channels: $\pi\eta$, $K\bar{K}$ and $\pi\eta'$. 
 The two-particle states with well defined isospin  for the isoscalar case read 
\begin{align}\label{ffisospin0norm}
| \pi\pi \ket_{I=0} &= -\frac{1}{\sqrt{2}} \frac{ |\pi^+ \pi^-\ket + |\pi^- \pi^+ \ket + |\pi^0 \pi^0 \ket}{\sqrt{3}}\,,\nonumber \\
| K\bar{K} \ket_{I=0} &= - \frac{ |K^+ K^-\ket + |K^0 \bar{K}^0 \ket }{\sqrt{2}}\,,\nonumber \\
| \eta\eta \ket_{I=0} &=  \frac{ |\eta \eta \ket }{\sqrt{2}}\,,\nonumber \\
| \eta\eta' \ket_{I=0} &= |\eta \eta' \ket \,,\nonumber \\
| \eta'\eta' \ket_{I=0} &= \frac{ |\eta' \eta' \ket }{\sqrt{2}}\,. 
\end{align}
 For the isovector case they are 
\begin{align}\label{ffisospin1norm}
| \pi\eta \ket_{I=1} &=  |\pi \eta \ket \,,\nonumber \\
| K\bar{K} \ket_{I=1} &= - \frac{ |K^+ K^-\ket - |K^0 \bar{K}^0 \ket }{\sqrt{2}}\,,\nonumber \\
| \pi\eta' \ket_{I=1} &= |\pi \eta' \ket \,.
\end{align}
We point out that the so-called unitary normalization for the identical particles, 
as proposed in Ref.~\cite{oller97npa}, has been used in Eq.~\eqref{ffisospin0norm} for the $\pi\pi$, 
$\eta\eta$ and $\eta'\eta'$ states. In this way, the normalization that we employ in this work coincides with the one 
used in \cite{guo11prd}. We can then construct the unitarized form factors by 
using the partial wave scattering amplitudes calculated in \cite{guo11prd} without any adjustment in the normalization. 

\begin{figure}[ht]
\begin{center}
\includegraphics[angle=0, width=0.6\textwidth]{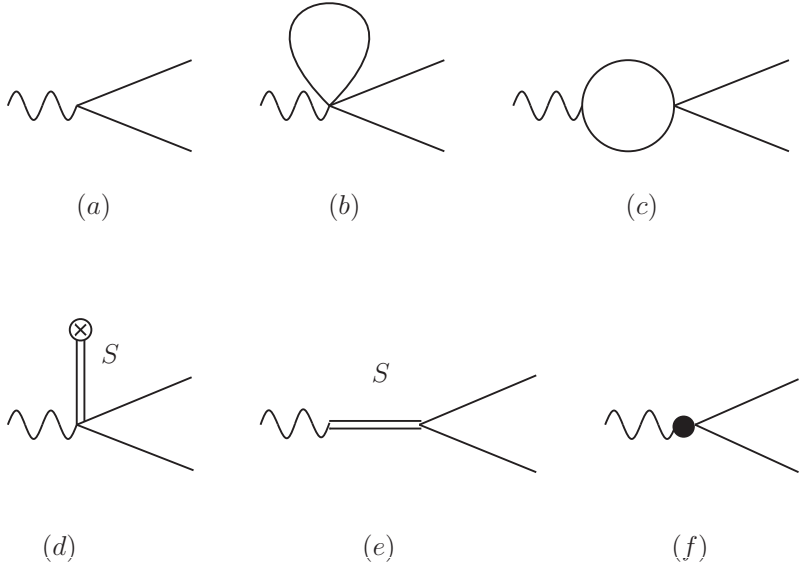}
\caption{{\small Feynman diagrams for the calculation of the scalar form factor in one-loop $U(3)$ $\chi$PT. The wavy line denotes the scalar external source. 
The cross in diagram (d) indicates the coupling between the scalar resonance and the vacuum. 
The filled circle in diagram (f) corresponds to the vertices that only involve the pseudo-Goldstone bosons and are 
beyond leading order from Eq.~\eqref{lolagrangian}. The different terms that contribute to diagram (f)  
are explained in detail in the text. }}
\label{fig.feynsff}
\end{center}
\end{figure}

The perturbative calculation of the scalar form factors of the pseudo-Goldstone pairs at the one-loop level consists of evaluating the 
Feynman graphs shown in Fig.~\ref{fig.feynsff}. 
As we have done for the scattering in Ref.~\cite{guo11prd}, the contributions beyond the leading order to the form factors are calculated 
in terms of $\overline{\eta}$ and $\overline{\eta}'$ fields that result by diagonalizing the fields $\eta_8$ and $\eta_1$ 
at leading order from Eq.~\eqref{lolagrangian}
\begin{align}\label{deflomixing}
\eta_8 & =  c_\theta \overline{\eta}+ s_\theta\overline{\eta}'\,, \qquad \nonumber\\
\eta_1 & = -s_\theta\overline{\eta} + c_\theta\overline{\eta}'\,,
\end{align}
with $c_\theta=\cos\theta$ and $ s_\theta=\sin\theta$. 
Compared to the basis with the fields $\eta_8$ and $\eta_1$,  the use of the fields  $\overline{\eta}$ and $\overline{\eta}'$ 
 allows us to include the relevant Feynman diagrams conveniently in the calculation. As discussed in Ref.~\cite{guo11prd} we avoid 
in this way having to include an arbitrary number of insertions of the leading order $\eta_8$--$\eta_1$ mixing. 
As a result, one can consider the $\eta$ and $\eta'$ mixing effects in the form factors 
order by order in the $\delta$ counting scheme.

  There are several sources that contribute to diagram (f) in Fig.~\ref{fig.feynsff}: 
$\overline{\eta}$--$\overline{\eta}'$ mixing (notice that it is different from $\eta_1$--$\eta_8$ mixing \cite{guo11prd}, as just discussed),
 and the local contributions from  
the last two terms in Eq.~\eqref{lagpscalarshift} and the operators in Eqs.~\eqref{lagdeltal8} and \eqref{laglam}. 
The contributions from the $\overline{\eta}$--$\overline{\eta}'$ mixing to the form factors are calculated 
in the same way as for the calculation of scattering amplitudes in Ref.~\cite{guo11prd}.
First, we substitute Eq.~(16) of Ref.~\cite{guo11prd}, which parameterizes the relations between
$\overline{\eta}\,,\overline{\eta}'$ and the physical states $\eta\,,\eta'$,
into the leading order Lagrangian Eq.~\eqref{lolagrangian}. Afterwards, we  calculate the form factors in terms of the physical 
states $\eta$ and $\eta'$. Due to the inclusion of the pseudoscalar resonances and the operator in Eq.~\eqref{lagdeltal8}, 
which contribute to the $\overline{\eta}-\overline{\eta}'$ mixing parameters $\delta_i$ in Eq.~(14) of Ref.~\cite{guo11prd}, 
we present the updated results in Appendix \ref{app.newmixing}.
In addition, the wave function renormalization of the pseudo-Goldstone bosons, which has been calculated
in Ref.~\cite{guo11prd}, also contributes to the scalar form factors. 
The explicit expressions for the scalar form factors are given in Appendix \ref{app.sff}.

In $U(3)$ $\chi$PT the resummation of the unitarity chiral loops is essential because of the large 
mass of the $s$ quark and the large anomaly mass in Eq.~\eqref{lolagrangian}.
Then, in many kinematical regions the resulting  pseudo-Goldstone thresholds 
are much larger than the three-momenta involved and this enhances the two-pseudo-Goldstone 
reducible loop contributions \cite{weinberg91npb}. In addition, we are 
also interested in the chiral dynamics involving the resonance region where the unitarity upper bound in partial waves could be 
easily found to be saturated. So that it is not meaningful to treat unitarity in a perturbative way 
as in $\chi$PT. As a result, one must resum the right-hand cut that stems from  unitarity and for that 
we employ Unitary $\chi$PT (U$\chi$PT), both for calculating the meson-meson partial 
wave amplitudes~\cite{oller99prd,jamin00npb,guo11prd} and the scalar form factors~\cite{meissner01npa,oller00prd,oller05prd}.

The unitarization method that we use to resum the unitarity cut for the scalar form factors was developed 
in Ref.~\cite{oller00prd,meissner01npa}. It is based on the $N/D$ method  
and was first applied to meson-meson scattering in Ref.~\cite{oller99prd}. 
For completeness, we recapitulate the essentials of this method here.

In the case of two-particle intermediate states, the absorptive part of the form factor obeys the  relation
\begin{equation}
 {\rm Im}\, F_j^I(s) =  \sum_{k=1}^Z {T^{IJ}_{jk}(s)}^{*} \, \rho_k(s)\, F_k^I(s)\,,
\end{equation}
where $T^{IJ}(s)$ denotes the partial wave scattering amplitudes with definite isospin number $I$ 
and angular momentum $J$ and $Z$ is the number of channels with the same quantum numbers $IJ$. The $T$-matrix 
$T^{00}(s)$, relevant to the scalar form factors with $a=0,8$, is a $5\times 5$ matrix and $T^{10}(s)$, 
relevant to $a=3$, is a $3\times 3$ matrix. Both of them are studied in detail in Ref.~\cite{guo11prd}. 
Nevertheless due to the fact that now we also include the pseudoscalar resonances and the $\delta L_8$ operator, 
we need to consider their contributions in the scattering amplitudes as well. 
The effect from the pseudoscalar resonances can be easily included, since after the field redefinition 
the only relevant terms are the last two operators in Eq.~\eqref{lagpscalarshift}. 
The pertinent expressions for the perturbative scattering amplitudes from $U(3)$ $\chi$PT including explicit exchanges of 
scalar and vector resonances \cite{guo11prd} is given in Ref.~\cite{mathcode}. The  
 new contributions from the $\delta L_8$ operator and pseudoscalar resonances (as well as from tensor resonances, introduced below) are given in \cite{mathcode2}, together with 
the expressions for the form factors. 

Following the method elaborated in Refs.~\cite{meissner01npa,oller00prd}, see Ref.~\cite{oller05prd} for a simplified discussion, the unitarized scalar form factor can be cast as
\begin{equation}\label{defunitarizedF}
F^I(s) =  \big[ 1 + N^{IJ}(s) \, g^{IJ}(s) \big]^{-1} R^I(s)\,,
\end{equation}
where 
\begin{align}\label{defNyR}
N^{IJ}(s) &= {T^{IJ}(s)}^{\rm (2)+Res+Loop} + T^{IJ}(s)^{\rm (2)} \,g^{IJ}(s)\, T^{IJ}(s)^{\rm (2)}\,,\nonumber \\
R^{I}(s)  &= {F^{I}(s)}^{\rm (2)+Res+Loop} + N^{IJ}(s)^{\rm (2)} \,g^{IJ}(s)\, F^{I}(s)^{\rm (2)}\,.
\end{align}
 The matrix $N^{IJ}(s)$  contains the crossed-channel cuts from the meson-meson scattering and the bare resonance poles. It was calculated in Ref.~\cite{guo11prd}, and extended here by including the pseudoscalar resonance exchanges and the $ \delta L_8$ operator. 
The superscripts (2), Res and Loop in Eq.~\eqref{defNyR} denote the perturbative calculations from 
the tree level result using the leading order Lagrangian~Eq.~\eqref{lolagrangian}, resonance contributions
(also including the operators in Eqs.~\eqref{lagdeltal8} and \eqref{laglam}) and pseudo-Goldstone loops, respectively. 
 On the other hand, 
$R^I(s)$ is a vector with $Z$ rows constructed from the $U(3)$ $\chi$PT form factors similarly as $N^{IJ}(s)$ is 
calculated for scattering. The vector $R^I(s)$ does not contain any cut singularity and is real \cite{meissner01npa,oller00prd}.  

The  matrix $g^{IJ}(s)$ in Eq.~\eqref{defNyR} is a diagonal $Z\times Z$  matrix, 
with its $i_{\rm th}$ non-vanishing matrix element given by 
\begin{align}\label{defgfuncionloop}
16\pi^2 g^{IJ}(s)_i &=  a_{SL}(\mu)+\log\frac{m_B^2}{\mu^2} 
-x_+\log\frac{x_+-1}{x_+}
-x_- \log\frac{x_--1}{x_-} \,, \\ 
x_{\pm} &=\frac{s+m_A^2-m_B^2}{2s}\pm\frac{1}{2s}\sqrt{ -4 s (m_A^2-i0^+)+(s+m_A^2-m_B^2)^2}\,, \nonumber 
\end{align}
where $a_{SL}(\mu)$ is a subtraction constant. The matrix $g^{IJ}(s)$ collects the discontinuity caused by the two-particle
 intermediate states along the right hand cut and plays a key role in the $N/D$ unitarization approach.
We refer to Section IV of Ref.~\cite{guo11prd}, and references therein, for detailed discussions on the calculation of the scattering $T$-matrix. The latter is given by a similar expression to Eq.~\eqref{defunitarizedF} in terms of $N^{IJ}(s)$ and $g^{IJ}(s)$,
\begin{align}\label{defunitarizedT}
T^{IJ}(s)&=\big[ 1 + N^{IJ}(s) \, g^{IJ}(s) \big]^{-1} N^{IJ}(s)~.
\end{align}

Before ending this section, we introduce the scalar form factors in the quark flavor basis 
instead of in the singlet-octet flavor basis. 
There are two kinds of isoscalar scalar densities in the quark flavor basis: $\bar{u}u+\bar{d}d$ and $\bar{s}s$. 
The relations between the form factors in the two different bases read
\begin{align}\label{defsffuuddss}
F^{\bar{u}u+\bar{d}d} &= \frac{1}{\sqrt{3}}F^{a=8} + \sqrt{\frac{2}{3}}F^{a=0}\,, \\ 
F^{\bar{s}s} &= -\frac{1}{\sqrt{3}}F^{a=8} + \frac{1}{\sqrt{6}}F^{a=0}\,.
\end{align}
The pion scalar radius $\bra r^2 \ket_S^\pi$, an important low energy observable, is defined through 
the low energy Taylor expansion of the pion scalar form factor in the quark flavor basis~\cite{gasser85ff,donoghue90npb}
\begin{equation}\label{defscradius}
F_{\pi\pi}^{\bar{u}u+\bar{d}d}(s) = F_{\pi\pi}^{\bar{u}u+\bar{d}d}(0) \bigg[  1 + \frac{1}{6}\bra r^2 \ket_S^\pi \,s   + ... \bigg]\,,
\end{equation}
with 
\begin{equation}
m_\pi^2 F_{\pi\pi}^{\bar{u}u+\bar{d}d}(s) \equiv 2B m \bra 0| \bar{u}u+\bar{d}d |\pi\pi\ket_{I=0}\,.
\end{equation}
In the above equation, $m$ stands for the up or down quark mass (we ignore isospin breaking) 
so that  $2Bm = \overline{m}_\pi^2$, 
with $\overline{m}_\pi$ the leading order pion mass from Eq.~\eqref{lolagrangian}. 
The relation between the physical mass square of the pion, $m_\pi^2$, and the leading order one, $\overline{m}_\pi^2$,  
was given in the Appendix of Ref.~\cite{guo11prd}. The updated version that includes  
the pseudoscalar resonances and the $\delta L_8$ effect is collected in Appendix~\ref{app.newmixing}. 

%%%%%%%%%%%%%%%%%%%%%%%%%%%%%%%%%%%%%%%%%%%%%%%%%%%%%%%%%%%%%%%%%%%%%%%%%%%%%%%%%%%%%%%%%%%%%%%%%%%%%%%%%
%%%%%%%%%%%%%%%%%%%%%%%%%%%%%%%%%%%%%%%%%%%%%%%%%%%%%%%%%%%%%%%%%%%%%%%%%%%%%%%%%%%%%%%%%%%%%%%%%%%%%%%%%
\subsection{Pseudoscalar sector}

For the pseudoscalar spectral function, the leading order contribution is due to
 the single pseudo-Goldstone boson exchange. The next non-vanishing contribution 
from the pure pseudo-Goldstone system requires at least three intermediate mesons,
which belongs to a two-loop calculation that  is beyond the scope of our 
current discussion. In order to take into account the chiral dynamics above 
1~GeV in the pseudoscalar spectral function, we include the pseudoscalar resonance exchanges 
explicitly making use of the chiral invariant Lagrangian, Eq.~\eqref{lagpscalarshift}. In light of the results of the work \cite{albaladejo10prd}, where the pseudoscalar resonances were 
dynamically generated through the scattering of  the scalar resonances 
and the pseudo-Goldstone bosons, we could instead consider the scalar resonances and the pseudo-Goldstone bosons 
in the intermediate states. This is left for future work.

The pseudoscalar spectral function is related to the pseudoscalar 
form factors calculated here in a simple way
\begin{equation}\label{defpsspecf}
{\rm Im}\, \Pi_{P^a}(s) = \sum_i \pi\, \delta(s-m_{P_i}^2)\, |H^a_i(s)|^2\,,
\end{equation}
where $\delta(x)$ is the standard Dirac $\delta$-function, 
$m_{P_i}$ denotes the masses of the pseudo-Goldstone bosons and the 
pseudoscalar resonances with the same isospin as the considered spectral function. Finally, 
 $H^a_i(s)$ is the corresponding pseudoscalar form factor
\begin{equation}\label{defpff}
 H_i^a(s) = \frac{1}{B} \bra 0|\, i \bar{q} \gamma_5  \lambda_a q \,|\, P_i \,\ket\,.
\end{equation}

\begin{figure}[ht]
\begin{center}
\includegraphics[angle=0, width=0.6\textwidth]{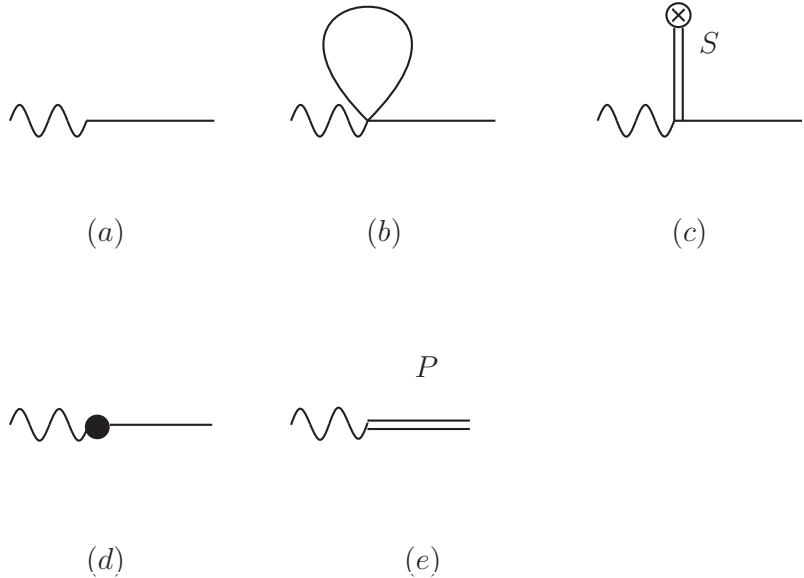}
\caption{{\small Pseudoscalar form factors of the pseudo-Goldstone bosons and pseudoscalar resonances. 
The symbols have the same meanings as those in Fig.~\ref{fig.feynsff}, except that the 
wavy line stands now for the pseudoscalar external source. See the text for further details.}}
\label{fig.feynpff}
\end{center}
\end{figure}

The calculation of the pseudoscalar form factor for the pseudo-Goldstone boson 
consists of evaluating the Feynman diagrams (a)--(d) in Fig.~\ref{fig.feynpff}.  
Only the Feynman diagram (e) in Fig.~\ref{fig.feynpff} is relevant for the calculation 
of the pseudoscalar form factor for the pseudoscalar resonances. This diagram stems 
from the operators in the second line of Eq.~\eqref{lagpscalarshift}. 
Moreover, we also consider the effects from the wave function renormalization of 
the pseudo-Goldstone bosons and express in the final results 
the pion decay constant $F$ in the chiral and large $N_C$ limits by the physical one $F_\pi$. The latter  
was calculated in Ref.~\cite{guo11prd}. This reshuffling in the expressions was also done for the scattering amplitudes 
in Ref.~\cite{guo11prd}.

%%%%%%%%%%%%%%%%%%%%%%%%%%%%%%%%%%%%%%%%%%%%%%%%%%%%%%%%%%%%%%%%%%%%%%%%%%%
%%%%%%%%%%%%%%%%%%%%%%%%%%%%%%%%%%%%%%%%%%%%%%%%%%%%%%%%%%%%%%%%%%%%%%%%%%%%%%%
\section{Remarks on semilocal duality} \label{sect.semilocalintro}

In this section, we introduce the basic idea of Regge theory in order to formulate the notation used in this work. 
For  a detailed account on Regge theory see e.g.  Ref.~\cite{collinsbook}.  
Regge theory relates the dynamics in the  $t$-channel with the $s$-channel for the scattering processes. In $\pi\pi$ scattering, the 
isospin well-defined amplitudes from both channels are related through~\cite{collinsbook}
\begin{align}\label{tsrelation}
T_{\rm t}^{(0)}(s,t) &= \frac{1}{3} T_{\rm s}^{(0)}(s,t) + T_{\rm s}^{(1)}(s,t) + \frac{5}{3} T_{\rm s}^{(2)}(s,t) \,, \nonumber \\
T_{\rm t}^{(1)}(s,t) &= \frac{1}{3} T_{\rm s}^{(0)}(s,t) + \frac{1}{2} T_{\rm s}^{(1)}(s,t) - \frac{5}{6} T_{\rm s}^{(2)}(s,t) \,, \nonumber \\
T_{\rm t}^{(2)}(s,t) &= \frac{1}{3} T_{\rm s}^{(0)}(s,t) - \frac{1}{2} T_{\rm s}^{(1)}(s,t) + \frac{1}{6} T_{\rm s}^{(2)}(s,t) \,, 
\end{align}
where the subscript of $T$ denotes the $t$- or $s$-channel and the superscript refers to the proper isospin quantum number $I$ in the $t$- or $s$-channel, respectively.   

The object of the Regge theory is the fixed-$t$ scattering amplitude and the proposed quantity to quantify the fulfillment of semilocal (or average)  
duality between Regge theory and hadronic degrees of freedom (h.d.f.) in Ref.~\cite{pelaez11prd} is 
\begin{equation}\label{defseml}
 \int_{\nu_1}^{\nu_2} \nu^{-n}\, {\rm Im}\, T_{\rm t, Regge }^{(I)}(\nu, t) \, d\nu =  
\int_{\nu_1}^{\nu_2} \nu^{-n}\, {\rm Im}\, T_{\rm t, Hadrons}^{(I)}(\nu, t) \, d\nu \,,
\end{equation}
where $\nu = \frac{s-u}{2}= \frac{2s + t-4m_\pi^2}{2}$ and $s,t,u$ are the standard Mandelstam variables. 
The ``averaging'' should take place over at least one resonance tower. Therefore, the integration region $\nu_2 - \nu_1$ is typically taken as 
a multiple of 1~GeV$^2$. In this work, we shall focus on the energy region below 2~GeV$^2$. 
Semilocal duality should be well satisfied for forward scattering, 
i.e. with $t=0$, where the leading Regge trajectory (the one taken here into account) dominates. Like in Ref.~\cite{pelaez11prd}, we also evaluate Eq.~\eqref{defseml} 
at the threshold point $t=t_{\rm th} = 4 m_\pi^2$  to show the robustness of the approach under changes of $t$ (still small compared with GeV$^{2}$). 
Different choices for $n$ enable us to test the different energy regions that dominate the integrals in Eq.~\eqref{defseml}. 
For negative values of $n$, the dynamics in the high energy region definitely plays a more important role and 
it is beyond the scope of the current work. For large positive values of $n$, the integrations in Eq.~\eqref{defseml} 
are then dominated by the very low energy physics, where resonances marginally contribute. 
In this work we test semilocal duality by taken $n$ from 0 to $3$, as suggested in Ref.~\cite{pelaez11prd}, 
which is an adequate choice for the intermediate energy region from  threshold up  to 2~GeV$^2$, where several resonances contribute.  

To evaluate the left hand side of Eq.~\eqref{defseml},  the Regge asymptotic results continued down to threshold for
 ${\rm Im}\, T_{\rm t, Regge }^{(I)}(\nu, t)$  at fixed $t$  are needed.  The explicit formulae  and detailed discussions  
can be found in Ref.~\cite{pelaez11prd} and references therein. We do not repeat here the formulation. 
 For the right hand side of Eq.~\eqref{defseml}, one can decompose the isospin amplitudes into a sum of partial waves 
\begin{equation}\label{pwdecompose}
  {\rm Im}\, T_{\rm s}^{(I)}(\nu, t) = \sum_J (2 J + 1)\,  {\rm Im}\, T^{IJ}(s) \, P_J(z_s)\,,
\end{equation}
where $z_s = 1 + 2 t /(s - 4m_\pi^2)$ is the cosine of the scattering angle in the $s$-channel 
center of mass frame and $P_J(z_s)$ denotes the Legendre polynomials. By substituting  Eq.~\eqref{pwdecompose} into 
Eq.~\eqref{tsrelation}  ${\rm Im}\, T_{t,{\rm Hadrons} }^{(I)}(\nu, t)$ is obtained and then the right hand side of Eq.~\eqref{defseml} can be 
compared with the results from Regge theory.

In order to quantify the fulfillment of semilocal duality we define, as in Ref.~\cite{pelaez11prd}, two types of ratios of integrals like these in  Eq.~\eqref{defseml}, instead of comparing directly the value of the integration from the Regge theory and the h.d.f.\,.  The first one, $R^I_n$  is defined as:
\begin{equation}\label{defRratio}
 R^I_n = \frac{ \int_{\nu_1}^{\nu_2} \nu^{-n}\, {\rm Im}\, T_{\rm t}^{(I)}(\nu, t)\, d\nu}
{\int_{\nu_1}^{\nu_3} \nu^{-n}\, {\rm Im}\, T_{\rm t}^{(I)}(\nu, t)\, d\nu}\,.
\end{equation}
 To make closer the comparison with Ref.~\cite{pelaez11prd}, we set  in the following $\nu_1$ at threshold, $\nu_2=1$~GeV$^2$ 
and $\nu_3 = 2$~GeV$^2$.  

Other interesting objects to consider are the Finite Energy Sum Rule (FESR) between different 
isospin amplitudes with the same upper integration limit 
\begin{equation}\label{defFratio}
 F_n^{I I'} = \frac{ \int_{\nu_1}^{\nu_{\rm max} } \nu^{-n}\, {\rm Im}\, T_{\rm t}^{(I)}(\nu, t)\, d\nu}
{\int_{\nu_1}^{\nu_{\rm max}} \nu^{-n}\, {\rm Im}\, T_{\rm t}^{(I')}(\nu, t)\, d\nu}\,,
\end{equation}
where $\nu_{\rm max}=$1~GeV$^2$ or 2~GeV$^2$ in the later discussions.

Among the various cases to investigate semilocal duality between h.d.f. 
and Regge theory in $\pi\pi$ scattering, the golden mode is the isotensor one in the $t$-channel, since then the Regge exchange 
is highly suppressed (as there are no $\bar{q}q$ states with $I=2$). 
As a result, the ratios $F_n^{21}$ and $F_n^{20}$ should tend to vanish in order to satisfy semilocal  duality.   
In contrast, the dual direct $s$-channel allows the exchanges of 
several resonances with isospin $I$ = 0,~1, see Eq.~\eqref{tsrelation}. 
Thus, if  semilocal duality is fulfilled, one should expect the cancellation between the different resonances exchanged in the $s$-channel, 
which sheds light on the resonance properties. Indeed, it establishes serious relations between the scalar and vector spectra, as we discuss later. 

Finally, for the evaluation of the quantities in Eqs.~\eqref{defRratio} and \eqref{defFratio} that quantify  semilocal duality, the key ingredients are the 
partial waves ${\rm Im}\, T^{IJ}(s)$ in Eq.~\eqref{pwdecompose}, which were studied by us in Refs.~\cite{guo11prd,guo12plb}. 
This method is different from the one used in Ref.~\cite{pelaez11prd}, where the resonances are 
regenerated by unitarizing the perturbative $\chi$PT amplitudes within the IAM~\cite{pelaezprl1,pelaezprl2}. 
An important difference is that instead of the explicit resonance states, as we employ here, 
the IAM depends on the LECs from $SU(2)$ or $SU(3)$ $\chi$PT \cite{weinberg79,gasser84,gasser85}. 
Nevertheless, including the explicit resonance states in the Lagrangian is not enough to guarantee that 
one can apply the theory to higher energies due to the important contribution from the 
pseudo-Goldstone boson loops. In this case, one needs to resum the unitarity chiral loops and a sophisticated 
approach based on the N/D method is formulated in Ref.~\cite{oller99prd}, and already used to construct the unitarized scalar 
form factors in the previous section where we also discussed why $U(3)$ $\chi$PT should be unitarized. 
Through the procedure to resum the chiral loops, in addition to extending the application energy region of 
R$\chi$T, one also generates physical resonances with finite widths \cite{oller97npa,oller99prd,oller00plb,oller00npa,oller00ppnp,jamin00npb,guo11prd}
in contrast to the zero-width resonances in the bare chiral Lagrangian. 
In the following, we employ the partial wave amplitudes from this procedure to analyze the semilocal duality. 

%%%%%%%%%%%%%%%%%%%%%%%%%%%%%%%%%%%%%%%%%%%%%%%%%%%%%%%%%%%%%%%%%
\section{Phenomenological Discussions}\label{sect.discussion}

In this section, we study the phenomenological results of the several different types of 
spectral-function sum rules presented in Eq.~\eqref{defweinbergsr} and 
the ratios in Eqs.~\eqref{defRratio} and \eqref{defFratio} that quantify the fulfillment of semilocal duality. 
In order to perform the analyses, we need to provide the values for the resonance parameters, 
the low energy constants in Eqs.~\eqref{lagdeltal8} and \eqref{laglam} and the subtraction constants in Eq.~\eqref{defgfuncionloop} 
introduced by the unitarization procedure. 

We want to stress that all the parameters in the form factors already appear in the meson-meson
scattering amplitudes and the pseudo-Goldstone masses. Thus, once we determine them by 
 fitting the scattering data and the $\eta$ and $\eta'$ masses, 
the form factors and the spectral functions are all predictions. 

%%%%%%%%%%%%%%%%%%%%%%%%%%%%%%%%%%%%%%%%%%%%%%%%%%%%%%%%%%%%%%%%%%%%%%%%%%%%%
\subsection{Former Fit}

\begin{figure}[ht]
\begin{center}
\includegraphics[angle=0, width=0.99\textwidth]{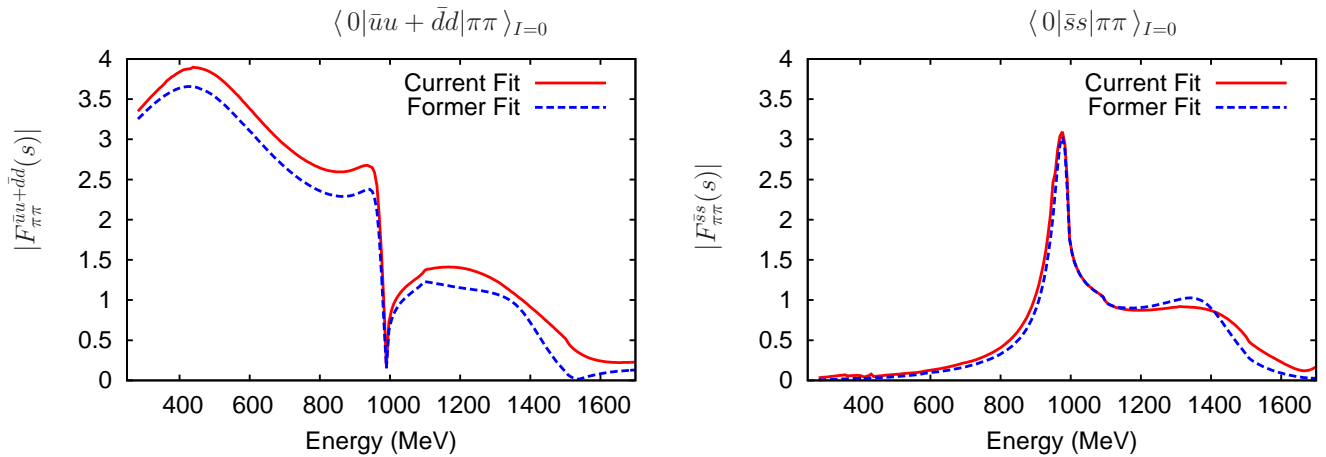}
\caption{{\small The $\bar{u}u+\bar{d}d$ (left panel) and $\bar{s}s$ (right panel) scalar form factors of the pion in the quark flavor basis. 
The (red) solid lines are for the current fit and the (blue) dashed ones for the former fit.}}
\label{fig.sffuuddss}
\end{center}
\end{figure}

\begin{figure}[ht]
\begin{center}
\includegraphics[angle=0, width=0.99\textwidth]{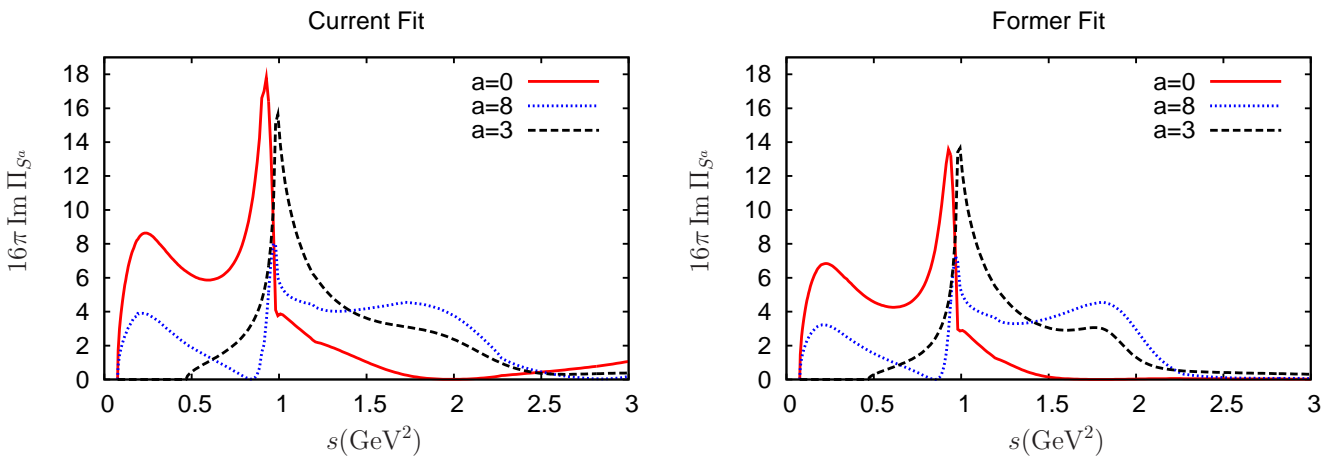}
\caption{{\small The scalar spectral functions for the $a=0$ (solid line), 8 (dotted line) and 3 (dashed line) correlators. 
The left panel is for the current fit and the right one for the former fit.}}
\label{fig.impi}
\end{center}
\end{figure}

We employ here the best fit in Eq.~(55) of Ref.~\cite{guo11prd}, which 
is referred as ``former fit" from now on, to calculate the 
form factors and spectral functions. In Figs.~\ref{fig.sffuuddss} 
and \ref{fig.impi}, we show in order the results for the pion scalar form factors 
in the quark flavor basis and the spectral functions with $a=0,$ $3$ and $8$. 

The quadratic pion scalar radius defined in Eq.~\eqref{defscradius} is found to be 
\begin{equation}\label{oldr2}
\bra r^2 \ket_S^\pi = 0.43 \pm 0.01 \,{\rm fm}^2\,.
\end{equation}
Compared to the dispersive result from Ref.~\cite{colangelo01npb} 
\begin{equation}
\bra r^2 \ket_S^\pi = 0.61 \pm 0.04\,{\rm fm}^2\,,
\end{equation}
our result is around a 30\% smaller than this well accepted value \cite{colangelo01npb,rocaoller}. 
The main reason for getting a smaller value for the pion scalar radius is 
that our predictions for the low energy constants $L_4$ and $L_5$ are
quite small by using the resonance parameters from the best 
fit in Ref.~\cite{guo11prd} 
\begin{align}\label{oldl4l5}
 L_4 (\mu = 770 \,{\rm MeV}) &= (0.03^{+0.08}_{-0.05}) \times 10^{-3}\,, \nonumber \\
\quad L_5 (\mu = 770 \,{\rm MeV}) &= (0.26^{+0.11}_{-0.18}) \times 10^{-3}\,.
\end{align}
These two low energy constants are important in the determination of the 
pion scalar radius~\cite{gasser85ff}. 
So in order to improve its determination, we need to increase the 
values of $L_4$ and $L_5$ from the resonance contribution. 
In Ref.~\cite{jamin02npb}, it is realized that around 50\% of $L_5$ is in 
fact contributed by a second nonet of scalar resonances. 
Thus we decide to include a second multiplet of scalar resonances in our discussion 
 so as to achieve a bigger pion scalar radius. 
This requires us to update the fit we did in Ref.~\cite{guo11prd}, which 
we discuss in detail in the next section. Another motivation to start a new fit is 
that we also include the pseudoscalar resonances in the present work, which 
turn out to be important in the spectral-function sum rules that we discuss below.

%%%%%%%%%%%%%%%%%%%%%%%%%%%%%%%%%%%%%%%%%%%%%%%%%%%%%%%%%%%%%%%%%%%%%%%%%%%%%%%%%%%%%%%%%%
\subsection{Current Fit}
\label{sec:newfit}

Since we do not include the 4$\pi$ channel which turns out 
to be important in the energy region above $1.3$~GeV in the isoscalar scalar case~\cite{albaladejo08prl}, 
we fit the scattering data in the $IJ=00$ channel up to $1.3$~GeV as done in Ref.~\cite{guo11prd}. 
In this way, the second multiplet of scalar resonances around $2$~GeV~\cite{jamin00npb,jamin02npb} 
behaves more like a background in our study and we cannot fit its parameters in a precise manner. 
 Instead, we take the values for the resonance parameters of the 
second multiplet of scalar resonances from the preferred fit of Ref.~\cite{jamin00npb}, given in its  Eq.~(6.10):\footnote{The bare mass of the second nonet of scalar resonances is not either well fixed from Ref.~\cite{jamin00npb}. E.g. in its fit (6.11) its bare mass is $\sim 2$~GeV. 
The range $2-2.5$~GeV is then a realistic one for the bare mass of the second nonet of scalar resonances as follows from Ref.~\cite{jamin00npb}.} 
\begin{align}
c_d' & = \sqrt{3} \widetilde{c}_d' = c_m' = \sqrt{3} \widetilde{c}_m' \simeq 40\,{\rm MeV}\,,\nn\\
M_{S_1'}& = M_{S_8'} = 2570\,{\rm MeV}\,.
\end{align}
In this reference the scattering data in the $IJ=\frac{1}{2} 0$ and $IJ=\frac{3}{2} 0$
channels were investigated in a similar approach as we use here up to around $2$~GeV. 
About the pseudoscalar resonances, we take 
$d_m= \sqrt{3}\,\widetilde{d}_m =30$~MeV and $M_{P_8} = M_{P_1} = 1350$~MeV, 
which lie in the commonly determined regions of these parameters~\cite{ecker89npb,golterman00prd,juanjo10prd}.

In Ref.~\cite{guo11prd} we found that the scalar resonance couplings 
from the fit perfectly obey the large $N_C$ relations 
\begin{align}\label{ncrelationcdtcmt}
\widetilde{c}_d &= \frac{c_d}{\sqrt3}\,,\nn\\
\widetilde{c}_m & = \frac{c_m}{\sqrt3}\,,
\end{align}
so that in the current fit we also impose them. 
For the subtraction constants  we always keep the isospin constraints on them \cite{guo11prd}.  
 Finally, in the updated fit  we have 18 free parameters and the fit results are 
\begin{equation}
\begin{array}{rlrl}
 c_d=                     & (19.8^{ +2.0}_{-5.2})~{\rm MeV}  &  c_m=     &  (41.9^{ +3.9}_{-9.2})~{\rm MeV}  \\
M_{S_8}=                  & (1397^{ +73}_{-61})~{\rm MeV}   & M_{S_1}=  &  (1100^{ +30}_{-63})~{\rm MeV}    \\
M_{\rho}=                 & (801.2^{ +8.2}_{-6.9})~{\rm MeV} & M_{K^*}=  & (910.0^{ +7.0}_{-9.1})~{\rm MeV}  \\
G_V=                      & (62.1^{ +1.9}_{-2.1})~{\rm MeV}   & a_{SL}^{1 \,0\,,\pi\eta}= & 2.0^{ +3.3}_{-4.5}  \\ 
a_{SL}^{00\,,\pi\pi}=     & -1.27^{ +0.12}_{-0.12}             & a_{SL}^{00\,,K\bar{K}}=   & -0.95^{ +0.33}_{-0.16} \\ 
a_{SL}^{\frac{1}{2} \,0\,, K\pi}=&  -1.12^{ +0.12}_{-0.17}     & a_{SL}^{\frac{1}{2} \,0\,, K\eta}=& -0.08^{ +0.38}_{-1.04} \\ 
a_{SL}^{\frac{1}{2} \,0\,, K\eta'}=& -1.25^{ +1.11}_{-1.23}    & \delta L_8= &  0.23^{ +0.29}_{-0.19} \times 10^{-3} \\ 
M_0 =                             & (951^{ +50}_{-50})~{\rm MeV} & \Lambda_2 =    &  -0.37^{ +0.19}_{-0.19} \\
{\cal N} =                        & (0.76^{ +0.36}_{-0.35})~{\rm MeV^{-2}} & c = & 1.05^{ +0.43}_{-0.33}  \\ 
\end{array}
\label{fitresultnew}
\end{equation}
 with $\chi^2/{\rm (degrees\,of \,freedom)} = 784/(348-18) \simeq 2.38 $. 
For the remaining subtraction constants, we impose the following relations in the fit 
\begin{align}
& a_{ SL}^{00\,,\, \eta\eta}
 = a_{ SL}^{00\,,\, \eta\eta'} = a_{ SL}^{00\,,\, \eta'\eta'} = a_{ SL}^{00\,,\, K\bar{K}}  \,, \nonumber \\ &
a_{ SL}^{2 0 \,,\, \pi\pi} = a_{ SL}^{00\,,\, \pi\pi} \,, \nonumber \\ &
a_{ SL}^{\frac{3}{2} \,0 \,, \,K\pi} = a_{ SL}^{\frac{1}{2} \,0\,,\, K\pi} \,, \nonumber \\ &
 a_{ SL}^{10 \,,\, \pi\eta'} = a_{SL}^{10\,,\, K\bar{K}} =  a_{ SL}^{00\,,\, K\bar{K}}\,,
\end{align}
and all of the subtraction constants in the vector channels (which are barely sensitive to them while they are of natural 
size \cite{ollermeissnerplb}) are set equal to $a_{ SL}^{00\,,\,\pi\pi}$. 
The parameters ${\cal N}$ and $c$ in Eq.~\eqref{fitresultnew} are introduced to describe the 
$\pi\eta$ distribution
\begin{eqnarray}
\frac{d N_{\pi\eta}}{ d E_{\pi\eta}} 
= q_{\pi\eta}\, {\cal N} \big| \,T^{10}_{ K \bar{K}\to \pi\eta}(s) + c\, T^{10}_{\pi\eta \to \pi\eta}(s)  \, \big|^2\,,
\label{a0.inv}
\end{eqnarray}
with $q_{\pi\eta}$ the three momentum of the $\pi\eta$ system in the center of mass frame.

\begin{figure}[H]
\begin{center}
\includegraphics[angle=0, width=0.99\textwidth]{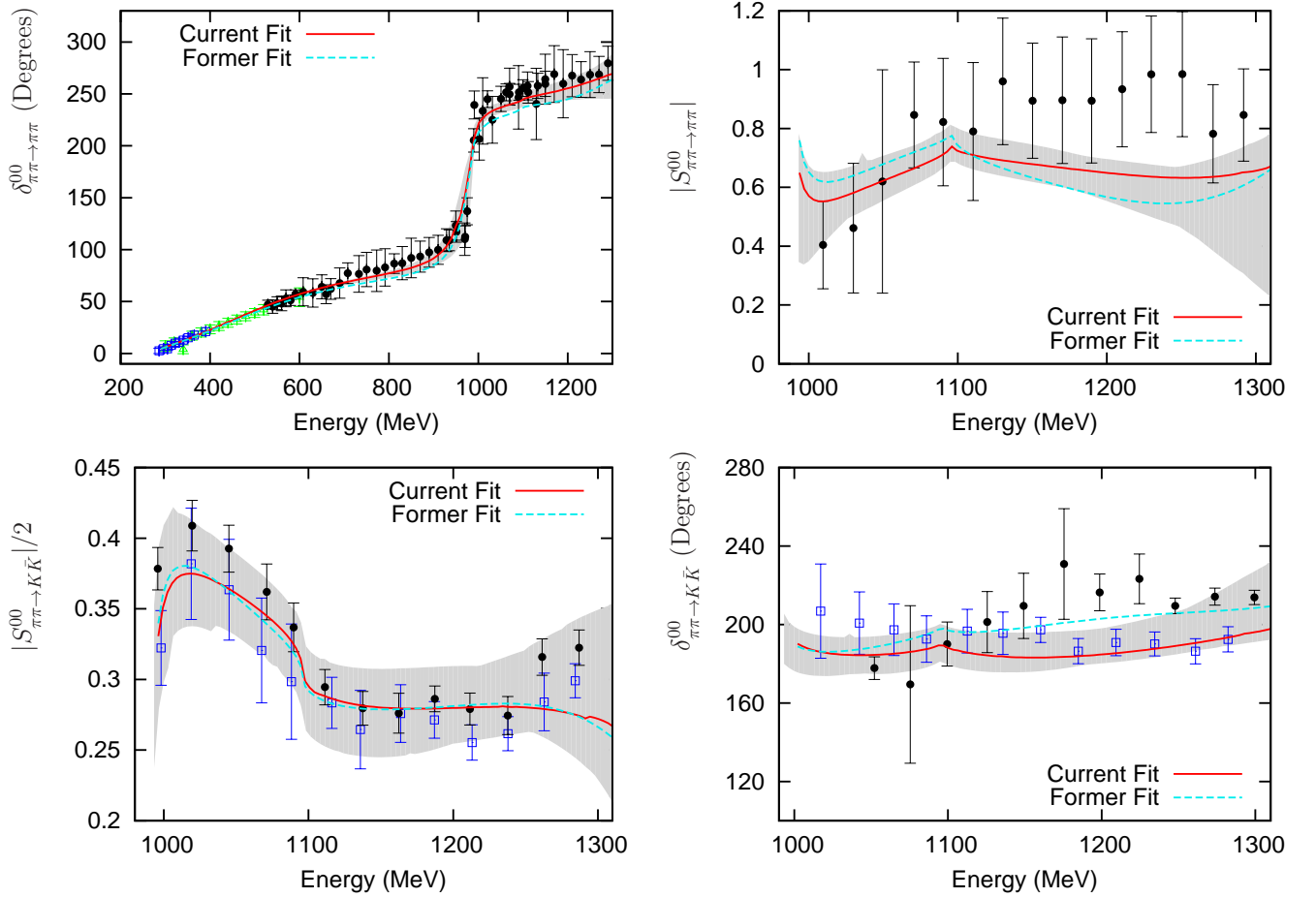}
\caption{{\small Plots for the $IJ = 00$ case from the current and former fits. From top to bottom and left to right: 
the $\pi\pi \to \pi\pi$ phase shifts ($\delta_{\pi\pi\to\pi\pi}^{00}$), 
the modulus of the $\pi\pi \to \pi\pi$ S-matrix element ($|S_{\pi\pi\to\pi\pi}^{00}|$), 
 one half of the modulus of the $\pi\pi \to K\bar{K}$ S-matrix element ($|S_{\pi\pi\to K\bar{K}}^{00}|/2$) 
and the $\pi\pi \to K\bar{K}$ phase shifts ($\delta_{\pi\pi\to K\bar{K}}^{00}$).
The solid (red) line corresponds to the current fit, Eq.~\eqref{fitresultnew}, and the dashed (blue) line 
represents the former fit of Ref.~\cite{guo11prd}. 
The error bands are represented by the shadowed areas, which are calculated using Eq.~\eqref{fitresultnew}. 
 The data for $\delta_{\pi\pi\to\pi\pi}^{00}$  are  from Refs.~\cite{frogratt77npb} (green triangle), \cite{na48}
(blue square) and the average data from Refs.~\cite{ochs74,refpipiphase,kaminski97zpc} (black circle), as used in Ref.~\cite{oller99prd}.  
The data for $|S_{\pi\pi \to \pi\pi}^{00}|$ are from Ref.~\cite{ochs74} while those for  
$|S_{\pi\pi \to K \bar{K}}^{00}|/2$ are  from Refs.~\cite{cohen80prd} (blue square) and \cite{martin79npb} (black circle). 
The phase shifts $\delta_{\pi\pi \to K \bar{K}}^{00}$ correspond to the data from Refs.~\cite{cohen80prd} (blue square) and \cite{etkin} (black circle). 
\label{fig.fitp1}}}
\end{center}
\end{figure}

\begin{figure}[H]
\begin{center}
\includegraphics[angle=0, width=0.99\textwidth]{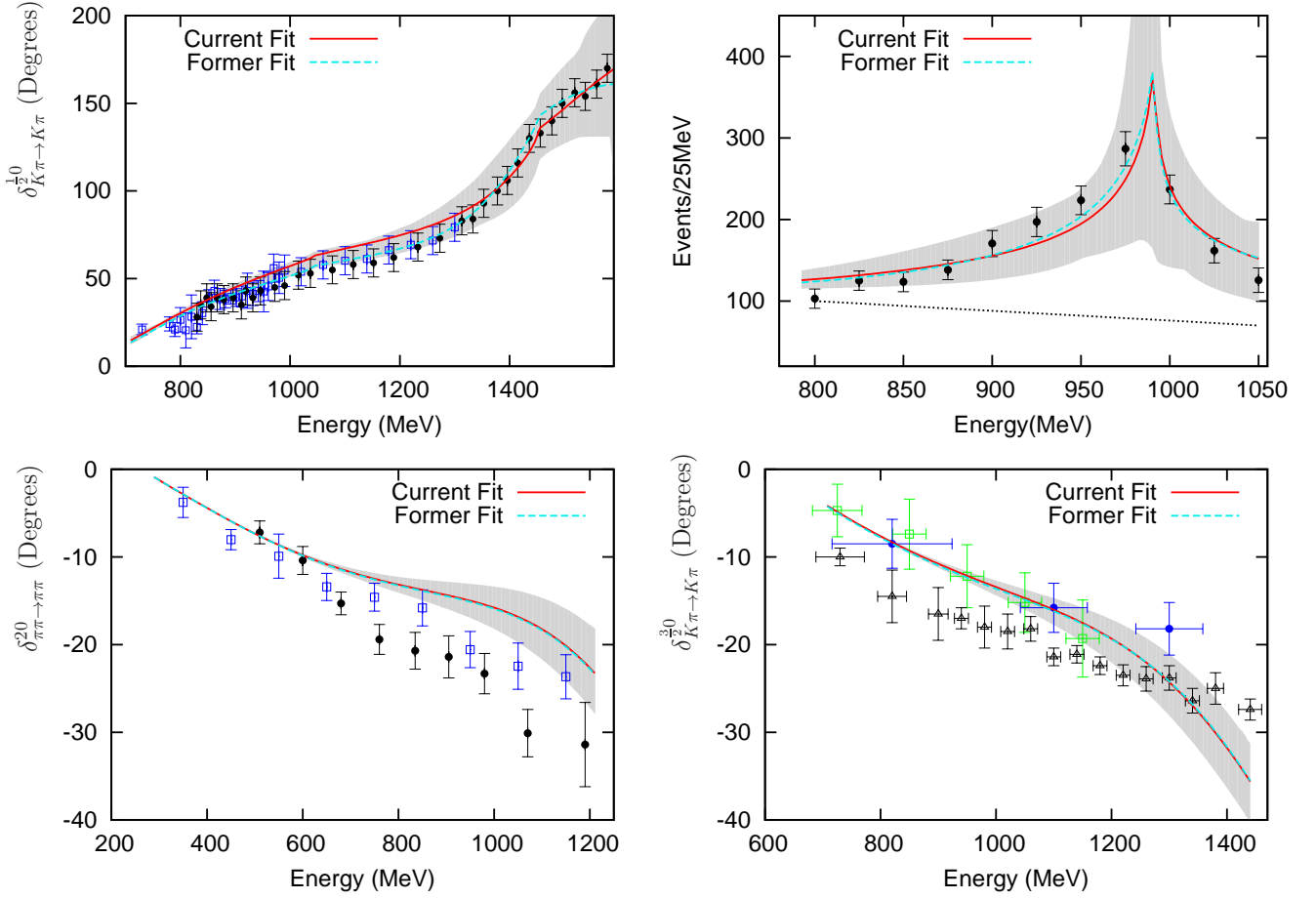}
 \caption{{\small From top to bottom and left to right: 
the $K \pi \to K\pi$ phase shifts with quantum numbers $IJ= \frac{1}{2}\,0$ ($\delta_{K \pi \to K\pi}^{\frac{1}{2} 0}$), 
the $\pi\eta$ event distribution with $IJ=1\,0$, the $\pi\pi \to \pi\pi$ 
phase shifts with $IJ= 2\,0$ ($\delta_{\pi\pi \to \pi\pi}^{2\,0}$) and the $K \pi \to K\pi$ 
phase shifts with $IJ= \frac{3}{2}\,0$ ($\delta_{K\pi \to K\pi}^{\frac{3}{2}0}$). 
The experimental points for $\delta_{K \pi \to K\pi}^{\frac{1}{2}0}$ correspond to the average data from 
Refs.~\cite{mercer71npb,estabrooks78npb,bingham72npb} (blue square), as employed in Ref.~\cite{oller99prd},
and Ref.\cite{aston88npb} (black circle).  
Data points for the $\pi\eta$ event distribution are taken from Ref.~\cite{armstrong91zpc} and  
the dotted line stands for the background \cite{oller99prd}. 
The data for $\delta_{\pi\pi \to \pi\pi}^{20}$  correspond to 
Refs.~\cite{hoogland77npb} (blue square) and  \cite{losty74npb} (black circle).
The experimental data of $\delta_{K \pi \to K\pi}^{\frac{3}{2}0}$ are from 
Refs.~\cite{bakker70npb} (green square), \cite{cho70plb} (blue circle) and \cite{estabrooks78npb} (black triangle). 
For the notation on the lines see Fig.~\ref{fig.fitp1}.
 } }
\label{fig.fitp2}
\end{center}
\end{figure}

\begin{figure}[H]
\begin{center}
\includegraphics[angle=0, width=0.99\textwidth]{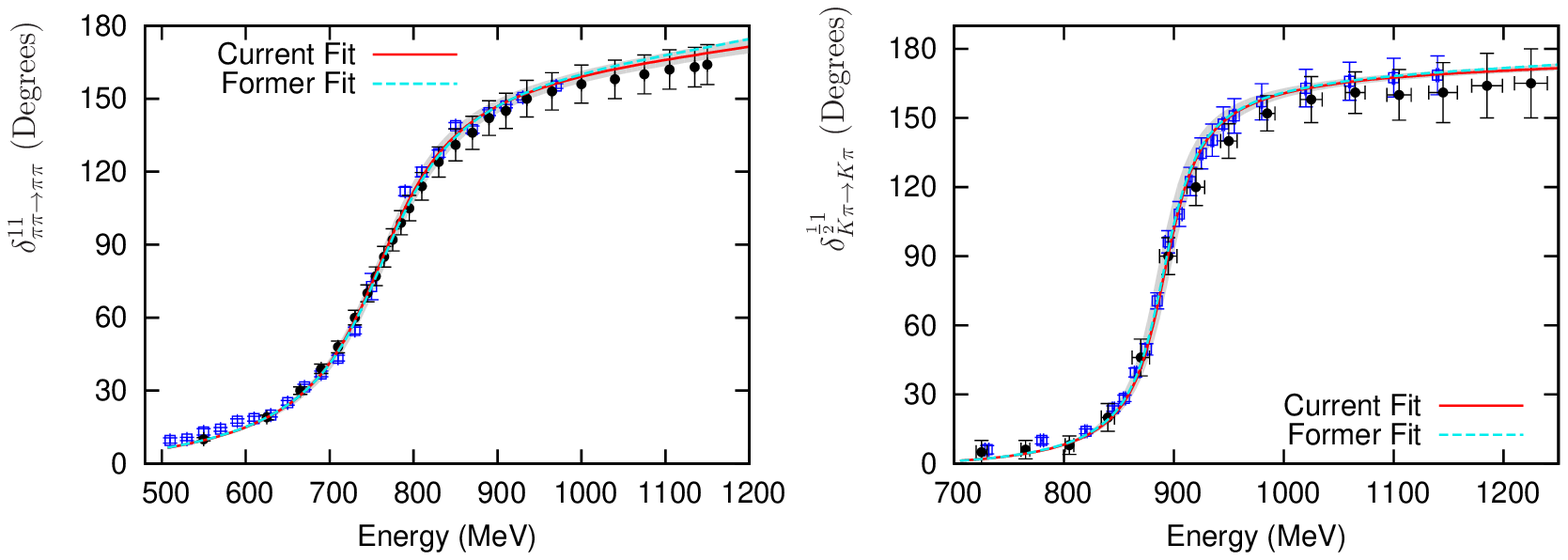}
 \caption{{\small The left panel displays the results for the $IJ=11$ channel 
and the right one is for the $IJ=\frac{1}{2}1$ case. 
The $\pi\pi \to \pi\pi$ phase shifts in the $IJ=11$ case, $\delta_{\pi\pi \to \pi\pi}^{11}$, correspond to  
Refs.~\cite{lindenbaum92plb} (blue square) and \cite{estabrooks74npb} (black circle). 
The $K\pi \to K\pi$ phase shifts with $IJ=\frac{1}{2}1$, $\delta_{K\pi \to K\pi}^{\frac{1}{2}1}$, are taken from 
Refs.~\cite{mercer71npb} (black circle) and \cite{estabrooks78npb} (blue square). For the notation on the lines see Fig.~\ref{fig.fitp1}.
} }
\label{fig.fitp3}
\end{center}
\end{figure}

The resulting plots from the fit in Eq.~\eqref{fitresultnew} are shown in Figs.~\ref{fig.fitp1}, \ref{fig.fitp2}
and ~\ref{fig.fitp3} by the solid curves, where we have also shown the best fit results from Ref.~\cite{guo11prd} by the dashed lines. 
The masses of $\eta$ and $\eta'$ from the new fit read
\begin{equation}
m_{\eta} = 536.7_{-39.6}^{+43.3}\,{\rm MeV}\,, \qquad m_{\eta'} = 956_{-30.4}^{+45.9}\,{\rm MeV}\,,
\end{equation}
which are clearly improved comparing with the ones from Ref.~\cite{guo11prd}.  
And the leading order mixing angle of $\eta_1$ and $\eta_8$ introduced in Eq.~\eqref{deflomixing} is 
\begin{equation}
\theta = -(15.1^{+2.4}_{-2.4})^\text{{\tiny o}}\,.
\end{equation}

Compared to the former fit from Ref.~\cite{guo11prd}, three more subtraction constants, 
$a_{SL}^{00\,,K\bar{K}}$, $a_{SL}^{\frac{1}{2} \,0\,, K\eta}$ and $a_{SL}^{\frac{1}{2} \,0\,, K\eta'}$ 
are set free in the current fit Eq.~\eqref{fitresultnew}, while we reduce now the scalar resonance parameters by
explicitly imposing the large $N_C$ relations of Eq.~\eqref{ncrelationcdtcmt}.  
 For the subtraction constants in the isoscalar scalar channel, $a_{SL}^{00\,,\pi\bar{\pi}}$ and $a_{SL}^{00\,,K\bar{K}}$ 
 are compatible with the former fit within  error bands. For the $IJ=\frac{1}{2}\,0$ channel, 
$a_{SL}^{\frac{1}{2} \,0\,, K\pi}$ and $a_{SL}^{\frac{1}{2} \,0\,, K\eta'}$ are quite similar and obey nonet symmetry approximately, 
while $a_{SL}^{\frac{1}{2} \,0\,, K\eta}$ is much more different. Nevertheless, one should notice 
that both $a_{SL}^{\frac{1}{2} \,0\,, K\eta}$ and $a_{SL}^{\frac{1}{2} \,0\,, K\eta'}$ carry especially large error bars. 
About the resonance parameter $c_d$, its value from the new fit is larger by around 30\% than the one from the former fit 
in Ref.~\cite{guo11prd}. This new value is   closer to those determined in other works~\cite{oller99prd,jamin00npb,guo09prd}. 
 The figure $c_d = 19.1^{+2.4}_{-2.1}$~MeV was reported in Ref.~\cite{oller99prd}, 
$c_d = 23.8$~MeV and $c_d = 26 \pm 7$~MeV were given in Refs.~\cite{jamin00npb,guo09prd}, respectively.   
For the parameter $c_m$, its value from the new fit also increases around 
30\% compared to that from the former fit in Ref.~\cite{guo11prd}. However, the error bars accompanying $c_m$ are now  considerably smaller. 
This is because in the new fit we impose the 
large $N_C$ relation for the singlet couplings, Eq.~\eqref{ncrelationcdtcmt}.  
Concerning the bare masses, both for the resonances (scalar and vector) and the singlet $\eta_1$, no appreciable
 changes in the new fit are seen, compared with the ones from the previous one ~\cite{guo11prd}. 
The central value of $\Lambda_2$ is now around a 60\% of that from the best fit in Eq.~(55) of Ref.~\cite{guo11prd}, though both 
determinations are affected by large uncertainties. 
For the vector resonance coupling $G_V$, the results from   
the current fit and the previous one in Ref.\cite{guo11prd} perfectly agree with each other. 
Concerning $\delta L_8$, which is introduced to compensate the uncertainties in the 
pseudoscalar resonance sector, it also carries large error bars.

The  quality of the current fit in Eq.~\eqref{fitresultnew} and 
the best fit in Ref.~\cite{guo11prd} is quite similar, 
as one can see from Figs.~\ref{fig.fitp1}, \ref{fig.fitp2} and ~\ref{fig.fitp3}. 
 Nevertheless, the prediction for $L_5$ from resonance contributions
using the new fit Eq.~\eqref{fitresultnew} is considerably increased, 
compared to the results in Eq.~\eqref{oldl4l5}, due to the inclusion 
of the second scalar multiplet. The resulting values for $L_4$ and $L_5$ are now
\begin{align}
 \label{newl4l5}
 L_4 (\mu = 770 {\rm MeV}) &= (0.09_{-0.04}^{+0.02} ) \times 10^{-3}~,\nn\\
 L_5 (\mu = 770 {\rm MeV}) &= (0.67_{-0.17}^{+0.04}) \times 10^{-3}~.
\end{align}
 They new value for $L_5$  agrees well with the recent determination $L_5=(0.58\pm 0.13) \times 10^{-3}$ from the latest ${\cal O}(p^6)$ $SU(3)$ $\chi$PT  fits of Ref.~\cite{jemos}. Concerning $L_4$ the latest reference cannot pin down a precise 
value, giving the result $L_4=(0.75\pm 0.75) \times 10^{-3}$. Our determination in Eq.~\eqref{newl4l5} is compatible 
with the latter number given its large uncertainty. 
Related to the larger value for  $L_5$ in Eq.~\eqref{newl4l5}, the quadratic pion scalar radius from the new fit is also improved  with the resulting value 
\begin{equation}\label{newr2}
\bra r^2 \ket_S^\pi = 0.49^{+0.01}_{-0.03} \,{\rm fm}^2\,,
\end{equation}
increasing around a 14\% compared to the value in Eq.~\eqref{oldr2} from the former fit, Eq.~(55) of Ref.~\cite{guo11prd}. 
The resulting scalar pion form factors and spectral functions are displayed 
in Figs.~\ref{fig.sffuuddss} and \ref{fig.impi} respectively, 
together with the results from the former fit \cite{guo11prd}. 

The resulting resonance pole positions on the complex plane from the new fit Eq.~\eqref{fitresultnew} are 
collected in Table \ref{tab:pole}. We refer to Ref.~\cite{guo11prd} for the discussions 
on how to perform the extrapolation from the physical sheet to the unphysical Riemann sheets. 
Around a  resonance pole $s_R$, corresponding to a resonance $R$, 
 the partial wave amplitude $T_J^I(s)_{i\to j}$ tends to 
\begin{align}
T_J^I(s)_{i\to j}\to -\frac{g_{R\to i}\, g_{R\to j}}{s-s_R}~.
\end{align}
 By calculating the residue of the resonance pole we then obtain the 
 product of the couplings  to the corresponding decay modes, $g_{R\to i}\, g_{R\to j}$.  
The pole positions for the vector resonances $\rho(770)$, $K^*(892)$ and $\phi(1020)$ agree perfectly 
with those in Ref.~\cite{guo11prd}. 
Only slight changes are observed for the $f_0(500)$, $f_0(980)$, $K^*_0(800)$ and $a_0(980)$ resonances. 
While all of the excited scalar resonances, such as the $f_0(1370)$, $K^{*}_0(1430)$ and $a_0(1450)$, have  
larger widths in the new fit Eq.~\eqref{fitresultnew} than in the previous one \cite{guo11prd}.

\begin{table}[ht]
\renewcommand{\tabcolsep}{0.05cm}
\renewcommand{\arraystretch}{1.2}
\begin{center}
{\small 
\begin{tabular}{|l|l|l|l|ll|}
\hline
%\\
R & M (MeV) & $\Gamma/2$ (MeV)&   $|$Residues$|^{1/2}$ (GeV) & Ratios &  
\\
[0.1cm] 
\hline
\hline
%\rule{0.1cm}{0.5cm}
%\\
$f_0(500)$& $442^{+4}_{-4}$ & $246^{+7}_{-5}$ & $3.02^{+0.03}_{-0.04}$ $(\pi\pi)$ 
&  $0.50^{+0.04}_{-0.08}$\,($K\bar{K}/\pi\pi$) &  $0.17^{+0.09}_{-0.09}$\,($\eta\eta/\pi\pi$) 
\\[0.1cm] 
& & & & $0.33^{+0.06}_{-0.10}$($\eta\eta'/\pi\pi$)\, & $0.11^{+0.05}_{-0.06}$($\eta'\eta'/\pi\pi$) 
\\[0.1cm]   
\hline 
$f_0(980)$& $978^{+17}_{-11}$ & $29^{+9}_{-11}$ & $1.8^{+0.2}_{-0.3}$($\pi\pi$)\,
& $2.6^{+0.2}_{-0.3}$($K\bar{K}/\pi\pi$)\,& $1.6^{+0.4}_{-0.2}$($\eta\eta/\pi\pi$)  
\\[0.1cm]
& & & & $1.0^{+0.3}_{-0.2}$($\eta\eta'/\pi\pi$)\,& $0.7^{+0.2}_{-0.3}$($\eta'\eta'/\pi\pi$)  
\\[0.1cm]  \hline 
$f_0(1370)$& $1360^{+80}_{-60}$ & $170^{+55}_{-55}$ & $3.2^{+0.6}_{-0.5}$($\pi\pi$)\, 
& $1.0^{+0.7}_{-0.3}$($K\bar{K}/\pi\pi$)\,& $1.2^{+0.7}_{-0.3}$($\eta\eta/\pi\pi$) 
\\[0.1cm]
& & & & $1.5^{+0.4}_{-0.5}$($\eta\eta'/\pi\pi$)\,& $0.7^{+0.2}_{-0.3}$($\eta'\eta'/\pi\pi$) 
\\[0.1cm]  
\hline
$K^*_0(800)$& $643^{+75}_{-30}$ & $303^{+25}_{-75}$ & $4.8^{+0.5}_{-1.0}$($K\pi$)\,
& $0.9^{+0.2}_{-0.3}$($K\eta/K\pi$)\,  & $0.7^{+0.2}_{-0.3}$($K\eta'/K\pi$)  
\\[0.1cm] 
 \hline  
$K^*_0(1430)$&  $1482^{+55}_{-110}$ & $132^{+40}_{-90}$ & $4.4^{+0.2}_{-1.1}$($K\pi$)\,
 & $0.3^{+0.3}_{-0.3}$ ($K\eta/K\pi$)\, & $1.2^{+0.2}_{-0.2}$($K\eta'/K\pi$) \\
[0.1cm]
  \hline  
$a_0(980)$
& $1007^{+75}_{-10}$ & $22^{+90}_{-10}$ & $2.4^{+3.2}_{-0.4}$($\pi\eta$)\, 
& $1.9^{+0.2}_{-0.5}$ ($K\bar{K}/\pi\eta$)\, & $0.03^{+0.10}_{-0.03}$($\pi\eta'/\pi\eta$)
 \\[0.1cm]  \hline 
$a_0(1450)$
& $1459^{+70}_{-95}$ & $174^{+110}_{-100}$ & $4.5^{+0.6}_{-1.7}$($\pi\eta$)\,
& $0.4^{+1.2}_{-0.2}$($K\bar{K}/\pi\eta$)\, & $1.0^{+0.8}_{-0.3}$($\pi\eta'/\pi\eta$) 
 \\[0.1cm]  \hline  
$\rho(770)$& $760^{+7}_{-5}$ & $71^{+4}_{-5}$ & $2.4^{+0.1}_{-0.1}$($\pi\pi$)\, 
& $0.64^{+0.01}_{-0.02}$($K\bar{K}/\pi\pi$) & 
 \\[0.1cm]  \hline 
$K^*(892)$& $892^{+5}_{-7}$ & $25^{+2}_{-2}$ & $1.85^{+0.07}_{-0.07}$($K\pi$)\,
& $0.91^{+0.03}_{-0.02}$($K\eta/K\pi$)\,& $0.41^{+0.07}_{-0.06}$($K\eta'/K\pi$) 
 \\[0.1cm] \hline 
$\phi(1020)$& $1019.1^{+0.5}_{-0.6}$ & $1.9^{+0.1}_{-0.1}$ & $0.85^{+0.01}_{-0.02}$($K\bar{K}$)\,
&   &  
 \\[0.1cm]  
\hline
\end{tabular}
}
 \caption{ {\small Pole positions for the different resonances in  $\sqrt{s} \equiv (\text{M},-i\frac{\Gamma}{2})$.
The mass (M) and the half width ($\Gamma/2$)  are given in units of MeV. 
The modulus of the square root of a residue is given in units of GeV, 
which corresponds to  the coupling of the resonance with the first channel 
(specified inside the parentheses). 
 The last two columns are the ratios of the coupling strengths of
the same resonance to the remaining channels with respect to the first one. 
Note that the residues for $\pi\pi$, $ \eta\eta$ and $\eta'\eta'$ 
are given in the unitary normalization, see Ref.~\cite{guo11prd}. 
} }  
\label{tab:pole}
\end{center}
\end{table}

%%%%%%%%%%%%%%%%%%%%%%%%%%%%%%%%%%%%%%%%%%%%%%%%%%%%%%%%%%%%%%%%%%%%%%%%%%%%%%%%%%%%%%%
\subsection{Phenomenological results of semilocal duality }

We show in Table \ref{tab:ratiosatnc3} the values of the ratios defined in Eqs.~\eqref{defRratio} 
and \eqref{defFratio} from the current fit, Eq.~\eqref{fitresultnew}.  
The dependence of these ratios with $N_C>3$ is discussed later. 
We also consider the contributions of the  $D$-waves to the integrals in Eqs.~\eqref{defRratio} and \eqref{defFratio}, 
 in addition to the $S$- and $P$-waves. For including the tensor resonances, which originate the $D$-waves,  we follow the formalism 
of Ref.~\cite{ecker07epjc} and take for the couplings the values determined there. 
The bare mass is adjusted such that the physical mass of the $f_2(1270)$ from the pole position agrees 
with the value in the PDG \cite{pdg}. To avoid interrupting the current discussion, 
we give the expressions for the tensor contributions to meson-meson scattering in Appendix~\ref{app.tensor}. 

In the leftmost column in Table~\ref{tab:ratiosatnc3} we indicate the partial waves involved in the evaluation of the integrals in Eqs.~\eqref{defRratio} and \eqref{defFratio}. The values of $n$ 
considered are given in the second column. In the rest of columns we give the quantities $R_n^I$ and $F_n^{21}$, as indicated in the top row. The values $t=0$ and $t_{{\rm}}=4 m_\pi^2$ are used in order to show the stability of the results under changes in $t$ that are small compared with GeV$^2$.
For the different quantities one should compare the numbers from Regge exchange and those obtained by including  only the $(S+P)$-waves or in addition 
including as well the $D$-waves. Our results in Table~\ref{tab:ratiosatnc3} quantitatively confirm the conclusions of Ref.~\cite{pelaez11prd}, that semilocal duality for $n=3$  with  $I_{\rm t}= 0$ and 1 
 can be perfectly satisfied by including only the $S$- and  $P$-waves, while 
the fulfillment for $n=2$ is already marginal. For smaller values of $n$, the higher partial waves and higher cut-offs are crucial 
in order to satisfy  semilocal duality. In this respect, we observe that once the $D$-waves are included 
 semilocal duality is satisfied better for all the $n$ values discussed, 
but particularly  for $n=0$ and 1. We have also considered the role of the $\rho(1450)$ but it is negligible if one takes 
the $\pi\pi$ branching decay ratio from the PDG~\cite{pdg}, which is only 6\%.

For $I_{\rm t}=2$ the situation is somewhat different. 
Before discussing the different numbers for $F_n^{21}$  in Table~\ref{tab:ratiosatnc3}, let us first comment on some specific 
values for the ratio $F_n^{21}$ in order to set up a criteria that allows one to consider a value small and then acceptable
 for satisfying semilocal duality. From Eq.~\eqref{tsrelation}, one has that $F_n^{21} \to -1$, if  the scalar 
contribution is dropped (the absorptive part of the $I=2$ channel should be negligible compared with that of the scalar and vector channels).
In contrast,  $F_n^{21} \to 1$ results by neglecting the vector contribution. As we commented before, the ratio 
of $F_n^{21}$ should vanish if semilocal duality works well. 
Taking this in mind we then see that with the $S$- and $P$-waves, we do not find any significant signal 
that semilocal duality is better satisfied for a specific value of $n$, even in some cases 
 duality is satisfied worse for a larger value of $n$, in contrast to the situations with $I_{\rm t}= 0$ and  
$I_{\rm t}= 1$. However, in all cases the numbers are much smaller than 1 in absolute value, so that semilocal duality seems to 
be fulfilled quite accurately. In our scattering amplitudes higher scalar resonances are generated, 
instead of only the $f_0(500)$ as in Ref.~\cite{pelaez11prd},  
which leads to an improvement for the $I_{\rm t}= 2$ channel by comparing the numbers for $F_n^{21}$ 
in Table \ref{tab:ratiosatnc3} with the ones in Table VI of  Ref.~\cite{pelaez11prd}. 
The masses of the heavier scalar resonances in our scattering amplitudes are close to or larger than 1~GeV$^2$, as shown in 
Table \ref{tab:pole}. Hence, only their effects can be taken into account in the discussion of semilocal duality when 
the integration upper limit $\nu_{\rm max}$ in Eq.~\eqref{defFratio} is larger than  1~GeV$^2$. 
Indeed, had we set instead $\nu_{\rm max}$=1~GeV$^2$ the fulfillment of semilocal duality would be much worse than 
for the $\nu_{\rm max}$=2~GeV$^2$ case, especially for $n=0$ and 1. Then, in the later discussions, 
we only consider the ratio $F_n^{21}$ in Eq.~\eqref{defFratio} calculated at $\nu_{\rm max}$=2~GeV$^2$. 
 On the other hand, we find that the introduction of the $D$-waves, instead of narrowing the gap between the Regge prediction and the h.d.f., 
 worsens the situation for $I_{\rm t}= 2$ in the $n=0$ case. It is then advisable to focus in this work on $n>0$ for $F_n^{21}$
 \cite{RuizdeElvira:2010an}.

\begin{table}[ht]
\renewcommand{\tabcolsep}{0.05cm}
\renewcommand{\arraystretch}{1.2}
\begin{center}
{\small 
\begin{tabular}{|l|l|llll|llll|llll|llll|}
\hline
%\\
 & n &&  $R_n^0$   && $R_n^0$  && $R_n^1$  && $R_n^1$ && $F_n^{21}$  && $F_n^{21}$  
\\
 &   &&  $t=t_{\rm th}$   && $t=0$   && $t=t_{\rm th}$  && $t=0$  &&  $t=t_{\rm th}$  && $t=0$
\\
 &   &&     &&   &&   &&   && {\footnotesize  $\nu_{\rm max} = $ } &&  2~GeV$^2$   
\\
[0.1cm] 
\hline
\hline
%\rule{0.1cm}{0.5cm}
%\\
Regge & 0 && 0.225 && 0.233 &&  0.325 &&  0.353  &&   $\sim 0$ &&  $\sim 0$   
\\[0.1cm] 
      & 1 && 0.425 && 0.452 &&  0.578 &&  0.642  &&   $\sim 0$ &&  $\sim 0$  
\\[0.1cm] 
      & 2 && 0.705 && 0.765 &&  0.839 &&  0.908  &&   $\sim 0$ &&  $\sim 0$   
\\[0.1cm] 
      & 3 && 0.916 && 0.958 &&  0.966 &&  0.990  &&   $\sim 0$ &&  $\sim 0$  
\\[0.1cm]   
\hline 
Ours  & 0   && 0.669 && 0.628 &&  0.836 &&  0.817 &&  -0.113 &&  0.040    
\\[0.1cm] 
$S+P$   & 1 && 0.837 && 0.812 &&  0.919 &&  0.908 && -0.230   &&  -0.087   
\\[0.1cm] 
Waves & 2   && 0.934 && 0.924 &&  0.966 &&  0.962 && -0.129   &&  0.028   
\\[0.1cm]  
      & 3   && 0.979 && 0.976 &&  0.989 &&  0.988 &&   0.169 &&   0.345  
\\[0.1cm]   
\hline 
Ours  & 0   && 0.410  && 0.400  &&  0.453  &&  0.468  &&  0.531  &&   0.587    
\\[0.1cm] 
$S+P+D$ & 1 && 0.653  && 0.643  &&  0.694  &&  0.706  &&  0.154  &&   0.236   
\\[0.1cm] 
Waves & 2   && 0.850  && 0.844  &&  0.875  &&  0.882  &&  0.027  &&   0.155    
\\[0.1cm] 
      & 3   && 0.954  && 0.953  &&  0.965  &&  0.968  &&  0.225  &&   0.388  
\\[0.1cm]  
\hline 
\hline
\end{tabular}
}
 \caption{{\small Current Fit: $R_n^I$ and $F_n^{II'}$ are defined in 
Eqs.~\eqref{defRratio} and \eqref{defFratio}, respectively. In the first column from the left the amplitudes involved in their 
evaluation are shown. The different values of $n$ are given in the second column. The rest of the columns correspond to $R_n^I$ and $F_n^{21}$ as 
indicated. Two values of $t$, 0 and $t_{{\rm th}}\equiv 4m_\pi^2$, are considered, as shown in the second row. }
 }  
\label{tab:ratiosatnc3}
\end{center}
\end{table}

The ratios $F_n^{20}$ are smaller in absolute value than $F_n^{21}$ because the coefficient multiplying $T_{\rm s}^{(1)}$ 
in Eq.~\eqref{tsrelation} is larger by a factor of two  for $T_{\rm t}^{(0)}$ than for  $T_{\rm t}^{(1)}$. 
In Table~\ref{tab:ratiosatnc3} we do not display their values since they can not reveal any new information comparing with $F_n^{21}$.

%%%%%%%%%%%%%%%%%%%%%%%%%%%%%%%%%%%%%%%%%%%%%%%%%%%%%%%%%%%%%%%%%%%%%%%%%%%%%%%%%%%%%%%%%%%%%%%%%%%%%%%%%%
\subsection{Study of spectral-function sum rules}

After fixing the unknown parameters through the fit to data, we are ready to investigate the 
spectral-function sum rules presented in Eq.~\eqref{defweinbergsr}.
 To study them one has to include not only nonperturbative QCD dynamics but also perturbative QCD and operator product expansion (OPE) \cite{wi69,pascual}. 
 In this way we split the integral in two parts 
\begin{equation}\label{defweinbergsr2}
\int_0^{s_0} \big[ {\rm Im}\,\Pi_R(s) - {\rm Im}\,\Pi_{R'}(s) \big]\,  d s
+ \int_{s_0}^{\infty} \big[ {\rm Im}\,\Pi_R(s) - {\rm Im}\,\Pi_{R'}(s) \big]\,d s  = 0\,.
\end{equation}
The first integral, that extends along the lower-energy regime, comprises the nonperturbative region 
and we use our results in terms of h.d.f. to evaluate it. For the second one, higher in energy, the results 
from OPE are employed to evaluate the theoretical spectral functions.  
According to the OPE study of Ref.~\cite{jamin92zpc} the different spectral functions with $R=S,~P$ and 
$R'=S,~P$ are equal in the asymptotic region in the chiral limit.\footnote{The calculation in Ref.~\cite{jamin92zpc} is done up to ${\cal O}(\alpha_s)$ 
and including up to dimension 5 operators.} 
As a result, the second integral in Eq.~\eqref{defweinbergsr2} is zero. Then, testing how well a spectral-function sum rule 
is satisfied reduces to evaluating the first integral, which extends along the energy region below $\sqrt{s_0}$. This is 
exactly the key object of our current study in this section.

As discussed above, we consider the strangeness conserving scalar 
and pseudoscalar spectral functions for  $a=0,8,3$. 
Hence there are 15 types of nontrivial spectral-function sum rules as those in Eq.~\eqref{defweinbergsr2}. 
In order to show the results in a compact way, we display the individual values for the integration up to $s_0$ for  
each of the phenomenological spectral functions in Table \ref{table.impi}, instead of the differences between the different spectral functions. 
Note that the second integral for $s>s_0$ is divergent, unless the difference between the spectral functions is taken as in Eq.~\eqref{defweinbergsr2}.  
This divergent behavior is not an issue for the first integral because, as shown in Fig.~\ref{fig.impi}, the phenomenological spectral functions 
are already very small for $s\gtrsim 2.5$~GeV$^2$. In this way, the results from the integration do not depend so much on $s_0$ as soon as they 
are larger than $\sim 2.5$~GeV$^2$. This vanishing behavior should be expected from Eq.~\eqref{defscspecf}. In the latter  only a finite number of 
two-body channels are considered so that if the form factors vanish for $s\to \infty$ (as expected from QCD counting rules \cite{lepa}) 
so their contribution to the spectral function does. Note also that $\rho_i(s)$, given in Eq.~\eqref{defkineticsigma}, tends to constant for $s\to \infty$. 
The definitions of the different quantities in Table \ref{table.impi} are  
\begin{align} \label{defwi}
 W_i &= 16\pi \int_0^{s_0}   {\rm Im}\,\Pi_i(s)  \,  d s\,, 
\\
\label{defwbar}
\overline{W} &= \frac{1}{3\times6}\sum_i W_{i}\,,
\\
\label{defvariance}
\sigma_{W}^2 &=  \sum_{i}  \frac{(W_{i}-\overline{W})^2}{17} \,,\quad  i=S^8,S^0,S^3,P^0,P^8,P^3\,,
\end{align} 
where we take three different values of $s_0$ to evaluate the integrations in order to show 
the dependences of the integrated results on $s_0$. 
The relative variance $\sigma_W/\overline{W}$ serves as a parameter to quantify how well the spectral-function  
sum rules in the scalar and pseudoscalar cases hold.

Two situations by  taking different masses for the pseudo-Goldstone bosons are investigated: 
the physical case and the chiral limit. In order to study the results at the chiral limit, 
it is necessary to perform the chiral extrapolation of our spectral functions. 
Though the resonance parameters in Eqs.~\eqref{lagscalar}-\eqref{kinersp} do not depend on the quark masses, 
the subtraction constants $a_{SL}$ in Eq.~\eqref{defgfuncionloop} introduced through the unitarization procedure 
 vary with them. In Ref.~\cite{Jido:2003cb}, it is demonstrated that in the $SU(3)$ limit (as also in chiral limit) 
all the subtraction constants should be equal for any pseudo-Goldstone pair made of the $\pi$, $K$ and $\eta_8$ mesons. 
Thus we need to extrapolate the subtraction constants from the fit, which are not necessarily equal to each other, to a common value. 
We find that at the chiral limit such a value indeed exists in a reasonable region (roughly from $-1$ to 0), where the results of the two-point 
correlators are stable and the spectral-function sum rules are better satisfied compared to the physical situation.  
In the following, we show the typical results in this region (with a common value taken for the subtraction constant at the chiral limit of $-0.5$).

\begin{figure}[H]
\begin{center}
\includegraphics[angle=0, width=0.99\textwidth]{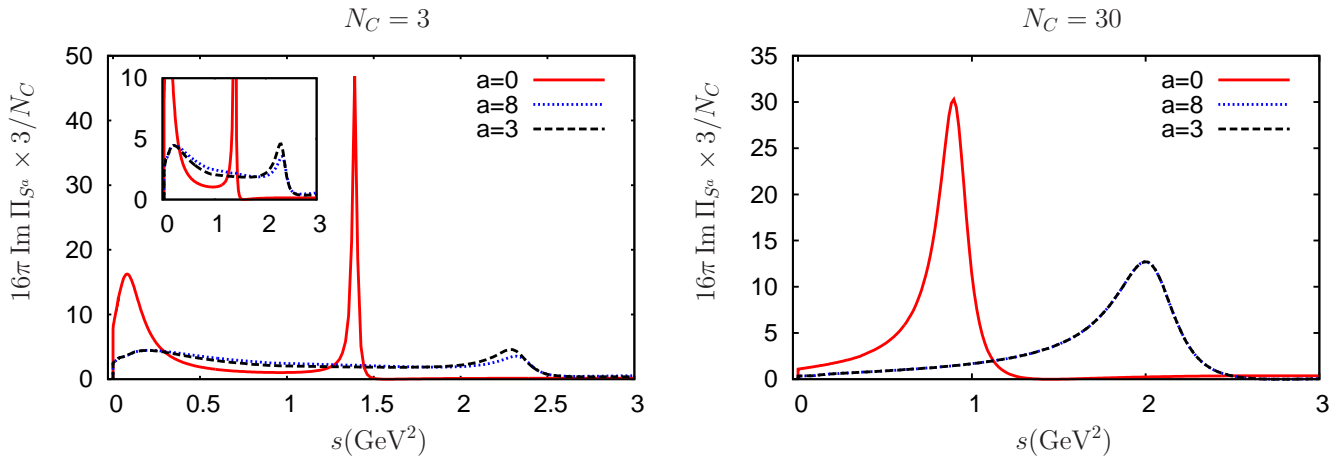}
\caption{  {\small The scalar spectral functions at the chiral limit for $N_C=3$ (left panel) and 
$N_C=30$ (right panel) with the new fit of Eq.~\eqref{fitresultnew}. The values $a=$0 (red solid line), 3 (black dashed line) 
and 8 (blue dotted line) are considered. The inset in the left panel shows the same figure with reducing  scale, so that the spectral functions 
with $a=3$ and 8 are better seen. }
\label{fig.impicl}}
\end{center}
\end{figure}

The corresponding scalar spectral functions at the chiral limit  are shown in Fig.~\ref{fig.impicl}.\footnote{We show for later convenience the same spectral functions   with $N_C=30$ in the right panel of Fig.~\ref{fig.impicl}.} 
It is easy to conclude from Table \ref{table.impi} that the spectral-function sum rules are much better fulfilled 
by the new fit in Eq.~\eqref{fitresultnew} than by the former one of  \cite{guo11prd}. 
The most significant changes in  $W_i$ from both fits happen for the pseudoscalar cases, with $i=P^0$, $P^8$ and $P^3$,  
which are caused by the pseudoscalar resonances that are now included in the new fit. 
We obtain that the smallest value for the violation of the spectral-function sum rules, around 10\%, takes place at chiral limit by 
using the new fit result. Nevertheless, the result for the physical case from the new fit is also quite similar, with a violation of around 16\%.

\begin{table}[H]
\renewcommand{\tabcolsep}{0.06cm}
\renewcommand{\arraystretch}{1.2}
\begin{footnotesize}
\begin{center}
\begin{tabular}{|c|c|c|c|c|c|c|c|c|c|}
\hline
 &  $W_{S^0}$ & $W_{S^8}$ & $W_{S^3}$ &  $W_{P^0}$ & $W_{P^8}$ & $W_{P^3}$& $\overline{W}$ & $\sigma_W$ & $\sigma_W/\overline{W}$ \\
\hline
Physical masses &&&&&&&&& \\
Current Fit    & 8.6\,\,\,9.0\,\,\,9.6 & 7.4\,\,\,7.5\,\,\,7.7   & 7.0\,\,\,7.2\,\,\,7.4 &  8.9 & 11.3 & 10.1 & 9.0 & 1.5 & 0.16\\
Former Fit & 6.1\,\,\,6.1\,\,\,6.1 & 6.1\,\,\,6.1\,\,\,6.1   & 5.8\,\,\,5.9\,\,\,6.1 &  1.8 & 5.0 & 5.1 & 5.0 & 1.5 & 0.31 \\
\hline
$m_q=0$ &&&&&&&&& \\
Current Fit    & 6.9\,\,\,7.0\,\,\,7.1 & 6.8\,\,\,7.0\,\,\,7.3   & 6.6\,\,\,6.8\,\,\,7.0 &  5.5 & 7.4 & 7.4 & 6.9 & 0.7 & 0.10 \\
Former Fit & 5.2\,\,\,5.3\,\,\,5.5 & 5.5\,\,\,5.7\,\,\,6.0   & 5.2\,\,\,5.3\,\,\,5.5 &  0.3 & 3.0 & 3.0 & 3.8 & 2.0 & 0.53 \\
\hline
\end{tabular}
\caption{\label{table.impi}{\small Results from the integration of the spectral 
functions from 0 up to $s_0$, Eq.~\eqref{defwi}. We show three results in the columns $W_{S^0}$, $W_{S^8}$
and $W_{S^3}$ by taking three different values for $s_0$: $ 2.5,\,3,\,3.5\,{\rm GeV}^2$. 
The results for the pseudoscalar cases are not changed for different $s_0$, since 
the pseudoscalar spectral functions are just some Dirac $\delta$-functions Eq.~\eqref{defpsspecf}.  
$W_i$ with {\footnotesize $i = S^0,S^8,S^3,P^0,P^8,P^3$ }, the mean value $\overline{W}$ and $\sigma_W$ are 
defined in Eqs.~\eqref{defwi}\eqref{defwbar}\eqref{defvariance}, which are given in units of GeV$^2$ in this table. 
In the last column we show the relative variance $\sigma_W/\overline{W}$. }}
\end{center}
\end{footnotesize}
\end{table}

In the left panel of of Fig.~\ref{fig.impicl} the singlet spectral function (solid line) is clearly dominated in the low energy region by a peak corresponding 
to a pole that evolves continuously with the pseudo-Goldstone boson masses from the  
$f_0(500)$ resonance pole at the physical case. This affinity of the $f_0(500)$ resonance to the singlet scalar source is 
in agreement with the study of Ref.~\cite{oller03npa}, which determined that the $f_0(500)$ meson was mostly a $SU(3)$ singlet. 
For the octet spectral function with $a=3$ one observes neatly another low energy peak (this is better seen in the inset of the left panel where the 
scale is changed to cover more adequately the values for $a=3$ and 8). This peak  is due to the $a_0(980)$ resonance in the chiral limit, 
as we have checked. The higher energy peak at around 2.3~GeV$^2$ for $a=3$,~8  is 
caused by the bare octet of scalar resonances with a common mass $M_{S_8}\simeq 1.4$~GeV, Eq.~\eqref{fitresultnew}, 
shifted to somewhat higher energies by interference with non-resonant dynamical contributions. Finally, the strong peak in the 
singlet spectral function at around 1.4~GeV$^2$ comes from several sources involving the bare singlet 
resonance $S_1$ with a mass $M_{S_1}\simeq 1.1$~GeV, Eq.~\eqref{fitresultnew}, and coupled channel dynamics 
with the $\eta\eta'$ state. The lightest scalar resonances in the chiral limit were studied earlier 
in Ref.~\cite{oller99prd} and the  pole positions for the $f_0(500)$ and $a_0(980)$ obtained there are in agreement with ours.

%%%%%%%%%%%%%%%%%%%%%%%%%%%%%%%%%%%%%%%%%%%%%%%%%%%%%%%%%%%%%%%%%%%%%%%%%%%%%%%%%%%%%%%%%%%%%%%%%%
\section{Results from the extrapolation of $N_C$ }
\label{sect.ncrunning}

Through the fit to experimental data in Sec.~\ref{sec:newfit}, we get the unknown couplings 
that appear in  the chiral Lagrangians as well as the  subtraction constants. 
One advantage to employ the chiral Lagrangian approach in the phenomenological study is 
that once the chiral couplings are determined from one or several sets of data, we 
completely predict the other quantities that can be calculated from the same theory. 
For example, the resonance pole positions, pion scalar radius, spectral functions,  
 spectral-function sum rules and semilocal duality that are discussed previously are 
all predictions from the fit to the scattering data.

Another advantage by using the chiral Lagrangian approach  is to 
study the behavior of the various quantities by extrapolating in the number of colors of QCD, $N_C$, 
and then to confront with the results from large $N_C$ QCD~\cite{largenc}. 
This is straightforward in the chiral Lagrangian approach once the $N_C$ behavior 
of the parameters from the chiral Lagrangians are known~\cite{ecker89npb,gasser85}. 
Moreover, $U(3)$ $\chi$PT  is more appropriate to discuss 
the large $N_C$ running compared to $SU(3)$ or $SU(2)$ $\chi$PT~\cite{gasser84,gasser85}, since the singlet 
$\eta_1$, explicitly included in the $U(3)$ chiral theory, becomes the ninth pseudo-Goldstone boson
in the large $N_C$ and chiral limits. Notice that this relevant degree of freedom is not treated as 
a dynamical active one in $SU(2)$ or $SU(3)$  $\chi$PT~\cite{gasser84,gasser85}. It is also worth stressing that  
the $\eta$ becomes much lighter with increasing $N_C$ \cite{wein2}, as explicitly shown in Ref.~\cite{guo11prd}, an effect disregarded 
in previous studies~\cite{nieves11prd,pelaezprl1,pelaezprl2,sun07mpl,xiao07mpa,guo07jhep,nieves09prd,Dai:2011bs}.

Only the leading order scaling with $N_C$ of the resonance parameters, such as couplings and masses,  is known 
without ambiguities~\cite{ecker89npb}.  
To show how robust is our knowledge on the $N_C$ behavior for the various quantities studied, 
the sub-leading orders for the parameters in the $N_C$ counting  are necessary and could also be important 
\cite{sun07mpl,nieves09prd,pelaez11prd,guo11prd,nieves11prd}. 
In Ref.~\cite{pelaez11prd}, this uncertainty induced by the sub-leading terms in the $1/N_C$ expansion 
of the LECs is estimated approximately by taking different values of the renormalization scale $\mu$. 
We adopt a direct way to estimate the sub-leading order of $1/N_C$ effects, 
that has been used in Ref.~\cite{nieves11prd}. The idea is that through the fit to data, one can determine the bare 
resonance couplings and masses from the Lagrangian, which represent their values at $N_C=3$. 
Once their values at large $N_C$ are known, we perform the most general smooth extrapolation from $N_C=3$ to the large $N_C$ values  up to 
and including $1/N_C$ suppressed corrections.   
 Of course, the values at large $N_C$ are not accessible directly by experiment, and can be ascertained only 
through theoretical considerations. In the last decades, a great progress along this line 
has been achieved in the analyses of short distance constraints of two- and three-point Green functions, 
form factors, $\tau$ decays and $\pi\pi$ scattering within R$\chi$T 
\cite{nieves11prd,jamin02npb,ecker89plb,pich02,cirigliano05jhep,pich11jhep,guo10prd,guo07jhep}. 

The pion decay constant $F_\pi$ is calculated from the one-loop $U(3)$ $\chi$PT in 
Ref.~\cite{guo11prd}, that includes also sub-leading terms in the $1/N_C$ expansion. Throughout we always consider 
both the leading and sub-leading $N_C$ scaling for $F_\pi$ when varying $N_C$ as given in Ref.~\cite{guo11prd}. 
The values for the fit parameters in Eq.~\eqref{fitresultnew} are the ones taken for $N_C=3$. 
On the other hand, due to the uncertainties of the values for the resonance 
parameters at large $N_C$, we consider several scenarios: 

\begin{itemize}
 \item  {\bf Scenario 1:} 
We take only the leading order running with large $N_C$ for all the resonance parameters,   
starting with their values at $N_C=3$. As discussed in more detail in Ref.~\cite{guo11prd,ecker89npb} the leading running with $N_C$ 
for the resonance parameters and meson-meson subtraction constants, $a_{SL}$, is given by:
\begin{align}
\label{ncrunning4res}
& \big\{c_d(N_C),c_m(N_C),G_V(N_C),d_m(N_C)\big\} \nn \\
&=\big\{c_d(3),c_m(3),G_V(3),d_m(3) \big\}\times \sqrt{\frac{N_C}{3}}~, 
\nn\\ &
\big\{ M_{S_1}(N_C),M_{S_8}(N_C),M_\rho(N_C),M_{K^*}(N_C),M_\omega(N_C),M_\phi(N_C),
\nn \\ & \quad M_{P_1}(N_C),M_{P_8}(N_C),a_{SL}(N_C) \big\} \nn\\& 
= \big\{ M_{S_1}(3),M_{S_8}(3),M_\rho(3),M_{K^*}(3),M_\omega(3),M_\phi(3),
\nn \\ & \qquad  M_{P_1}(3),M_{P_8}(3),a_{SL}(3)\big\}.
\end{align}
For the singlet couplings $\widetilde{c}_d$, $\widetilde{c}_m$ and $\widetilde{d}_m$, we take the 
large $N_C$ constraints in Eqs.~\eqref{ncrelationcdtcmt} and \eqref{lnc4dmt}. 

 About the $N_C$ running of the subtraction constant $a_{SL}$, we argue that it is natural to assume 
its constant behavior at large $N_C$, though some sub-leading $N_C$ corrections may exist. This is based on 
the fact that the unitarized scattering amplitude, defined in Eq.~\eqref{defunitarizedT}, is in fact 
the sum of a series of bubble diagrams with the kernel $N^{IJ}(s)$, since Eq.~\eqref{defunitarizedT} can 
be expanded as 
\begin{equation} \label{expandunitariedT}
T^{IJ}(s)=N^{IJ}(s) - N^{IJ}(s) g^{IJ}(s) N^{IJ}(s)+ N^{IJ}(s) [g^{IJ} (s)N^{IJ}(s)]^2 + ....  
\end{equation}
Within large $N_C$ QCD it is well known that the leading $N_C$ behavior of a meson-meson scattering amplitude, ${N_C}^{\alpha}$, corresponds to $\alpha=-1$ and it can also 
contain other sub-leading pieces with $\alpha=-2, -3,\ldots$ \cite{largenc}. This feature for  meson-meson scattering 
is inherited by the construction of $\chi$PT \cite{gasser85}. 
 Each single diagram in the geometric series expansion in powers of $g^{IJ}(s)$ 
 of the unitarized amplitude in Eq.~\eqref{expandunitariedT} should decrease with $N_C$ at least as $1/N_C$. 
 Focusing on the first term in Eq.~\eqref{expandunitariedT}, i.e. the kernel $N^{IJ}(s)$, 
it represents the perturbative results calculated from $\chi$PT and hence regardless of the resummation it should 
inherit the $N_C$ behavior of meson-meson scattering amplitudes dictated by  large $N_C$ QCD. Its calculation within $\chi$PT tells us that   
 it scales as $N_C^\alpha$ with $\alpha= -1$, including typically other sub-leading components. 
 An immediate conclusion that follows is that the $N_C$ scaling index $\alpha$ 
for $g^{IJ}(s)$ can be only an integer, following the above arguments. Moreover, $\alpha \geqslant 2$ can be also simply 
excluded otherwise the terms with $g^{IJ}(s)$ in Eq.~\eqref{expandunitariedT} could violate the large $N_C$ QCD prediction to the scattering amplitudes. 
The case $\alpha=0$, i.e. $g^{IJ}(s)$ behaves as a constant at large $N_C$, is indeed the natural choice for the following reasons:
\\
i) The relative size between a term and the next one in the expansion of  Eq.~\eqref{expandunitariedT} is 
$g^{IJ}(s) N^{IJ}(s)$. At leading order $N^{IJ}(s)$ behaves as $p^2/F^2$, being $p^2$ a typical soft external four-momentum squared attached to the pseudo-Goldstone bosons. From the subtraction constant $a_{SL}$ in Eq.~\eqref{defgfuncionloop} we then have the suppression factor
\begin{align}
\frac{a_{SL}\,p^2}{(4\pi F)^2}~.
\label{fact.supp}
\end{align}
For subtraction constants $a_{SL}$ of ${\cal O}(1)$ size, as the fitted values shown in Eq.~\eqref{fitresultnew}, one then has the typical suppression for unitarity loops in $\chi$PT, given in terms of the chiral symmetry breaking scale $\Lambda_{\chi P T}=4\pi F$ \cite{georgi}. In order to keep this interpretation with running $N_C$ it is necessary that every unitarity loop is suppressed by an extra power of $1/N_C$ and, for that, the subtraction constants $a_{SL}$ should be ${\cal O}(N_C^0)$.

ii) The combination 
\begin{align}
a_{SL}(\mu)-\log \mu^2
\end{align}
in Eq.~\eqref{defgfuncionloop} is independent of the renormalization scale $\mu$. Let us consider another value $\mu'$ for which $a_{SL}(\mu')=0$. From the previous equation it follows that
\begin{align}
{\mu'}=\mu\, e^{-a_{SL}(\mu)/2}~.
\label{newscale}
\end{align}
From here it is obvious that if $|a_{SL}(\mu)|$ is too different from 1 then $\mu'$ exponentially diverges for $a_{SL}(\mu)\ll -1$ 
or tends to 0 for $a_{SL}\gg 1$. In both cases one has too different values from the typical one for 
a renormalization scale in $\chi$PT, $\mu \sim 0.5-1$~GeV, of the similar size to the previously introduced chiral symmetry 
breaking scale $\sim 1$~GeV or the mass of the $ \rho$ resonance. The fitted values for the subtraction constants in Eq.~\eqref{fitresultnew} 
have the right size so as to keep an adequate value for $\mu'$ which does not scale with $N_C$ (nor should the $a_{SL}(\mu)$ 
so that Eq.~\eqref{newscale} is meaningful).

As commented in Eq.~\eqref{relatel8chptanddeltal8}, since we explicitly include 
the resonance contributions to $L_8$, that grows like $N_C$ in the $1/N_C$ expansion~\cite{ecker89npb}, 
 we consider that $\delta L_8$ is just some remnant piece  sub-leading in $N_C$. So we take 
\begin{equation}\label{ncrunning4deltal8}
 \delta L_8(N_C) = \delta L_8(3) \,,
\end{equation}
throughout the following discussion. 
Concerning the parameters $\Lambda_2$ and $M_0$, 
their leading $N_C$ scaling reads \cite{kaiser00epjc}
\begin{equation}\label{ncrunning4m0lam2}
\left\{\Lambda_2(N_C), M_0^2(N_C) \right\} = \left\{\Lambda_2(3), M_0^2(3) \right\}\times \frac{3}{N_C}\,.  
\end{equation}

\item {\bf Scenario 2:}  Comparing with Scenario 1, we include the sub-leading $N_C$ scaling for the vector 
resonance parameter $G_V$ in Eq.~\eqref{lagvector}, which describes the 
interaction between the vector resonances and the pseudo-Goldstone boson pairs, e.g. the $\rho(770)\pi\pi$ coupling. 
The original type of Kawarabayashi-Suzuki-Riazuddin-Fayyazuddin (KSRF) relation~\cite{ksrf} predicts $G_V = F/\sqrt{2}$.  
This relation was also derived from the high energy constraint of the pion vector form factor at tree level \cite{ecker89plb}. 
An updated study of the vector form factor at the one-loop level \cite{pich11jhep} revealed a new version for the constraint: 
\begin{equation}\label{gvatnc}
G_V = \frac{F}{\sqrt{3}}\,.  
\end{equation}
This modified KSRF-like relation has also been confirmed in various contexts: 
partial wave $\pi\pi$ scattering~\cite{guo11prd,guo07jhep}, radiative tau decay~\cite{guo10prd} and 
extra-dimension model for $\pi\pi$ scattering~\cite{chivukula07prd}. 
The large $N_C$ value for the pion decay constant in the chiral limit can be deduced from the $U(3)$ $\chi$PT study of Ref.~\cite{guo11prd} 
with the current fit results in Eq.~\eqref{fitresultnew}, leading to  $F \simeq 80\sqrt{\frac{N_C}{3}}$~MeV. 

We impose the constraint for $G_V$ given in Eq.~\eqref{gvatnc} at large $N_C$ and 
the extrapolation function between $N_C=3$ and $N_C\to \infty$ with $1/N_C$ suppressed corrections included is: 
\begin{align}\label{gvsubnc}
 G_V(N_C) &=  G_V(N_C=3)\sqrt{\frac{N_C}{3}} \nonumber \\&
\times \left[ 1 + \frac{ G_V(N_C=3) -G_V^{\rm Nor}(N_C\to\infty) }{G_V(N_C=3)}\left( \frac{3}{N_C} - 1 \right)  \right]\,, 
\end{align}
where $G_V^{\rm Nor}( N_C \to \infty )  = G_V(N_C\to\infty)\sqrt{\frac{3}{N_C}}$ and $G_V(N_C\to\infty)$ is given by Eq.~\eqref{gvatnc}. 
Notice that $G_V^{\rm Nor}( N_C \to \infty )$ is finite in the large $N_C$ limit. 
With the running of $F_\pi$ as a function of $N_C$ from Ref.~\cite{guo11prd} we have the numerical value
\begin{align}
G_V^{\rm Nor}( N_C \to \infty )&\simeq 46~\hbox{MeV}~.
\end{align}
Were the sub-leading $N_C$ scaling for $G_V$ not considered, i.e. if  
\begin{align}
G_V^{\rm Nor}( N_C \to \infty ) = G_V(N_C=3)~,
\end{align} 
then Eq.~\eqref{gvsubnc} reduces to the leading behavior $G_V(N_C) = G_V(N_C=3)\sqrt{\frac{N_C}{3}}$, as in Eq.~\eqref{ncrunning4res}. 
We stress that the extrapolation function in Eq.~\eqref{gvsubnc} is unique if one considers only 
the next-to-leading order in the $1/N_C$ scaling for the considered parameter. 
Similar extrapolation functions are also used for the other resonance parameters when needed, as specified below. 
For the other parameters, we keep the same setups from Scenario 1. 

\item {\bf Scenario 3:} Here, in addition to Eq.~\eqref{gvsubnc} of Scenario 2, we also assume that the bare masses 
of the $\rho(770)$ resonance and the singlet scalar resonance $S_1$ (an important component of the $f_0(980)$ at $N_C=3$) approach 
to the same value at large $N_C$. We can realize this scenario by increasing the bare  $\rho(770)$ mass by a $16\%$ and decrease the bare $S_1$ mass by another $16\%$, so that their  
 large $N_C$ masses meet around 930~MeV. This value is indeed quite close to the preferred one for the $\rho(770)$ in Ref.~\cite{pelaez11prd} in the large $N_C$ limit. 
We take as extrapolation function the analogous one to  Eq.~\eqref{gvsubnc}:
\begin{align}
M^2(N_C)&=M^2(N_C=3)\left[
1+\frac{M^2(N_C=3)-M^2(N_C\to\infty)}{M^2(N_C=3)}\left(\frac{3}{N_C}-1\right)
\right]~,
\label{mass2run}
\end{align}
with $M$ either the $\rho(770)$ or $S_1$ bare mass.

\item {\bf Scenario 4:}  On top of the considerations in Scenario 3 we now consider the effects of the $D$-waves,
which include additionally the contributions from the tensor resonances. 
 For their  resonance parameters in Eq.~\eqref{lagtensor}, we take  the leading order scaling with $N_C$,
\begin{equation}
 g_T(N_C) = g_T(3) \times \sqrt{\frac{N_C}{3}}\,, \qquad M_{T}(N_C) = M_T(3)\,.
\end{equation}

\end{itemize}

We summarize the different  Scenarios 1--4 in Table \ref{tab:scenarios}. 
We  also considered another situation in which together with the characteristics of Scenario 3 we take at large $N_C$ the mass of the 
octet of scalar resonances to be the same as that of the $S_1$ and $\rho(770)$. 
However, we  checked that this new addition produces negligible contributions 
to the ratios $F_n^{II'}$ and $R_n^I$. The reason is because the coupling of the octet of scalar 
resonances to $\pi\pi$ is suppressed numerically compared with that of the singlet scalar resonance. 
Due to the fact that at large $N_C$ the $\bar{q}q$ resonances fall down to the bare mass position in the real axis, 
it will cause noticeable changes for the octet resonance pole trajectories. 
Nevertheless since the reason is obvious, we do not discuss any further this scenario.

\begin{table}[ht]
\begin{center}
\begin{tabular}{|p{1.9cm}|p{1.5cm}| p{1.5cm}| p{1.5cm} | }
\hline
 & $G_V$ & $M_\rho$, $ M_{S_1}$ &  $D$-wave
\\
\hline
Scenario 1 & $-$ & $-$ & $-$  
\\
\hline
Scenario 2 & $\surd$ & $-$ & $-$  
\\
\hline
Scenario 3 & $\surd$ & $\surd$ & $-$  
\\
\hline
Scenario 4 & $\surd$ & $\surd$ & $\surd$ 
\\
\hline 
\end{tabular}
 \caption{{\small Description of Scenarios 1--4. In the second and third columns the symbol $-$ 
denotes that the sub-leading $N_C$ scaling for the corresponding parameters (indicated in the first row)  
is not considered. In turn, $\surd$ denotes that the sub-leading $N_C$ scaling is taken into account. 
In the last column, the symbol $-$ means that we do not consider the contribution from the $D$-waves 
and $\surd$ indicates that the latter are taken into account.  
 } }\label{tab:scenarios}
\end{center}
\end{table}

%%%%%%%%%%%%%%%%%%%%%%%%%%%%%%%%%%%%%%%%%%%%%%%%%%%%%%%%%%%%%%%%%%%%%%%%%%%%%%%%%%%%%%%%%%%%%%
\subsection{Semilocal duality for $N_C>3$}
\label{semilocal}

\begin{figure}[ht]
\includegraphics[angle=0, width=0.95\textwidth]{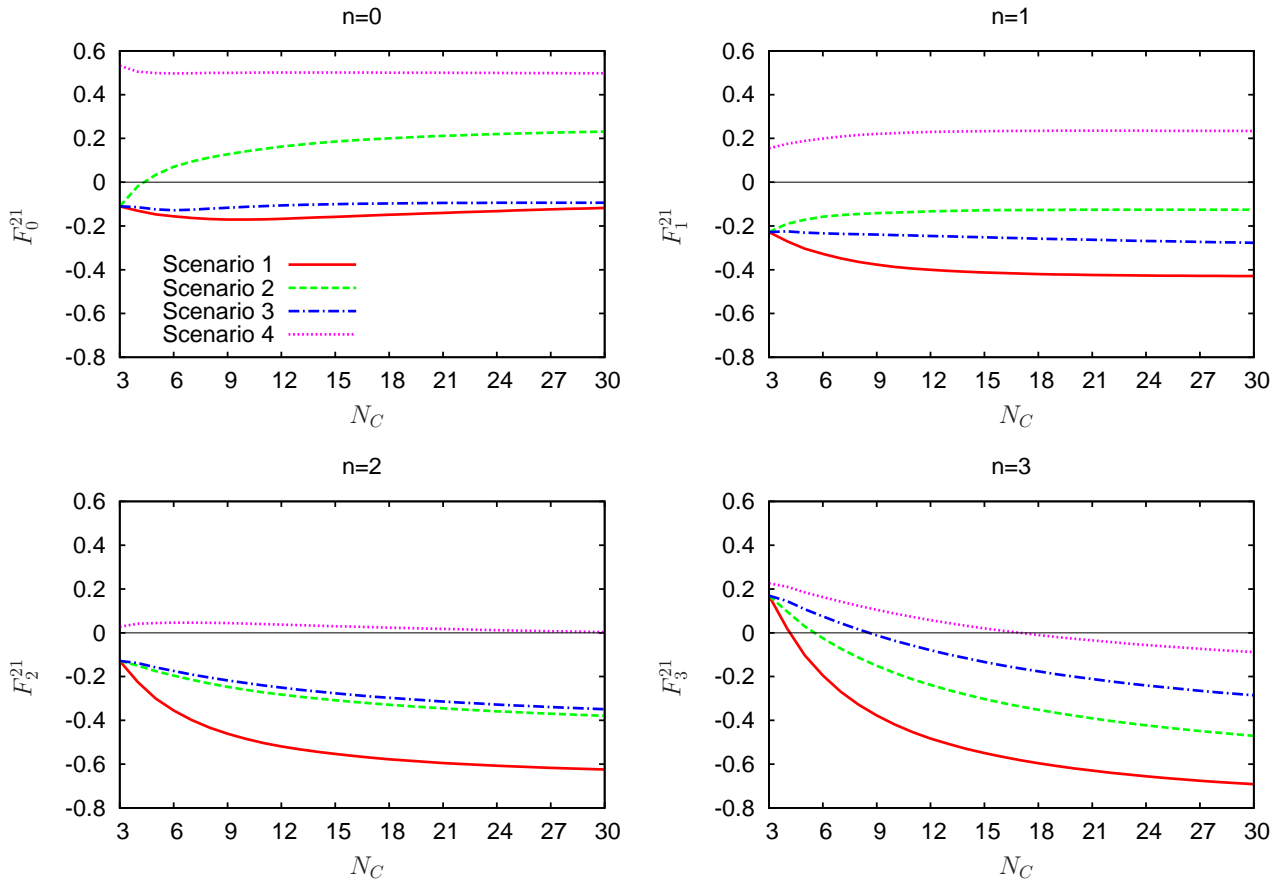}
\caption{{\small $F_n^{21}(t=4m_\pi^2)$. The solid (red) lines  correspond to the current fit and only 
the leading order of the $N_C$ scaling for the resonance parameters  
is considered, i.e. Scenario 1. The dashed (green) lines  additionally include the sub-leading $N_C$ scaling for $G_V$, i.e. Scenario 2.
The dot-dashed (blue) lines  correspond to take into account the sub-leading $N_C$ scaling for $G_V$, $M_\rho$ and $M_{S_1}$, i.e. Scenario 3. 
The doted (magenta) lines Scenario 4 so that the $D$-wave contribution is included as well.}}
\label{fig.ncdualityf21}
\end{figure}

\begin{figure}[ht]
\includegraphics[angle=0, width=0.95\textwidth]{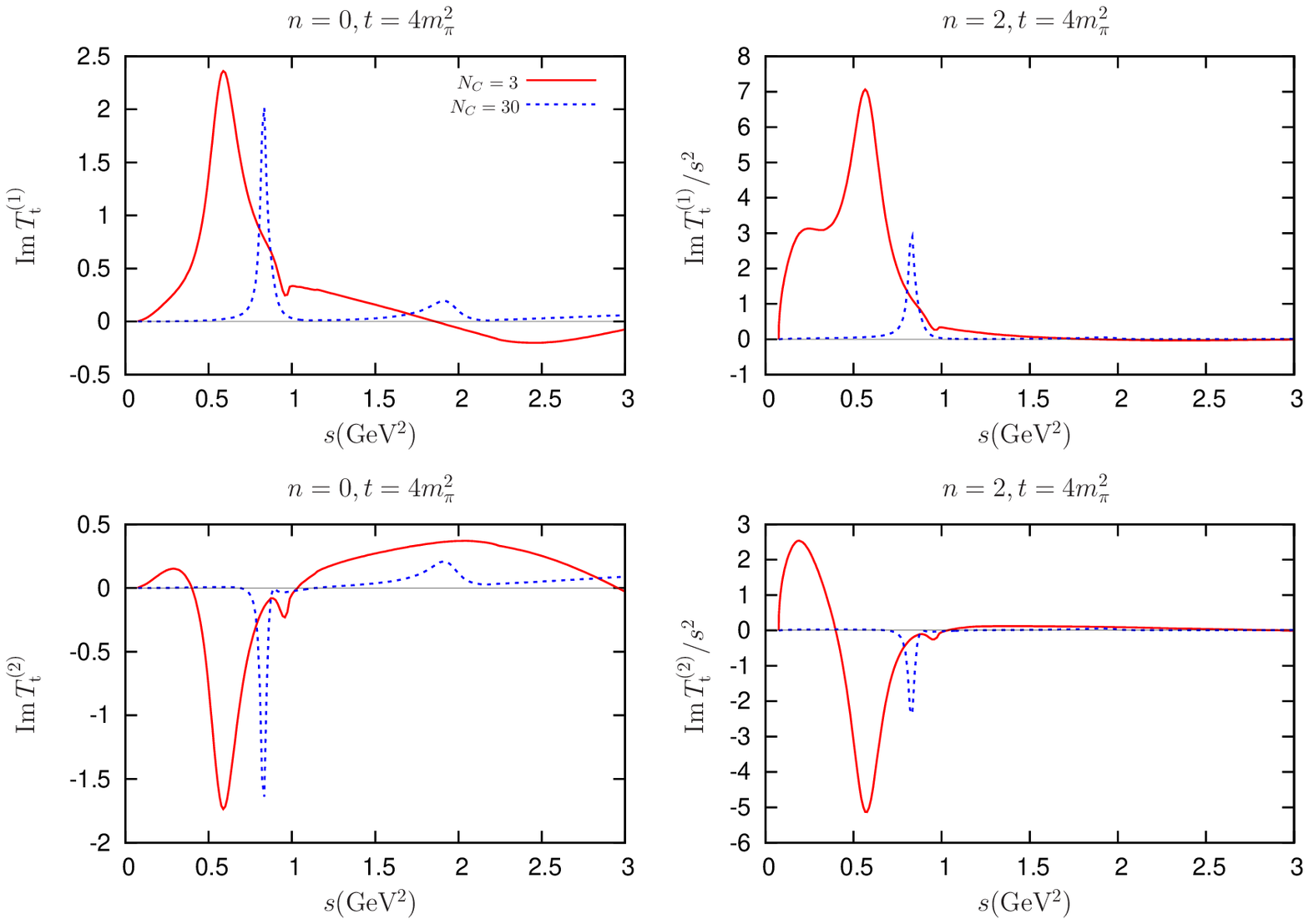}
\caption{{\small For Scenario 3, we show ${\rm Im}\,T_{\rm t}^{(1)}(t=4m_\pi^2)$ 
in the two panels of the top row and ${\rm Im}\,T_{\rm t}^{(2)}(t=4m_\pi^2)$ 
in the ones at the bottom. The results from $N_C=3$ (red solid lines) and 30 (dashed blue lines) are displayed for all the four figures. }}
\label{ima2nc-s3}
\end{figure}

For all the scenarios we plot in Fig.~\ref{fig.ncdualityf21} the $N_C$ trajectories of the ratio $F_n^{21}$ 
with $t=4m_\pi^2$, defined in Eq.~\eqref{defFratio}. We verify that the results for $t=0$ are quite similar. The 
 (red) solid line is for Scenario 1, the  (green)  dashed line is for Scenario 2, 
the (blue)  dot-dashed line corresponds to Scenario 3 and the (magenta) dotted   line does to Scenario 4. 
Between the first three scenarios  the best is the third one since then  the curves have the smallest absolute values 
for most of the $N_C$ axis, as required by the Regge theory and semilocal duality. The only exception is $F_1^{21}$ where Scenario 2 gives  smaller values.  
Concerning Scenario 4, though it gives better results for $n=2,3$ than the others, 
it leads to too large values for the $n=0$ case. 
In Fig.~\ref{ima2nc-s3}, we show  the imaginary part of the amplitudes with 
well defined  isospin in the $t$-channel for Scenario 3 at $N_C=3$ and $N_C=30$. 
Their integration, according to Eq.~\eqref{defFratio}, gives $F_{n}^{21}$.   
For $N_C=3$ one can clearly see in the bottom-left plot of Fig.~\ref{ima2nc-s3} a resonant bump in the 1-3~GeV$^2$ region, which is 
absent in Ref.~\cite{pelaez11prd}. This is mainly contributed by the higher scalar resonance $f_0(1370)$, and it
plays an important role to balance the contribution of the $\rho(770)$ resonance. Nonetheless, its contribution becomes less important 
with increasing $N_C$ and for $n\geq 2$ it only has a marginal contribution for all the $N_C$ values.  
When $N_C$ grows, the $f_0(500)$ resonance pole obtained in unitarized $U(3)$ R$\chi$T moves deeper and deeper in the complex energy plane and 
thus barely contributes. The role played in Ref.~\cite{pelaez11prd} by the subdominant $\bar qq$  component for the $f_0(500)$ 
with a mass around 1 GeV to cancel the $\rho(770)$ contribution for $n=2,~3$, is played in this work by the $f_0(980)$ resonance, 
which gradually evolves to the scalar singlet $S_1$ resonance and starts behaving as a $\bar{qq}$ state for $N_C>6$. 
In both works, a $\bar qq$ scalar state with a mass around 1 GeV  is needed in order to satisfy local duality. 
The evolution of the resonance poles with increasing $N_C$ will be discussed in detail in the next section.

Focusing on the  solid (red) lines in the four panels of Fig.~\ref{fig.ncdualityf21}, resulting from Scenario 1, 
one can immediately conclude that semilocal duality, though satisfied at $N_C=3$ for all the values of $n$, 
it is not well satisfied at large $N_C$ for $n\geq 1$.  
Indeed, the situation taking place for Scenario 1 is quite similar to the one-loop IAM case of Ref.~\cite{pelaez11prd} 
where the $f_0(500)$ does not show a sub-leading $\bar qq$ component.  For $n=0$, both 
the one- and two-loop IAM results satisfy approximately semilocal duality at large $N_C$,   
while only the results which include the $\bar qq$ sub-leading component for the $f_0(500)$ \cite{pelaez11prd}, 
show a clear sign of duality for $n=1,2$ and $3$. The appearance of this $\bar{q}q$ subdominant component, 
which approaches to the real axis in the complex energy plane at large $N_C$,
plays the fundamental role in the fulfillment of duality in Ref.~\cite{pelaez11prd}. 
However, one should also keep in mind that  the $N_C$-behavior for the $\rho(770)$ resonance  
in the two-loop result, with its pole position 
at $\sqrt{s_\rho}= M_\rho - i \Gamma_\rho/2  =(950.0-i\,34.8)$~MeV for $N_C=12$, is quantitatively different  compared with the one-loop result 
for the same value of $N_C$, $\sqrt{s_\rho} =(710.5-i\,18.4)$~MeV~\cite{pelaez11prd}.  
Due to the small ratios of $\Gamma_\rho / M_\rho$ for $N_C=12$ (in both cases it is less than 0.1), 
 a reasonable approximation consists of using the narrow resonance approximate formula to estimate the $\rho(770)$ contribution in both cases to 
${\rm Im}T^{IJ=11}(s)$. It reads \cite{collinsbook}
\begin{equation}\label{narrowwidthapprox}
 {\rm Im}\, T^{11}(s)  = 16\,\pi^2\,\frac{  M_\rho \Gamma_\rho }{  \bar{\sigma}(M_\rho^2)}\delta(s-M_\rho^2)\,,
\end{equation}
where $\bar{\sigma}(s) = \sqrt{1 - 4m_\pi^2/s}$. We have adjusted the normalization of the partial wave amplitude 
in Eq.~\eqref{narrowwidthapprox} to the one defined in Ref.~\cite{guo11prd}, and used in this work. Now, one can analytically calculate  
the $\rho(770)$ contribution to the right hand side of Eq.~\eqref{defseml}, 
which finally enters in the ratios of Eqs.~\eqref{defRratio} and \eqref{defFratio}. 
By applying Eq.~\eqref{narrowwidthapprox}, the $\rho(770)$ contribution to the FESR for $I_{\rm t} =2 $  
in terms of its width and mass is
\begin{equation}
\label{narho}
 \int_{\nu_1}^{\nu_{\rm max}} \nu^{-n}\, {\rm Im}\, T_{t, { \rho} }^{I=2}(\nu, t) d\nu 
= -24\,\pi^2\frac{\Gamma_\rho\,M_\rho^{1-2n}}{\bar{\sigma}(M_\rho^2)}\,.
\end{equation}
where the integration region between $\nu_1$ and $\nu_{\rm max} $ always covers the $\rho(770)$ mass. 
 For simplicity we show in the previous equation the result at $t=0$ and similar results can be straightforwardly
deduced for other values of $t$. The ratio of the $\rho(770)$ contribution between the two- and one-loop cases from \cite{pelaez11prd} 
as follows from Eq.~\eqref{narho} is 
\begin{equation}
\frac{\Gamma_{\rho,{\rm two-loop}}}{\Gamma_{\rho,{\rm one-loop}}} \,
\bigg(\frac{M_{\rho,{\rm two-loop}}}{M_{\rho,{\rm one-loop}}} \bigg)^{1-2n} \,
\frac{\bar{\sigma}(M_{\rho,{\rm one-loop}}^2)}{\bar{\sigma}(M_{\rho,{\rm two-loop}}^2)}~.
\end{equation}
Taking into account the $\rho(770)$ pole positions for $N_C=12$ from Ref.~\cite{pelaez11prd}, that were explicitly shown above, 
the ratio in the previous equation 
is $0.43$ for $n=3$. This implies that the $\rho(770)$ contribution is reduced by more than $50\%$ in the two-loop result 
compared with the one-loop case in the IAM study of Ref.~\cite{pelaez11prd}. When increasing $N_C$ this ratio will stay put since the $\rho(770)$ resonance 
already starts to behave as a standard $\bar{q}q$ resonance at $N_C=12$, with its mass approaching to a constant and 
its width decreasing as $1/N_C$ \cite{pelaezprl1,pelaezprl2,pelaez11prd}. 
Though this $N_C$ behavior of the $\rho(770)$ pole is not essential in Ref.~\cite{pelaez11prd} to satisfy local duality, 
it definitely helps to improve it. 

This reduction of the $\rho$ signal with increasing $N_C$ is explained by our present approach in a quite transparent way. 
For that one needs to take into account the sub-leading $N_C$ scaling of the resonance parameters. It is also the case that when the latter are 
taken into account  semilocal duality is also better fulfilled. Among the estimates of the sub-leading scaling for various resonance parameters,
 $G_V$ is the most reliable one, since it can be directly derived by requiring a proper high energy behavior of the partial 
wave amplitudes, which are the key input in the study of semilocal duality. 
In addition, this constraint has also been confirmed in different processes, as already 
discussed above \cite{pich11jhep,guo11prd,guo07jhep,guo10prd,chivukula07prd}. It turns out that the large $N_C$ condition Eq.~\eqref{gvatnc} considerably 
improves the fulfillment of semilocal duality. Thus, we provide another hint to confirm this constraint. 
This improvement is  displayed  in Fig.~\ref{fig.ncdualityf21} by the difference between 
the  solid (red) lines and the  dashed (green) lines. 
 The slight readjustment of the bare masses for the $\rho(770)$ and $S_1$ at large $N_C$ seems 
to improve the situations for $n=0$ and $n=3$, which can be seen from the differences between the   dashed (green) lines and 
the  dot-dashed (blue) ones. It causes significant effects for $n=0$ by alternating the sign of the ratio, 
nevertheless the magnitudes are always less than 0.2. 

Finally, we comment on the $D$-wave effects. Though the $D$-wave amplitudes can gain contributions from many sources, such as chiral loops,
scalar, pseudoscalar and vector resonances exchanged in crossed channels, one expects from phenomenological reasons that the most important ones correspond to the 
tensor resonances \cite{gammagamma,dobapela}.  From the change between the dot-dashed (blue) lines and the dotted (magenta) ones in Fig.~\ref{fig.ncdualityf21}, 
one can discern the role that the $D$-waves play in the FESR. 
As one can see, it is significant for $n=0$, 1, 2 and slight for $n=3$ (and it should be even smaller for larger values of $n$ since then 
low-energy physics is enhanced, as already commented).   
For $n=1$, though the tensor resonances alternate the sign of the ratios, the magnitudes are still quite small when varying $N_C$.  
It also clearly improves the condition for $n=2$ and $n=3$, though for $n=0$ it clearly deteriorates the fulfillment of semilocal duality. 
We observe that the $D$-waves give rise to a large contribution that overbalances the one of the $\rho(770)$. 
This hints that higher vector resonances are necessary to cancel the $D$-wave contributions. 
In order to give a rough idea on whether the $\rho(1450)$ resonance in PDG~\cite{pdg} can counteract the $D$-wave contribution, 
we include  another heavier vector resonance in $\pi\pi$ scattering following the formula in Eq.~\eqref{narrowwidthapprox}. 
We refer this resonance as $\rho'$ in the following. The only difference now is that $\Gamma_{\rho'}$ in Eq.~\eqref{narrowwidthapprox} should be 
understood as the partial decay width $\Gamma_{\rho' \rightarrow \pi\pi}$.  Moreover, since we focus on  semilocal duality below 2~GeV$^2$, 
we simply set the $\rho'$ mass as 1350~MeV in order to cover its peak for the integral in Eq.~\eqref{defseml}. 
We find the $\rho(1450)$-type resonance gives a negligible contribution for all the values of $n$, 
because of its too small decay branching ratio to $\pi\pi$ (around a 6\%), 
as follows from the present information in the PDG~\cite{pdg}. 
We verify that the decay branching ratio to $\pi\pi$ of the hypothetical resonance $\rho'$ needs to be around 40\% in order to decrease the large 
 contribution by the $D$-wave for the $n=0$ case down to 0.2.  This is a hint in favor for the existence of a heavy 
$\rho'$ with a significant branching decay ratio to $\pi\pi$. For  $n=1,2,$ $3$  no cancellations or just small ones are in fact needed 
in order to have a rather suppressed FESR with $I=2$ 
in the $t$ channel. Nevertheless, if included for these values of $n$ the resulting curves are equally satisfactory (even better for $n=1$).

%%%%%%%%%%%%%%%%%%%%%%%%%%%%%%%%%%%%%%%%%%%%%%%%%%%%%%%%%
\subsection{Resonance pole trajectories with varying $N_C$}
\label{rptwvn}

Unless the opposite is stated, all the $N_C$ pole-trajectory evolutions studied from now on correspond to the Scenario 3 
in Table \ref{tab:scenarios}. This is also the scenario that satisfies best semilocal duality, as seen in Sec.~\ref{semilocal}.

\begin{figure}[H]
\begin{center}
\includegraphics[angle=0, width=0.6\textwidth]{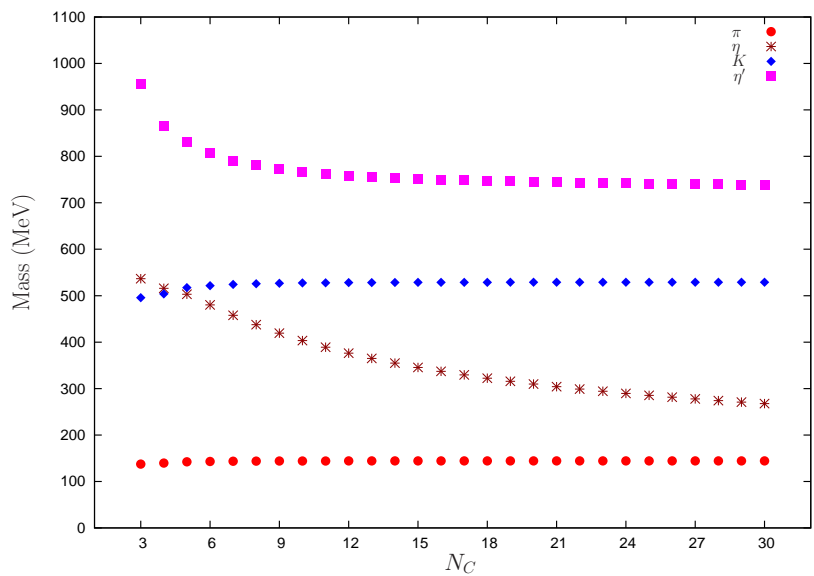}
\caption{{\small $N_C$ running for the masses of pseudo-Goldstone bosons.}}
\label{fig.ncpsmass}
\end{center}
\end{figure}

Due to the $N_C$ behavior of the singlet $\eta_1$ mass $M_0$ in Eq.~\eqref{ncrunning4m0lam2}, a novel  
feature of $U(3)$ $\chi$PT, as compared with the $SU(3)$ version, is that the masses of the pseudo-Goldstone bosons 
at large $N_C$, especially for $\eta$ and $\eta'$, can be significantly different from their physical values at $N_C=3$,  as shown in Ref.~\cite{guo11prd}. 
On the other hand, the masses of the pion and kaon barely change when varying $N_C$. Similar results are obtained from the new fit in Eq.~\eqref{fitresultnew}, 
that are depicted in Fig.~\ref{fig.ncpsmass}.  
 One can see there that for $N_C=30$, the $\eta$ and $\eta'$ masses decrease from their physical values down to around 300~MeV and 700~MeV, respectively.  
We point out the $\eta'$ retains a somewhat heavy mass at large values of $N_C$ mainly caused by the kaon mass \cite{guo11prd}. This is easily seen by the 
leading order expression of the $\eta$ and  $\eta'$ masses from $U(3)$ $\chi$PT \cite{guo11prd}. They result from the tree level calculation using  
the leading order Lagrangian Eq.~\eqref{lolagrangian} at the large $N_C$ limit, i.e. $M_0 \to 0$:  
\begin{align} \label{metaloLnc}
\overline{m}^2_\eta &= \overline{m}_\pi^2\,, \qquad \nn\\
\overline{m}^2_{\eta'} &= 2 \overline{m}_K^2 - \overline{m}_\pi^2\,,
\end{align}
where $\overline{m}_\eta$ and $\overline{m}_{\eta'}$ stand for the $\eta$ and $\eta'$ masses, respectively.

\begin{figure}[H]
\begin{center}
\includegraphics[angle=0, width=0.6\textwidth]{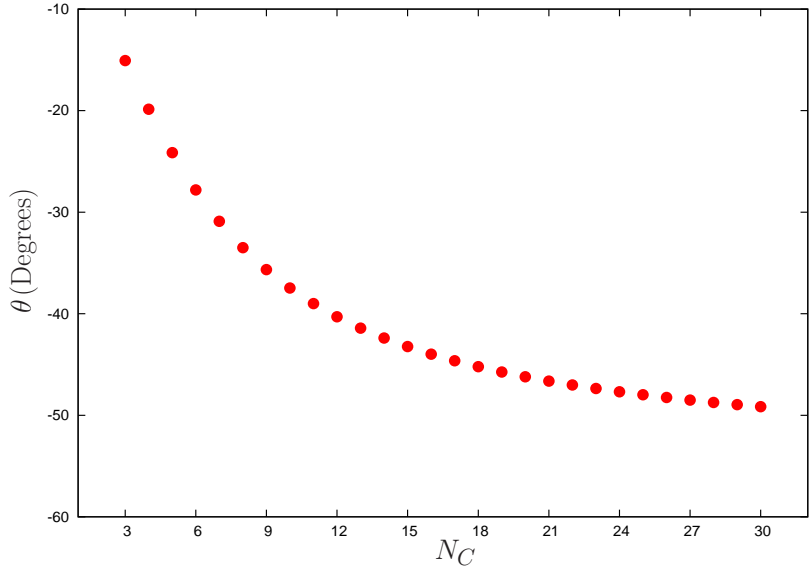}
\caption{{\small $N_C$ running for the leading order $\eta-\eta'$ mixing angle $\theta$ introduced in Eq.~\eqref{deflomixing}.}}
\label{fig.nctheta}
\end{center}
\end{figure}

The expressions that relate $\overline{m}_\pi^2$ and $\overline{m}_K^2$ with $m_\pi^2$ and $m_K^2$ 
at the one-loop level are given by Eqs.~(A1), (B2), (B4) and (B7)  of Ref.~\cite{guo11prd} to which one 
must add the new contribution  Eq.~\eqref{newmpi2}, due to $\delta L_8$ and the exchange of the pseudoscalar resonances. 
The numerical values from the new fit in Eq.~\eqref{fitresultnew} for the pion and kaon leading-order masses extracted at the one-loop level are  
\begin{align} 
\label{pikbmass}
\overline{m}_\pi= 136.4^{+2.3}_{-1.7}~{\rm MeV}\,, \qquad 
\overline{m}_K  = 499.6^{+30.6}_{-30.4}~{\rm MeV}\,,
\end{align}
which leads to 
\begin{eqnarray} 
\overline{m}_\eta     = 136.4^{+2.3}_{-1.7}~{\rm MeV}\,, \qquad 
\overline{m}_{\eta'}   = 693.3^{+43.7}_{-43.8} ~{\rm MeV}\,,
\end{eqnarray} 
according to Eq.~\eqref{metaloLnc}. 
We point out that at the chiral and large $N_C$ limits, all the masses of the nonet of 
 Goldstone bosons vanish. 
The $N_C$ evolution of the leading order mixing angle in 
Eq.~\eqref{deflomixing}, given explicitly in Eq.~(B7) of Ref.\cite{guo11prd}, is displayed in Fig.~\ref{fig.nctheta}.
At large $N_C$ it corresponds to ideal mixing, as it should.

In the study of the evolutions of the resonance pole positions we always take the $N_C$ running 
masses for the pseudo-Goldstone bosons as shown in Fig.~\ref{fig.ncpsmass}.  
The resulting $N_C$ trajectories of the various resonance poles and their residues were studied with detail in the Ref.~\cite{guo11prd}. 
Qualitatively speaking we do not find any significant changes in the trajectories by using the new fit in Eq.~\eqref{fitresultnew} 
and thus we confirm the conclusions obtained in this reference: the $f_0(500)$, $K^*_0(800)$ and $a_0(980)$ resonances go deeper in the 
complex energy plane when increasing the values of $N_C$. For the other resonances, $f_0(980)$, $f_0(1370)$, $K^{*}_0(1430)$, 
$a_0(1450)$, $\rho(980)$, $K^*(892)$ and $\phi(1020)$, approach to the real axis and behave like the standard $\bar{q}q$ resonances 
at large $N_C$. Notice that the $f_0(980)$ for not so large values of $N_C$ does not follow the standard $\bar{q}q$ 
pattern (compare  with the pole trajectories for the vector resonances in Fig.~\ref{fig.ncv}). 
E.g. its width indeed clearly increases up to  $N_C\simeq 7$. 
This is a signal of the fact that the $f_0(980)$ has also a strong contribution to its nature as a $K\bar{K}$ bound state \cite{oller97npa}.

Two variant approximations,  named as {\it vector  reduced } and {\it mimic $SU(3)$}, were studied to explore the $N_C$ trajectories 
of the resonance pole positions in Ref.~\cite{guo11prd}. In the {\it vector  reduced } case, 
we freeze out the full propagators of the vector resonances in the scattering amplitudes and only keep the leading local 
terms generated from them, which are $\cO{(p^4)}$ in the chiral counting (or ${\cal O}(\delta^3)$). 
The purpose for introducing this approximation is to highlight the difference between using the tree level resonance exchanges and 
the local LECs in the meson-meson scattering amplitudes. As discussed in Ref.~\cite{guo11prd}, we find 
that freezing the bare scalar resonances in the amplitudes does not lead to significant effects for the $f_0(500)$ resonance. 
This is why we only discuss the {\it vector reduced} case. 
In the approximation of {\it mimic $SU(3)$}, instead of taking the 
$N_C$ running masses for the pseudo-Goldstone bosons, we freeze them  throughout and, in addition, we do not consider 
any $\eta-\eta'$ mixing terms. Thus, in this case, the $\eta$ and $\eta'$ correspond
to the octet $\eta_8$ and singlet $\eta_1$ respectively. The masses of $\pi, K$ and $\eta$ are fixed at the physical values 
and the $\eta'$ mass is taken from the leading order prediction of Eq.~\eqref{lolagrangian}, which is around 1040~MeV. This 
mimics the conditions of $SU(3)$ $\chi$PT that we consider for comparison with our $U(3)$ $\chi$PT results.

\begin{figure}[H]
\begin{center}
\includegraphics[angle=0, width=0.99\textwidth]{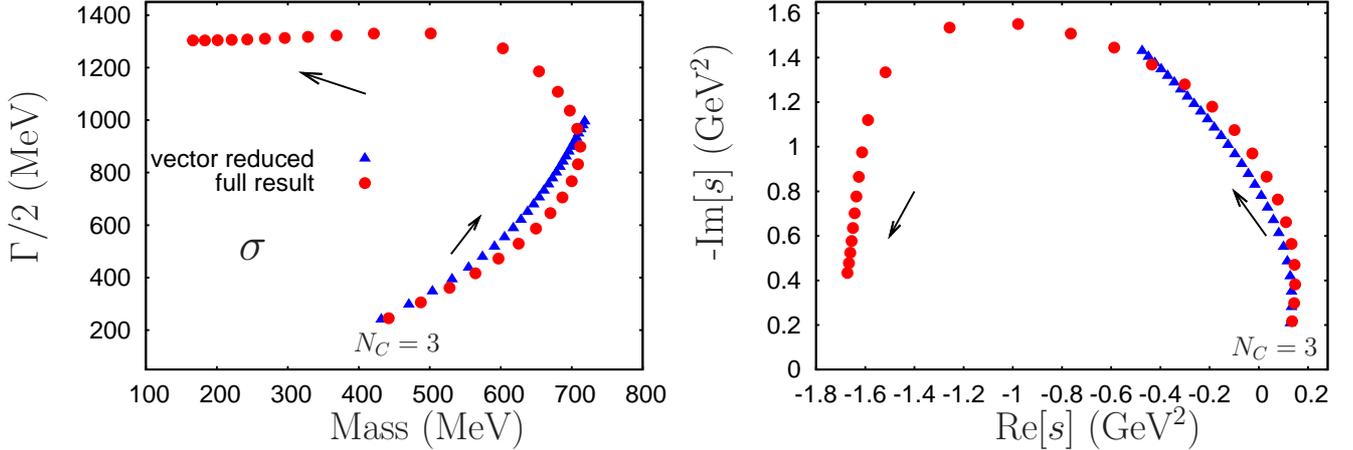}
\caption{{\small $N_C$ running of the pole positions for the $f_0(500)$ (or $\sigma$) resonance. We give the results for $N_C$ from $3$ to $30$ with 
one unit step. Both the full results and the ones from the {\it vector reduced} approximation are displayed. The left panel shows $\sqrt{s_\sigma}$ 
and the right one $s_\sigma$. }}
\label{fig.ncsig}
\end{center}
\end{figure}

\begin{figure}[ht]
\begin{center}
\includegraphics[angle=0, width=0.95\textwidth]{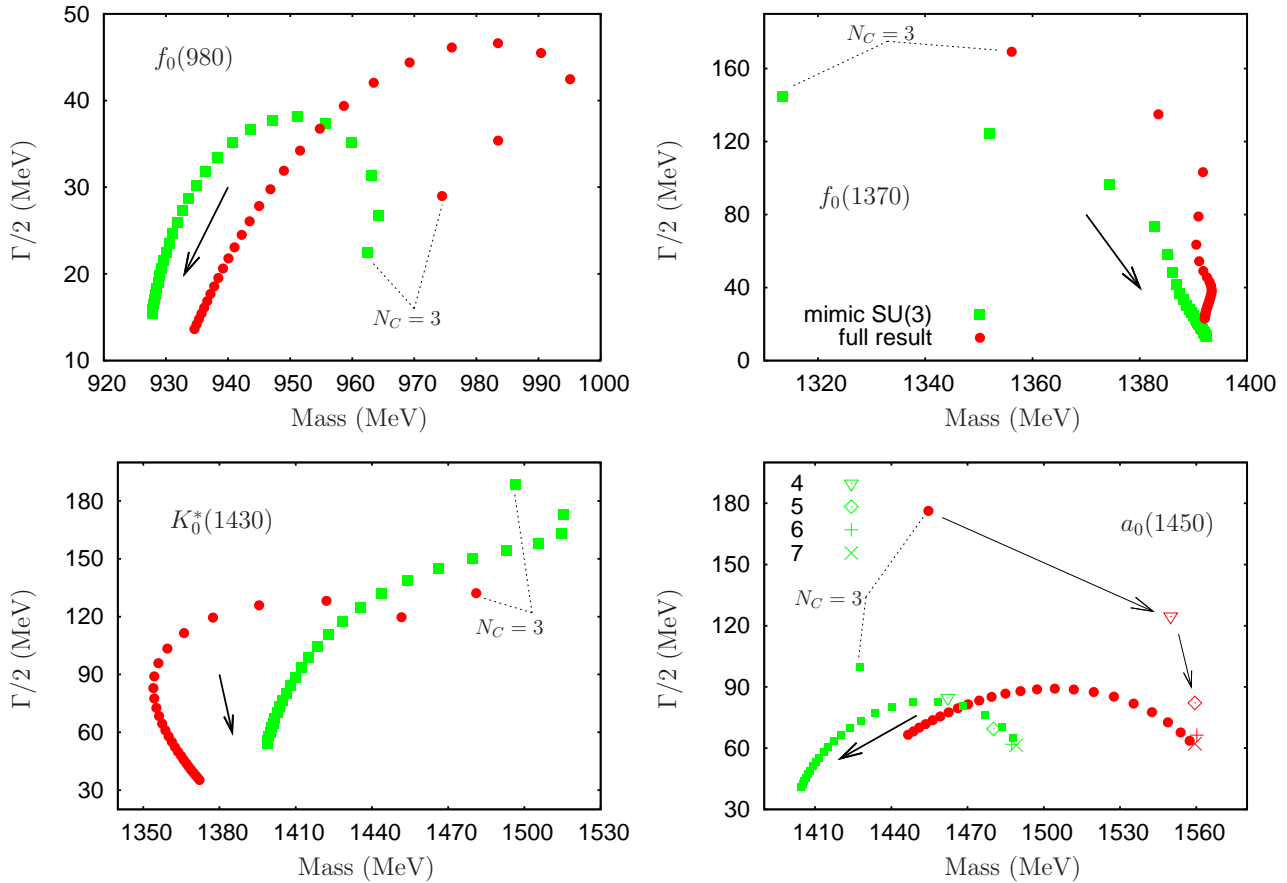}
\caption{{\small $N_C$ trajectories of the scalar resonances $f_0(980)$, $f_0(1370)$, $K^*_0(1430)$ and $a_0(1450)$. 
We show the results for $N_C$ from $3$ to $30$ with one unit step.  The results for both the full  (circles) and the {\it mimic $SU(3)$} (squares) calculations 
 are shown.  
The differences between the two sets of trajectories show the sensitivity of the corresponding resonances to 
the $\eta$ and $\eta'$ mesons. Specific symbols are assigned to the $a_0(1450)$ resonance for $N_C=4,5,6,7$ in order to show more clearly 
 its pole trajectory.  }}
\label{fig.ncs1}
\end{center}
\end{figure}

In Fig~\ref{fig.ncsig}, we explicitly show the full (circles) and {\it vector reduced} (triangles) results for the $f_0(500)$
  resonance pole ($s_\sigma$) from   
$N_C=3$ up to $N_C=30$ in one unit step (as it will be also the case for the rest of pole trajectories). 
Though different strategies to estimate the sub-leading $N_C$ scaling  
for the resonance parameters have been used in the current work, 
the $N_C$ trajectories for the $f_0(500)$ are qualitatively consistent with those found in Ref.~\cite{guo11prd}. 
That is, the $f_0(500)$ resonance from the full calculation tends to fall down to the negative real axis in the $s_\sigma$-complex plane and
  moves further and further away  along the nearby of the $s_\sigma$ negative axis as $N_C$ increases. For the  {\it vector reduced} case this is not the case. 
The outcome from the latter resembles the  $f_0(500)$ pole trajectory from 
 the one-loop IAM \cite{pelaezprl1,sun07mpl}, 
and the full result trajectory is also one of the different $f_0(500)$ pole trajectories \cite{pelaez11prd,Pelaez:2005fd}. 
It is important to remark that both results, as well as the ones of  Ref.~\cite{pelaez11prd}, are compatible for values of $N_C<10$, 
and confirm again the results
obtained in \cite{pelaezprl1,pelaezprl2}, which predict a non dominant $\bar
qq$ behavior for the $f_0(500)$. 
The outcome from the {\it mimic $SU(3)$} approximation, which is not explicitly shown in 
the figure, is quite close to the full results, indicating the insensitivity of the $\sigma$ resonance to 
the $\eta$ and $\eta'$ mesons, even when quite different masses result for these two particles as a function of $N_C$. 
As explained in Ref.~\cite{guo11prd} this relative insensitivity is due to the small couplings of the $f_0(500)$ to the $\eta\eta$, 
$\eta\eta'$ and $\eta'\eta'$ , even though these couplings are somewhat larger in the new fit than in the previous one. 
However, the largest couplings to $\pi\pi$ and $K\bar{K}$ are almost the same as in Ref.~\cite{guo11prd}.   This pole trajectory 
clearly indicates that the $f_0(500)$ resonance has no significant $q\bar{q}$ or glueball components and it is in agreement 
with its dynamical generation from the isoscalar scalar $\pi\pi$ interactions. Ref.~\cite{albaladejo12} calculates the quadratic scalar radius
of the $\sigma$, $\langle r^2\rangle_\sigma^s=(0.19 \pm 0.02)-i\, (0.06 \pm 0.02)$~fm$^2$, so that it is concluded
 that this resonance is a compact one and the two pions merge inside it. As a result, a four-quark picture is more favorable than the molecular description.

On the other hand, the resulting pole trajectories of the heavier scalar resonances for the full result and
the {\it mimic $SU(3)$} approximation are quite different between each other, as one can see in Fig~\ref{fig.ncs1}. This tells 
us that the scalar resonances $f_0(980)$, $f_0(1370)$, $K^{*}_0(1430)$ and $a_0(1450)$ are sensitive to the 
$\eta$ and $\eta'$ states, to which they couple strongly as shown in Table~\ref{tab:pole}. 
 For the vector resonances, we show the $N_C$ trajectories in Fig \ref{fig.ncv},  
where the different treatments of the $\eta$ and $\eta'$ mesons barely change the $\rho(770)$ resonance 
and only a little the $K^*(892)$. A kink structure has been found for the coupling strength between the $K^*(892)$ resonance   
and the $K\eta$ channel in the full result around $N_C=14$~\cite{guo11prd}, which is caused by the crossover of the $K^*(892)$ mass 
by the $K\eta$ threshold. This is mainly due to the variation of the $\eta$ mass. We find the structure from the new fit 
is quite similar to the one in Ref.~\cite{guo11prd}. 
For the $\phi(1020)$ and $\omega(780)$ resonances, we verify that it is also well behaved as a $\bar{q}q$ 
resonance at large $N_C$. The situation is even simpler in this case, since it only involves the $K\bar{K}$ channel in our approach.\footnote{In the present study we are missing the important $3\pi$ channel for a realistic treatment of the $\omega(780)$ resonance. We can then only study its mass which can be fixed to its experimental value by tuning its bare mass, as similarly done for the other vector 
resonances. }  
For the $K^*_0(800)$ and $a_0(980)$ resonances, we re-confirm as in Ref.~\cite{guo11prd} that their pole positions  
go deeper and deeper in the complex energy plane when increasing $N_C$, which are displayed in Fig \ref{fig.nca0kappa}.

\begin{figure}[ht]
\begin{center}
\includegraphics[angle=0, width=0.95\textwidth]{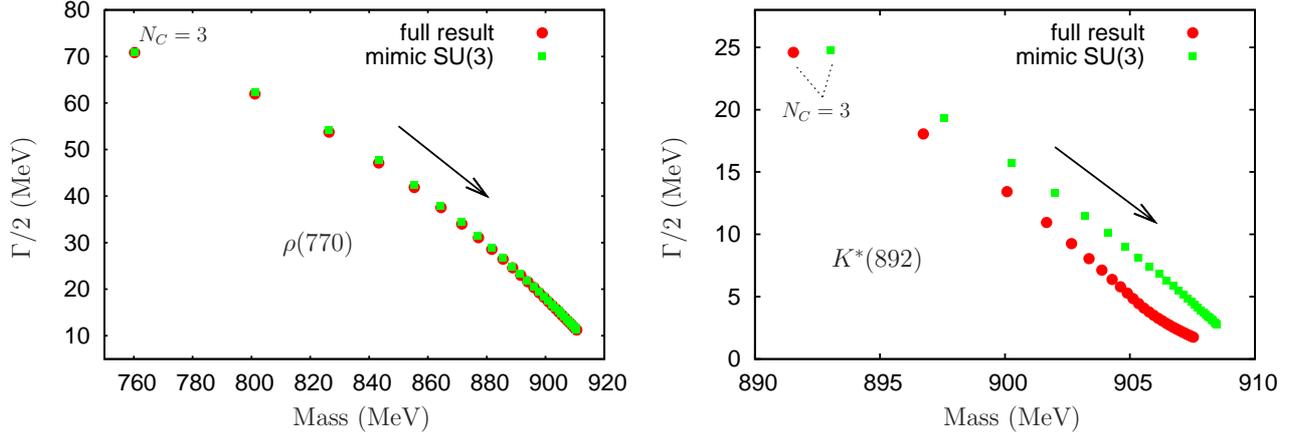}
\caption{{\small $N_C$ running of the pole positions for the $\rho(770)$ and $K^*(892)$ resonances. For notation see Fig.~\ref{fig.ncs1}.}}
\label{fig.ncv}
\end{center}
\end{figure}

\begin{figure}[ht]
\begin{center}
\includegraphics[angle=0, width=0.95\textwidth]{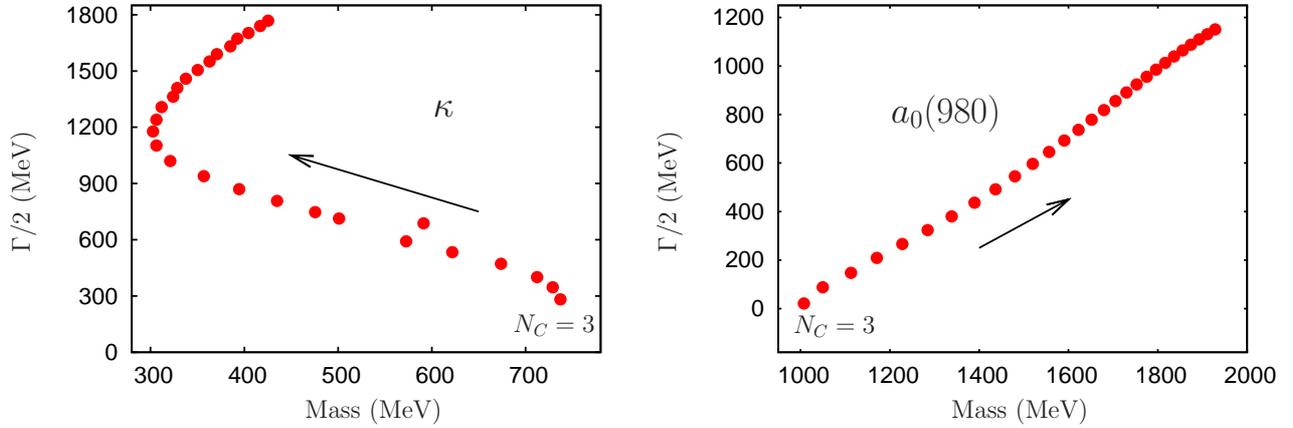}
\caption{{\small $N_C$ dependences of the $K^*_0(800)$ and $a_0(980)$ resonances both in the fourth 
Riemann sheets from $N_C=3$ to 30 with one unit step. See Ref.~\cite{guo11prd} for details on the definitions of the different 
Riemann sheets.} }
\label{fig.nca0kappa}
\end{center}
\end{figure}

%%%%%%%%%%%%%%%%%%%%%%%%%%%%%%%%%%%%%%%%%%%%%%%%%%%%%%%%%%%%%%%%%%%%%%%%%%%%%%%%%%%%%%%%%%%%%%%%%%%%
%%%%%%%%%%%%%%%%%%%%%%%%%%%%%%%%%%%%%%%%%%%%%%%%%%%%%%%%%%%%%%%%%%%%%%%%%%%%%%%%%%%%%%%%%%%%%%%%%%%%%%%
\subsection{Spectral-function  sum rules for $N_C>3$}

As a further application, we study the $N_C$ evolution of the spectral functions and the spectral-function sum rules in this section. 
The scaling with $N_C$ for the two-point correlators, i.e. $W_i$ in Eq.~\eqref{defwi}, is proportional to $N_C$ in  
the large $N_C$ QCD~\cite{largenc}. The situation for the pseudoscalar correlators is obvious, since after 
substituting Eqs.~\eqref{defpsspecf}, \eqref{pssffpsgoldstone} and \eqref{pssffpsres} into Eq.~\eqref{defwi} one has
\begin{equation}\label{nclowi4ps}
 W_{i} \propto F_\pi^2 + 8d_m^2 + \cdots\,, \quad i = P^0,~ P^8,~ P^3\,,
\end{equation}
where the ellipsis stand both for  sub-leading $N_C$ or suppressed pieces in the chiral counting.  
From Eq.~\eqref{ncrunning4res} and the fact that  $F_\pi$ scales as $\sqrt{N_C}$, it results that 
the leading $N_C$ scaling of the pseudoscalar correlators is proportional to $N_C$.

However, the situation for the scalar correlators is not so obvious. In fact, if we simply use 
the tree level results in Eqs.~\eqref{analylosffs0}, \eqref{analylosffs8} and \eqref{analylosffs3} 
from the Lagrangian in Eq.~\eqref{lolagrangian} 
and substitute them into Eq.~\eqref{defscspecf} to calculate the correlators in Eq.~\eqref{defwi}, one would 
 conclude that the latter behave as a constant at large $N_C$, instead of running like $N_C$. 
The subtlety, as shown in detail below, comes from the fact that at large $N_C$  the width of a $\bar{q}q$-like resonance tends to zero and the imaginary part of  
its propagator behaves then as a Dirac $\delta$-function, which is the dominant contribution to 
the correlators in Eq.~\eqref{defwi}.

In the following discussion, we demonstrate that by properly taking into account the resonance contributions 
in the unitarized scalar form factors Eq.~\eqref{defunitarizedF} for the pseudo-Goldstone boson pairs, 
the scalar correlators in our formalism also scale like $N_C$, as required by large $N_C$ QCD. 
In order to make the analytical discussion neatly, we illustrate the proof in the chiral limit and 
consider the single channel case, i.e. we only include the $\pi\pi$ channel both for the 
scalar form factor and the scattering. The relevant parts $R^I(s)$, $N^{IJ}(s)$ and $g^{IJ}(s)$
in Eq.~\eqref{defunitarizedF} are just functions not matrices for the single channel. 
 Notice also that in the large $N_C$ limit, only the tree level contributions, i.e. the (2)+Res parts in Eq.~\eqref{defNyR}, can survive. 
We focus in the case $a=0$, while the others  can be obtained analogously. 
The corresponding expressions in the chiral and large $N_C$ limits are: 
\begin{align} 
\label{lncclR}
R^{a=0}(s) &= -2 - \frac{8 c_d c_m}{F^2}\frac{s}{M_{S}^2 -s }\,, \\ 
\label{lncclN}
N^{00}(s) &= \frac{s}{F^2} 
 - \frac{G_V^2(2M_\rho^2 + 3 s)}{F^4} + \frac{2G_V^2 M_\rho^2(M_\rho^2 + 2 s)}{F^4 s}\log[1 + \frac{s}{M_\rho^2}]
\nonumber \\ & 
+ \frac{2c_d^2 M_S^4}{F^4 s}\log\big[1 + \frac{s}{M_S^2}\big] 
+ \frac{c_d^2 }{F^4}\frac{(2s-M_S^2)(s+2M_S^2)}{M_{S}^2 -s } \,,
\end{align}
where we assume exact large $N_C$ nonet  symmetry for the scalar resonances 
in Eqs.~\eqref{lagscalar} and \eqref{kinersp}, i.e. imposing that the masses for singlet and 
octet scalar resonances are equal and taking the large $N_C$ relations in Eq.~\eqref{ncrelationcdtcmt}. In the 
previous equation $M_S$ denotes 
the bare mass of the scalar nonet. In fact for the scalar resonances, 
only the combination  $\sqrt{\frac{1}{3}}S_8 + \sqrt{\frac{2}{3}}S_0 $ is the one relevant in the large $N_C$ limit
 for the $\pi\pi$ form factor and scattering. The interaction between
the other orthogonal combination $\sqrt{\frac{1}{3}}S_0-\sqrt{\frac{2}{3}}S_8 $ and $\pi\pi$ is $1/N_C$ suppressed. 

Let us concentrate in the $s$ region around  the bare resonance pole at $s=M_S^2$ 
 in Eqs.~\eqref{lncclR} and \eqref{lncclN}. Substituting Eqs.~\eqref{lncclR} and \eqref{lncclN}   
into Eq.~\eqref{defunitarizedF}, we have the simple expression around the resonance pole $s \to M_S^2$ for 
the unitarized form factor  
\begin{eqnarray}\label{unitarizedFinnccl}
F_{\pi\pi}^{a=0}(s) = -\frac{8 c_d c_m M_S^2}{F^2}\frac{1}{M_S^2 - s - i\frac{3c_d^2 M_S^4}{16\pi F^4} }\,,
\end{eqnarray}
where we have dropped the tiny contribution that stems from the real part of the function $g^{00}(s)$ (that would give rise to a 
self-energy contribution to the resonance bare mass), and explicitly show the  contribution from the imaginary part in the chiral limit 
\begin{equation}
 {\rm Im}\,g^{00} = - \frac{1}{16\pi}\,.
\end{equation}

By substituting the tree level decay width of $S\equiv\sqrt{\frac{1}{3}}S_8 + \sqrt{\frac{2}{3}}S_0 $ 
to $\pi\pi$ calculated from the Lagrangian in Eq.~\eqref{lagscalar} at the chiral limit \cite{guo07jhep} 
\begin{equation}\label{decaywidthgammas}
\Gamma_{S} = \frac{3c_d^2 M_S^3}{16\pi F^4}\,,
\end{equation}
into Eq.~\eqref{unitarizedFinnccl}, we get the standard Breit-Wigner propagator for the scalar resonance
\begin{eqnarray}\label{unitarizedFinnccl2}
F_{\pi\pi}^{a=0}(s) = -\frac{8 c_d c_m M_S^2}{F^2}\frac{1}{M_S^2 - s - i M_S \Gamma_S }\,. 
\end{eqnarray}
Due to the fact that $\Gamma_S$ behaves as $1/N_C$ in the large $N_C$ limit, we can use the 
standard narrow width approximation to write the form factor squared in terms of the Dirac $\delta$-function
\begin{align}\label{F2indeltafunc}
|F_{\pi\pi}^{a=0}(s)|^2 &= \frac{64 c_d^2 c_m^2 M_S^4}{F^4} \frac{1}{(M_S^2-s)^2 + (M_S\Gamma_S)^2} 
\nonumber \\
& \Rightarrow  \frac{1024\pi^2}{3} c_m^2 \delta(s-M_S^2)\,,
\end{align}
where we have used Eq.~\eqref{decaywidthgammas} and the following way to approach the Dirac $\delta$-function: 
\begin{equation}
\frac{1}{\pi}\frac{M_S \Gamma_S}{(M_S^2-s)^2 + (M_S\Gamma_S)^2} \Rightarrow \delta(s-M_S^2)\,, 
\quad {\rm when}\,\,\, M_S\Gamma_S \rightarrow 0\,.
\end{equation}
By combining Eqs.~\eqref{defscspecf}, \eqref{F2indeltafunc}, \eqref{defwi} and \eqref{ncrunning4res}
we obtain the leading $N_C$ behavior for the singlet two-point scalar correlator 
\begin{equation}
W_{S^0} \propto c_m^2 \propto N_C\,,
\end{equation}
which has the same scaling as the pseudoscalar case Eq.~\eqref{nclowi4ps}. This is important because otherwise we would run 
into the contradiction that it would not make sense to consider the  spectral-function sum-rules as a function of $N_C$.  
 The generalization to the cases with $a=3$ and $8$ is straightforward. 

\begin{figure}[H]
\begin{center}
\includegraphics[angle=0, width=0.99\textwidth]{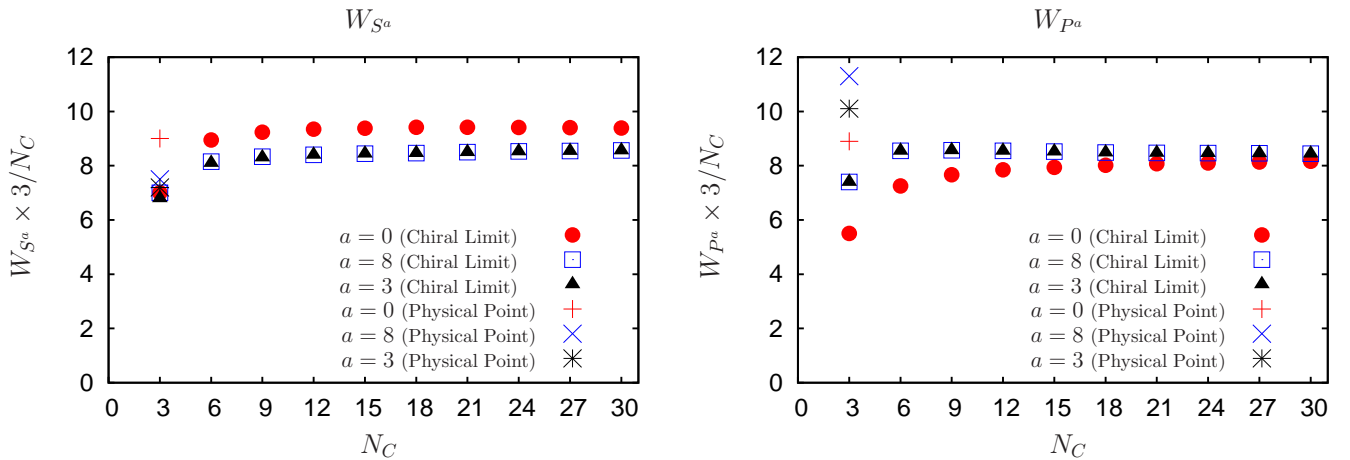}
\caption{ {\small $W_i\times 3/N_C$ as a function of $N_C$ within Scenario 3. 
We distinguish between the chiral limit and the physical case as indicated 
in the panels. We also show the corresponding value of $a$ for each point. } }
\label{fig.winc}
\end{center}
\end{figure}

We stress that we can make a close analytical discussion only at the leading order in $N_C$ and in the single channel case. 
Beyond these two conditions, it is rather difficult to perform 
the discussions analytically even in the chiral limit. Instead, our full calculations for the two-point correlators 
which include both the sub-leading $N_C$ parts, such as the chiral loops, and the multiple-channel dynamics, 
are carried out numerically. We show $W_i\times 3/N_C$ in Fig.~\ref{fig.winc} for $i=S^0,$ $S^8,$ $S^3$ (left panel) and $P^0,$ $P^8$, $P^3$ (right panel). 
The flat behavior at large $N_C$ clearly shows that all the two-point correlators $W_i$ are proportional to $N_C$ for large values of $N_C$. 
In addition, the asymptotic value is reached rather quickly with $N_C$, establishing already for  $N_C\gtrsim 10$. 
The $W_i$ results are nearly the same at large values of $N_C$ either for the $SS$ correlators or the $PP$ ones as it is required by 
the spectral-function sum rules, Eqs.~\eqref{defweinbergsr} (the discussion after Eq.~\eqref{defweinbergsr2} should be taken into account).  
To be more precise, we find that the relative variance $\sigma_W/\overline{W}$ for the six points at $N_C=30$ in Fig.~\ref{fig.winc} 
is only 5\%. This indicates that the spectral-function sum rules are better satisfied in large $N_C$ than for the $N_C=3$ case, 
reported in Table \ref{table.impi} with a relative variance of $10\%$. 

The reason behind this improvement can be attributed to the fact that at large $N_C$ the correlators reduce to the single resonance or pseudo-Goldstone exchanges 
and the large $N_C$ constraints from the $SS-PP$ sector requires that (as previously worked out in Ref.~\cite{golterman00prd})
\begin{equation}
c_m^2 = d_m^2 + \frac{F^2}{8}\,.
\end{equation}
This is perfectly satisfied by the results from our research, with $c_m$ determined from the fit 
in Eq.~\eqref{fitresultnew},  $d_m=30$~MeV adopted in this work \cite{ecker89npb,golterman00prd,juanjo10prd} and $F\simeq80$~MeV predicted from the scalar resonance parameters in the fit Eq.~\eqref{fitresultnew} using the $U(3)$ $\chi$PT one-loop calculation of  Ref.~\cite{guo11prd}. 

It is also worth considering the evolution with $N_C$ of the scalar spectral functions. In the right panel of Fig.~\ref{fig.impicl} 
the latter are shown for $N_C=30$. It is clear that the spectral functions are then completely dominated by the bare scalar pole $S_1$ 
and the octet of scalar resonances at around $\sqrt{s}=1.4$~GeV. The low energy peaks in the left panel of Fig.~\ref{fig.impicl} disappear for large $N_C$ due to 
its meson-meson dynamically generated nature, in agreement with what was already shown for the pole trajectories with $N_C$ in Sec.~\ref{rptwvn}. 

All the points in Fig.~\ref{fig.winc} are calculated by setting 
the upper limit of the integral in Eq.~\eqref{defwi} to $s_0= 3$~GeV$^2$. We verify  
that the results are quite stable for $s_0=2.5$~GeV$^2$ and $3.5$~GeV$^2$, as it was already discussed for $N_C=3$ 
in Table~\ref{table.impi}.

%%%%%%%%%%%%%%%%%%%%%%%%%%%%%%%%%%%%%%%%%%%%%%%%%%%%%%%%%%%%%%%%%%%%%%%%%%%%%%%%%%%%%%%%%%%%%%%%%%%
%%%%%%%%%%%%%%%%%%%%%%%%%%%%%%%%%%%%%%%%%%%%%%%%%%%%%%%%%%%%%%%%%%%%%%%%%%%%%%%%%%%%%%%%%%%%%%%%%%%
\section{Conclusions}
\label{sect.conclusion}
                                                                                                                                                          
In this work, we perform a complete one-loop calculation of the strangeness conserving scalar and pseudoscalar form factors 
within  $U(3)$ $\chi$PT. We employ the resonance chiral Lagrangian to include the tree level resonance exchanges, 
instead of using the LECs from the local operators at higher chiral orders (which are supposed to be saturated by the resonance 
exchanges within the assumption of resonance saturation \cite{ecker89npb}). We extend the previous one-loop calculation 
on  meson-meson scattering~\cite{guo11prd} by including the pseudoscalar resonance exchanges. 
 The full results for the scalar form factors are calculated by using the unitarization method in Refs.~\cite{oller00prd,meissner01npa}, 
 based on the N/D method, taking as input the perturbative calculations of the scalar form factors  
and the partial wave scattering amplitudes. Then, we employ different form factors to calculate the various spectral 
functions, which allows us to study the spectral-function sum rules in the scalar and pseudoscalar sectors.  

The unknown parameters in our theory are fitted to the experimental data. Comparing with the discussion in Ref.~\cite{guo11prd},
a new fit is carried out in the present work, due to the inclusion of new ingredients, 
such as the pseudoscalar resonances and the second multiplet of scalar resonances. 
Various poles in the complex energy plane for the $f_0(500)$, $f_0(980)$, $f_0(1370)$, $a_0(980)$,
$a_0(1450)$, $K^*_0(800)$, $K^*_0(1430)$, $\rho(770)$, $K^*(892)$ and $\phi(1020)$ resonances are found and
they agree well with the values from PDG~\cite{pdg}. The coupling strengths of the various resonances to 
the pseudo-Goldstone boson pairs are also calculated. 
The resonance content of the new fit presented here and that of Ref.~\cite{guo11prd} is quite similar, both at the qualitative and quantitative levels. 
 However, the new fit gives a better numerical value for the  pion scalar radius than the one in Ref.~\cite{guo11prd}. The former also contains 
the pseudoscalar resonance exchanges which implies that the spectral-function sum rules in the scalar and pseudoscalar cases are satisfactorily fulfilled at the level of around a 10\%. 
 In addition, the new fit results are employed to investigate  semilocal duality in  $\pi\pi$ scattering 
between Regge theory and the dynamics in terms of hadronic degree of freedom. We conclude that semilocal duality is well satisfied in general terms for $n\geq 1$.

An important advantage of working within the chiral Lagrangian approach is that it allows us 
to study the $N_C$ evolution of the various quantities calculated, once the $N_C$ scaling behavior of the parameters in the theory is settled.  
The leading evolution with $N_C$ is known unambiguously, while this is not the case for the sub-leading one.  
We propose three scenarios to take into account the uncertainties of the sub-leading $N_C$ scaling 
for the resonance parameters and another one to include the tensor resonances in the study of  semilocal duality. 
Interestingly, we find that  semilocal duality  at large $N_C$ 
imposes strong constraints on the resonance parameters and Scenario 3 turns out to be the best one. 
Under this scenario we then study the extrapolation of $N_C>3$ for many other quantities: the masses of the pseudo-Goldstone bosons,
the leading order $\eta-\eta'$ mixing angle, the various resonance poles and the two-point correlators. 
We find that semilocal duality and the spectral-function sum rules in the scalar and pseudoscalar cases are 
well satisfied by using the parameters from the fit to experimental data, both  for the physical case at $N_C=3$,  
 and  for varying $N_C>3$. 
 It is important to stress that the fulfillment of  semilocal duality and the spectral-function sum rules in the 
scalar and pseudoscalar sectors for both  the physical  and the large $N_C$ cases is not trivial and gives a deep insight in the 
isoscalar scalar spectrum for $\sqrt{s}\lesssim 1$~GeV. 
In the physical case, the $f_0(500)$  resonance plays an important role for both semilocal duality and spectral-function sum rules, 
dominating the strength in the low-energy scalar isoscalar channel. 
However, when $N_C$ increases this resonance moves deeper in the complex energy plane and its contribution tends to vanish. 
Instead, the $f_0(980)$ resonance, which gradually evolves to the singlet scalar bare resonance $S_1$ when increasing $N_C$,
plays the key role for the fulfillments of  semilocal duality and  spectral-function sum rules.  It is then clear that the 
physical picture for the scalar isoscalar sector in the real and large $N_C$ worlds are very different. The former case is dominated by the 
$f_0(500)$ resonance, which should not have a significant $\bar{q}q$ or glueball components as it is clear from its $N_C$ pole trajectory.
 Within our approach it is dynamically generated 
by the $\pi\pi$ self-interactions. On the other hand, in the large $N_C$ limit the lightest resonance is the singlet $S_1$ bare or pre-existing resonance 
with a mass around 0.9-1.0 GeV, which for $N_C=3$ is a component of the $f_0(980)$.

%%%%%%%%%%%%%%%%%%%%%%%%%%%%%%%%%%%%%%%%%%%%%%%%%%%%%%%%%%%%%%%%%%%%%%%%%%
\section*{Acknowledgements}

We would like to thank J.~R.~Pel\'aez for interesting discussions. 
This work is partially funded by the grants MEC  FPA2010-17806 and the Fundaci\'on S\'eneca 11871/PI/09.
 We also thank the financial support from  the BMBF grant 06BN411, the EU-Research Infrastructure
Integrating Activity ``Study of Strongly Interacting Matter" (HadronPhysics2, grant No. 227431)
under the Seventh Framework Program of EU and the Consolider-Ingenio 2010 Programme CPAN (CSD2007-00042). 
Z.H.G. acknowledges CPAN postdoc contract in the Universidad de Murcia and 
financial support from the grants National Natural Science Foundation of China (NSFC) under contract No. 11105038, 
Natural Science Foundation of Hebei Province with contract No. A2011205093 and Doctor Foundation of Hebei Normal 
University with contract No. L2010B04.

\vspace{1.0cm}

\appendix{\huge \bf Appendix}

%%%%%%%%%%%%%%%%%%%%%%%%%%%%%%%%%%%%%%%%%%%%%%%%%%%%%%%%%%%
\section{Masses and mixing parameters }
\def\theequation{\Alph{section}.\arabic{equation}}
\setcounter{equation}{0}
\label{app.newmixing}

In this section, we provide the contributions to the self-energies of the pseudo-Goldstone bosons  
that stem from the exchange of the pseudoscalar resonances, i.e. the operators in the last line of Eq.~\eqref{lagpscalarshift}, and the $\delta L_8$ 
operator in Eq.~\eqref{lagdeltal8}. 
About the second multiplet of scalar resonances, its contributions share the same  form as the lowest multiplet given in \cite{guo11prd} with the obvious 
changes in couplings and masses.
Hence, we do not reproduce them here.   
Notice that one needs to combine the results in the present section  
and those in the Appendices of Ref.~\cite{guo11prd} to get the full expressions. 

We point out that the exchange of pseudoscalar resonances and the $\delta L_8$ operator do not contribute to the wave function renormalization 
 of the pseudo-Goldstone bosons and the kinetic terms of $\overline{\eta}$-$\overline{\eta}'$ mixing parameterized by  
$\delta_{\eta}$, $\delta_{\eta'}$ and $\delta_{k}$ in Eq.~(14) of Ref.~\cite{guo11prd}. 
For the pion and kaon masses these new contributions  read
\begin{align} 
\label{newmpi2}
m_\pi^2 &= \overline{m}_\pi^2 + \frac{16 m_\pi^4}{F_\pi^2 }\big( \delta L_8 - \frac{d_m^2}{2 M_{P_8}^2}  \big)\,,  \\
\label{newmk2}
m_K^2 &= \overline{m}_K^2 + \frac{16 m_K^4}{F_\pi^2 }\big( \delta L_8 - \frac{d_m^2}{2 M_{P_8}^2}  \big)\,. 
\end{align}
The mass terms in the $\overline{\eta}$-$\overline{\eta}'$ mixing are: 
\begin{align}
\label{newmeta2}
 \delta_{m_{\overline{\eta}}^2} &=  - \frac{8 d_m^2[ c_\theta^2 (m_\pi^2 - 4m_K^2)^2
+ 4\sqrt{2} c_\theta s_\theta(4m_K^4 - 5m_K^2 m_\pi^2 + m_\pi^4) + 8s_\theta^2 (m_\pi^2 - m_K^2)^2 ] }{9 F_\pi^2 M_{P_8}^2} 
\nonumber \\ &
- \frac{8 \widetilde{d}_m^2[ 8c_\theta^2 (m_\pi^2 - m_K^2)^2  
+ 4\sqrt{2} c_\theta s_\theta(2m_K^4 - m_K^2 m_\pi^2 - m_\pi^4) + s_\theta^2 ( m_\pi^2 + 2m_K^2)^2 ] }{3 F_\pi^2 M_{P_1}^2} 
\nonumber \\ &
+ \frac{16 \,\delta L_8}{3 F_\pi^2}\bigg[  c_\theta^2 (8m_K^4 - 8 m_K^2 m_\pi^2 +3 m_\pi^4) 
+ 8\sqrt{2} c_\theta s_\theta m_K^2(m_K^2 - m_\pi^2) 
\nonumber \\ & \qquad \qquad \,\,
+  s_\theta^2 (4m_K^4 - 4 m_K^2 m_\pi^2 +3 m_\pi^4)   \bigg]\,, 
\end{align}
\begin{align}
 \label{newmetap2}
 \delta_{m_{\overline{\eta}'}^2} &=  - \frac{8 d_m^2[ 8c_\theta^2 (m_\pi^2 - m_K^2)^2
- 4\sqrt{2} c_\theta s_\theta(4m_K^4 - 5m_K^2 m_\pi^2 + m_\pi^4) + s_\theta^2 (m_\pi^2 - 4m_K^2)^2 ] }{9 F_\pi^2 M_{P_8}^2} 
\nonumber \\ &
- \frac{8 \widetilde{d}_m^2[ c_\theta^2 (m_\pi^2 + 2m_K^2)^2  
- 4\sqrt{2} c_\theta s_\theta( 2m_K^4 - m_K^2 m_\pi^2 - m_\pi^4) + 8s_\theta^2 ( m_\pi^2 - m_K^2)^2 ] }{3 F_\pi^2 M_{P_1}^2} 
\nonumber \\ &
+ \frac{16 \,\delta L_8}{3 F_\pi^2}\bigg[  c_\theta^2 (4m_K^4 - 4 m_K^2 m_\pi^2 +3 m_\pi^4) 
- 8\sqrt{2} c_\theta s_\theta m_K^2(m_K^2 - m_\pi^2) 
 \nonumber \\ & \qquad \qquad \,\,
+  s_\theta^2 (8m_K^4 - 8 m_K^2 m_\pi^2 +3 m_\pi^4)   \bigg]\,,
\end{align}
\begin{align}
\label{newm2}
 \delta_{m^2} &=   \frac{8 d_m^2}{9 F_\pi^2 M_{P_8}^2}  \bigg[ 2\sqrt{2}c_\theta^2 (4m_K^4 - 5m_K^2 m_\pi^2 + m_\pi^4)
+ c_\theta s_\theta( -8m_K^4 - 8m_K^2 m_\pi^2 + 7m_\pi^4) 
 \nonumber \\ & \qquad \qquad \,\,
-2\sqrt{2} s_\theta^2 (4m_K^4 - 5m_K^2 m_\pi^2 + m_\pi^4) \bigg] 
\nonumber \\&
+ \frac{8 \widetilde{d}_m^2}{3 F_\pi^2 M_{P_1}^2} \bigg[ 2\sqrt{2}c_\theta^2 (2m_K^4 - m_K^2 m_\pi^2 - m_\pi^4)
+ c_\theta s_\theta( -4m_K^4 +20m_K^2 m_\pi^2 - 7m_\pi^4) 
 \nonumber \\ & \qquad \qquad \,\,
+ 2\sqrt{2} s_\theta^2 (-2m_K^4 + m_K^2 m_\pi^2 + m_\pi^4) \bigg] 
\nonumber \\ &
+ \frac{64 \,\delta L_8 m_K^2 (m_\pi^2 - m_K^2) (\sqrt{2}c_\theta^2 -c_\theta s_\theta -\sqrt{2}s_\theta^2 )}{3 F_\pi^2}\,,
\end{align}
where $\delta_{m_{\overline{\eta}}^2}$, $\delta_{m_{\overline{\eta}'}^2}$ and $\delta_{m^2}$ are 
defined in Eq.~(14) of Ref.~\cite{guo11prd}. Notice that the two operators in the last line of the pseudoscalar resonance Lagrangian in Eq.~\eqref{lagpscalarshift} 
and the $\delta L_8$ operator of Eq.~\eqref{lagdeltal8} do not contribute to the pion decay constant $F_\pi$. 

%%%%%%%%%%%%%%%%%%%%%%%%%%%%%%%%%%%%%%%%%%%%%%%%%%%%%%%%%%%%%%%%%%%%%%%%%%%%%%%%%%%%%%%
%%%%%%%%%%%%%%%%%%%%%%%%%%%%%%%%%%%%%%%%%%%%%%%%%%%%%%%%%%%%%%%%%%%%%%%%%%%%%%%%%%%%%%%%%%%%%%%
\section{Analytic expressions for scalar and pseudoscalar form factors}
\label{app.sff}

In order to calculate the scalar form factors in the isospin basis of Eqs.~\eqref{ffisospin0norm} and \eqref{ffisospin1norm},  
 we need to evaluate for the isoscalar case eight form factors in the charged bases, namely: 
$F_{\pi^0\pi^0}^{0,8}$, $F_{\pi^+\pi^-}^{0,8}$, $F_{\pi^-\pi^+}^{0,8}$, 
$F_{K^+K^-}^{0,8}$, $F_{K^0 \overline{K}^0}^{{0,8}}$, $F_{\eta\eta}^{0,8}$, 
$F_{\eta\eta'}^{0,8}$ and $F_{\eta'\eta'}^{0,8}$. For the isovector case, one needs: $F_{\pi^0\eta}^{3}$, $F_{K^+K^-}^{3}$, $F_{K^0 \overline{K}^0}^{3}$ 
and $F_{\pi^0\eta'}^{3}$. 

The leading order contributions to the scalar form factors defined in Eq.~\eqref{defsff} stem 
from the Lagrangian in Eq.~\eqref{lolagrangian}. 
The expressions for $a=0$ read: 
\begin{align}\label{analylosffs0}
F_{\pi^0\pi^0}^{0} &= F_{\pi^+\pi^-}^{0} = F_{\pi^-\pi^+}^{0} = 2 \sqrt{\frac{2}{3}}\,, \nonumber \\
F_{K^+ K^-}^{0} &= F_{K^0 \overline{K}^0}^{0} = 2 \sqrt{\frac{2}{3}} \,, \nonumber \\
F_{\eta \eta}^{0} &= 2 \sqrt{\frac{2}{3}}\,, \nonumber \\
F_{\eta \eta'}^{0} &= 0\,,  \nonumber \\
F_{\eta' \eta'}^{0}  &= 2 \sqrt{\frac{2}{3}}\,. 
\end{align}
For  $a=8$ one has:
\begin{align}\label{analylosffs8}
F_{\pi^0\pi^0}^{8} &= F_{\pi^+\pi^-}^{8} = F_{\pi^-\pi^+}^{8} = \frac{2}{\sqrt{3}}\,, \nonumber \\
F_{K^+ K^-}^{8} &= F_{K^0 \overline{K}^0}^{8} = -\frac{1}{\sqrt{3}} \,, \nonumber \\
F_{\eta \eta}^{8} &=  -\frac{2 c_\theta(c_\theta + 2\sqrt{2}s_\theta) }{\sqrt{3}}\,, \nonumber \\
F_{\eta \eta'}^{8} &= \frac{ 2 (\sqrt{2}c_\theta^2 - c_\theta s_\theta -  \sqrt{2}s_\theta^2 ) }{\sqrt{3}}\,,  \nonumber \\
F_{\eta' \eta'}^{8}  &= \frac{2 s_\theta( 2\sqrt{2}c_\theta - s_\theta) }{\sqrt{3}}\,. 
\end{align}
And the results for $a=3$ are:
\begin{align}\label{analylosffs3}
F_{\pi^0\eta}^{3} &=  \frac{2(c_\theta - \sqrt{2}s_\theta)}{\sqrt{3}}\,, \nonumber \\
F_{K^+ K^-}^{3} &= - F_{K^0 \overline{K}^0}^{3} = 1 \,, \nonumber \\
F_{\pi^0\eta'}^{3} &=  \frac{2(\sqrt{2}c_\theta + s_\theta)}{\sqrt{3}}\,. 
\end{align}

The scalar form factors in the quark flavor basis defined in Eqs.~\eqref{defsffuuddss} at leading order are 
\begin{align}
F_{\pi^0\pi^0}^{\bar{u}u+\bar{d}d} &= F_{\pi^+\pi^-}^{\bar{u}u+\bar{d}d} = F_{\pi^-\pi^+}^{\bar{u}u+\bar{d}d} = 2 \,, \nonumber \\
F_{K^+ K^-}^{\bar{u}u+\bar{d}d} &= F_{K^0 \overline{K}^0}^{\bar{u}u+\bar{d}d} = 1  \,, \nonumber \\
F_{\eta \eta}^{\bar{u}u+\bar{d}d} &= -\frac{2}{3}(c_\theta^2 + 2\sqrt{2}c_\theta s_\theta -2 )\,, \nonumber \\
F_{\eta \eta'}^{\bar{u}u+\bar{d}d} &= \frac{2}{3}( \sqrt{2}c_\theta^2 - c_\theta s_\theta - \sqrt{2} s_\theta^2 )\,, \nonumber  \\
F_{\eta' \eta'}^{\bar{u}u+\bar{d}d}  &= \frac{2}{3}( -s_\theta^2 + 2\sqrt{2}c_\theta s_\theta + 2 )\,, 
\end{align}
and
\begin{align}
F_{\pi^0\pi^0}^{\bar{s}s} &= F_{\pi^+\pi^-}^{\bar{s}s} = F_{\pi^-\pi^+}^{\bar{s}s} = 0 \,, \nonumber \\
F_{K^+ K^-}^{\bar{s}s} &= F_{K^0 \overline{K}^0}^{\bar{s}s} = 1  \,, \nonumber \\
F_{\eta \eta}^{\bar{s}s} &= \frac{2}{3}(c_\theta^2 + 2\sqrt{2}c_\theta s_\theta + 1 )\,, \nonumber \\
F_{\eta \eta'}^{\bar{s}s} &= -\frac{2}{3}( \sqrt{2}c_\theta^2 - c_\theta s_\theta - \sqrt{2} s_\theta^2 )\,,  \nonumber \\
F_{\eta' \eta'}^{\bar{s}s}  &= \frac{2}{3}( s_\theta^2 - 2\sqrt{2}c_\theta s_\theta + 1 )\,. 
\end{align}

At the same order, the pseudoscalar form factors for the pseudo-Goldstone bosons defined in Eq.~\eqref{defpff} are 
\begin{eqnarray}\label{pssffpsgoldstone}
H^{0}_{\pi^0} =  0\,, \quad  &
H^{0}_{\eta}  =  - 2 F_\pi \, s_\theta\,, \quad  &
H^{0}_{\eta'} =  2 F_\pi \, c_\theta\,, \nonumber \\ 
H^{8}_{\pi^0} =  0\,, \quad  &
H^{8}_{\eta}  =   2 F_\pi \, c_\theta\,,  \quad &
H^{8}_{\eta'} =  2 F_\pi \, s_\theta\,, \nonumber \\ 
H^{3}_{\pi^0} =  2 F_\pi \,,\quad &
H^{3}_{\eta}  =  0\,,  \quad &
H^{3}_{\eta'} =   0\,, 
\end{eqnarray}
and for the pseudoscalar resonances they read 
\begin{align}\label{pssffpsres}
H^{0}_{\rm Pseudoscalar\,Resonances} &=  - 4\sqrt{6} \widetilde{d}_m\,, \nonumber \\ 
H^{8}_{\rm Pseudoscalar\, Resonances} &=  - 4\sqrt{2} d_m\,, \nonumber \\ 
H^{3}_{\rm Pseudoscalar\, Resonances} &=  - 4\sqrt{2} d_m\,.  
\end{align}

We provide  the remaining expressions, such as those from the chiral loops, scalar resonances, pseudoscalar resonances, $\Lambda_2$ and 
$\delta L_8$,  in the Mathematica code \cite{mathcode2}. 

%%%%%%%%%%%%%%%%%%%%%%%%%%%%%%%%%%%%%%%%%%%%%%%%%%%%%%%%%%%%%%%%%%%%%%%%%%%%%%%%%%%%%%%
%%%%%%%%%%%%%%%%%%%%%%%%%%%%%%%%%%%%%%%%%%%%%%%%%%%%%%%%%%%%%%%%%%%%%%%%%%%%%%%%%%%%%%%%%%%
\section{Tensor resonances in meson-meson scattering} 
\label{app.tensor}

We follow the framework proposed in Ref.~\cite{ecker07epjc} to include the tensor resonances in meson-meson scattering. 
The relevant Lagrangian reads 
\begin{equation}\label{lagtensor}
 \mathcal{L}_{T} = -\frac{1}{2} \bra T_{\mu\nu} D_{\rm T}^{\mu\nu,\rho\sigma} T_{\rho\sigma} \ket
+ g_T \bra T_{\mu\nu} \{u^\mu, u^\nu\} \ket + \beta \bra T_{\mu}^{\mu} u_\nu u^\nu \ket  
+ \gamma \bra T_{\mu}^{\mu} \chi_+ \ket\,,
\end{equation}
where the nonet of the tensor resonances are collected in the matrix 
\begin{align}\label{tensor9}
T_{\mu\nu} &= \left(
                                        \begin{array}{ccc}
                                          \frac{a_2^0}{\sqrt2}+\frac{f_2^8}{\sqrt6} +\frac{f_2^1}{\sqrt3} & a_2^+ & K_2^{*+}  \\
                                          a_2^- & -\frac{a_2^0}{\sqrt2}+\frac{f_2^8}{\sqrt6} +\frac{f_2^1}{\sqrt3}  & K_2^{*0}  \\
                                          K_2^{*-} & \bar{K}_2^{*0} & -\frac{2 f_2^8}{\sqrt6} +\frac{f_2^1}{\sqrt3} \\
                                        \end{array}
                                       \right)_{\mu\nu} \,.
\end{align}
See Ref.~\cite{ecker89npb} for the definition of the remaining chiral building blocks. 
Like in the vector resonance case~\cite{guo11prd}, ideal mixing is also assumed for the tensor resonances: 
\begin{align}
 f_2(1270) &= \sqrt{\frac{2}{3}}f_2^1 + \sqrt{\frac{1}{3}}f_2^8~,\nonumber \\
 f_2' &= \sqrt{\frac{1}{3}}f_2^1 - \sqrt{\frac{2}{3}}f_2^8 ~.
\end{align}
In the present work, the purpose to introduce the tensor resonances is to discuss  semilocal duality 
in $\pi\pi$ scattering. So we will set equal all the tensor masses, which is determined by the most relevant  resonance $f_2(1270)$. 
 The first operator in Eq.~\eqref{lagtensor} corresponds to the kinetic term while the remaining terms describe the interactions
 between the tensor resonances and the pseudo-Goldstone boson pairs. 
The terms proportional to $\beta$ and $\gamma$ do not contribute to the on-shell decay of the tensor resonances  because the tensor field is
traceless in the space-time indices~\cite{ecker07epjc}. Nevertheless, they do contribute to meson-meson scattering. 
It is argued in Ref.~\cite{ecker07epjc} that though  
the final result is independent of the choice of $\beta$, it is convenient to set $\beta = - g_T$ to avoid the inclusion 
of the $\cO(p^6)$ LECs in order to fulfill the high energy constraints for the forward $\pi\pi$ scattering.   
 The $\gamma$ operator is always accompanied by quark masses and its effects should be much less important than 
the $\beta$ term. So we will omit the former term throughout as done in Ref.~\cite{ecker07epjc}.

Last but not least, one should guarantee the right high energy behavior of  meson-meson scattering in the presence of tensor resonances resulting from the Lagrangian in Eq.~\eqref{lagtensor}. 
The high energy constraint imposed in Ref.~\cite{ecker07epjc} concerns the fulfillment of a once subtracted forward dispersion relation 
 for $\pi^+\pi^0$ elastic scattering, 
which turns out to play a crucial role to get the correct prediction for the LECs. 
We follow the same approach here to satisfy this high energy constraint. Moreover we calculate all the relevant 
meson-meson coupled channels for scattering in $U(3)$ $\chi$PT, not just  $\pi\pi$, since the other processes 
can enter through the unitarization procedure~\cite{guo11prd} that couples the different states with the same quantum numbers.   

The final results on meson-meson scattering contributed by the tensor resonances and the high 
energy constraints are given in the Mathematica code \cite{mathcode2}.   
Due to the consideration of the scattering processes involving $\eta$ and $\eta'$, additional LECs are needed 
to guarantee the proper short distance constraint imposed by the forward scattering. 
The short distance constraints that we find here are 
\begin{eqnarray}
\beta_{13}^{SD} = \frac{4 g_T^2}{M_T^2}\,, \quad \beta_{14}^{SD} = -\frac{g_T^2}{M_T^2}\,,  \quad
\beta_{15}^{SD} = -\frac{2 g_T^2}{M_T^2}\,, \quad \beta_{16}^{SD} = -\frac{g_T^2}{2M_T^2}\,, 
\label{u3.const}
\end{eqnarray}
in addition to the ones determined in Ref.~\cite{ecker07epjc}
\begin{eqnarray}
\beta_{1}^{SD} = -\frac{g_T^2}{2M_T^2}\,, \quad \beta_{2}^{SD} = -\frac{g_T^2}{M_T^2}\,, \quad
\beta_{3}^{SD} = \frac{2 g_T^2}{M_T^2}\,. 
\end{eqnarray}
 The convention to label the LECs in Eq.~\eqref{u3.const} is the same as in Ref.~\cite{herrera97npb}, where one can also 
find the corresponding monomials multiplied by the coefficients with  $i = 1,$ $2,$ $3,$ $13,$ $14,$ $15,$ and $16$.  
We checked that only after the  short distance constraints are fulfilled, 
the introduction of the tensor resonances does not spoil the fit that we obtained in Ref.~\cite{guo11prd} and 
the new one in Eq.~\eqref{fitresultnew}. 

For the value of $g_T$, we take the result $g_T=28$~MeV from Ref.~\cite{ecker07epjc}. 
 As in the vector channels~\cite{guo11prd}, due to the dominant role played by the tree level tensor resonances, 
 one does not expect the subtraction constants arising from the unitarization procedure to  
play an important role in $IJ=02$. We simply fix their values  to the $IJ=00$ channel and checked that the results are stable 
under ${\cal O}(1)$ changes  in these numbers, so that the subtraction constants keep natural values in Eq.~\eqref{fitresultnew}.   
The bare mass of the tensor resonances is adjusted to $M_T=1300$~MeV, leading to the pole position $\sqrt{s}=(1275.2-i\,75.8)$~MeV, 
which is close to the values of the $f_2(1270)$ resonance given in the PDG~\cite{pdg}.

\end{document}